\documentclass[notitlepage]{revtex4-1}
\usepackage{setspace}
\usepackage[utf8]{inputenc}
\usepackage[dvipsnames]{xcolor}
\usepackage[colorlinks=true,citecolor=RedViolet,urlcolor=BlueGreen,linkcolor=RedViolet]{hyperref}
\usepackage[normalem]{ulem}
\usepackage{url}
\usepackage{graphicx,wrapfig,float,slashed,cancel}
\usepackage{amsmath,amssymb,epsfig,xcolor,stmaryrd}
\usepackage{bm}
\usepackage{enumitem}
\usepackage{hhline,multirow,tabularx} 
\usepackage{graphics}
\usepackage{latexsym}
\usepackage{rotating}
\usepackage{subfigure}
\usepackage{color}
\usepackage{blindtext}
\usepackage{lipsum}
\usepackage{multirow}
\usepackage{orcidlink}

	\allowdisplaybreaks
\begin{document}
	
	
	\title{\textcolor{BlueViolet}{Properties of the ground and excited states of triply heavy spin-1/2 baryons}}

	\author{Z.~Rajabi Najjar$^{a}$\orcidlink{0009-0002-2690-334X}}
	\email{rajabinajar8361@ut.ac.ir }
	
	\author{K.~Azizi$^{a,b}$\orcidlink{0000-0003-3741-2167}}
	\email{kazem.azizi@ut.ac.ir}
	\thanks{Corresponding author}

	\author{H.~R.~Moshfegh$^{a,c}$\orcidlink{0000-0002-9657-7116}}
	\email{hmoshfegh@ut.ac.ir }

	\affiliation{
		$^{a}$Department of Physics, University of Tehran, North Karegar Avenue, Tehran 14395-547, Iran\\
			$^{b}$Department of Physics, Do\v{g}u\c{s} University, Dudullu-\"{U}mraniye, 34775
		Istanbul, T\"{u}rkiye\\
    	$^{c}$Departamento de F\'{i}sica, Pontif\'{i}cia Universidade Cat\'{o}lica do Rio de Janeiro, Rio de Janeiro 22452-970, Brazil	
	}

	\date{\today}
	
	\preprint{}
	
	\begin{abstract}
We study  the triply heavy spin-1/2 baryons with quark contents $ ccb $ and $ bbc $, and  calculate their mass and residue using QCD sum rules. In the calculations, we consider the ground (1S), first orbitally excited (1P)  and first radially excited (2S) states. Aiming to achieve higher accuracies in the results, we perform  the computations by taking into account the non-perturbative operators  up to eight mass dimensions. We compare our results with the predictions of  other theoretical studies existing in the literature. The obtained results may help experimental groups in their search for these yet unseen, but previously predicted by the quark model, interesting particles.

	\end{abstract}
	
	
	\maketitle
	
	\renewcommand{\thefootnote}{\#\arabic{footnote}}
	\setcounter{footnote}{0}
	
	\section {Introduction}\label{sec:one}
	The investigation of baryons consist of heavy quarks has been one of the important directions of research in non-perturbative quantum chromodynamics (QCD). The prosperous quark model predicts the existence of three types of heavy baryons comprising single, double, or triple heavy quarks. The accessible literature predominantly focuses on the single heavy baryons. In the last couple of decades, the various experimental groups such as CDF, CLEO, BABAR, BELLE, BESIII and LHCb have discovered many ground and excited states of single  heavy baryons like $\Lambda_{b(c)}$, $\Sigma_{b(c)}^{(*)}$, $\Xi_{b(c)}^{(*,\prime)}$ and $\Omega_{b(c)}^{(*)}$  listed in particle data group (PDG) summary tables  \cite{ParticleDataGroup:2022pth}. The discovery of $\Xi_{cc}$, by the SELEX collaboration \cite{SELEX:2002wqn,SELEX:2004lln}  and its confirmation by   LHCb  \cite{LHCb:2017iph,LHCb:2018pcs}, marked a significant breakthrough in the pursuit of doubly heavy baryons. Thus far, none of the triply heavy baryons representing the last group of standard heavy baryons have been reported experimentally. Though they are in the focus of some experimental groups, compared to the single and doubly heavy baryons, there is less consideration dedicated to the identification of these states. Motivated by the opportunity of detecting triply heavy baryons in the experiment on the horizon, the various theoretical approaches are carried out on the properties of triply heavy baryons such as non-relativistic quark model \cite{Roberts:2007ni,Patel:2008mv,Vijande:2015faa,Shah:2017jkr,Shah:2018div,Shah:2018bnr,Liu:2019vtx}, relativistic quark model \cite{Migura:2006ep,Martynenko:2007je,Yang:2019lsg,Faustov:2021qqf}, lattice QCD \cite{Meinel:2010pw,Meinel:2012qz,Briceno:2012wt,Padmanath:2013zfa,Brown:2014ena,Mathur:2018epb,Can:2015exa}, the QCD sum rules \cite{Zhang:2009re,Wang:2011ae,Aliev:2012tt,Aliev:2014lxa,Wang:2020avt}, Bag model \cite{Hasenfratz:1980ka,Bernotas:2008bu}, the Regge trajectories \cite{Wei:2015gsa,Wei:2016jyk,Oudichhya:2021yln,Oudichhya:2021kop,Oudichhya:2023pkg}, Fadeev equation \cite{Radin:2014yna,Gutierrez-Guerrero:2019uwa,Qin:2019hgk,Yin:2019bxe}, Hyperspherical Harmonics method \cite{Zhao:2023qww}, various potential models \cite{Silvestre-Brac:1996myf,Jia:2006gw,Brambilla:2009cd,Llanes-Estrada:2011gwu,Flynn:2011gf,Thakkar:2016sog,Serafin:2018aih,Shah:2023zph}, etc. Taking into consideration all the theoretical approaches that have been pointed so far, the QCD sum rule method is considered as one of the most powerful and predictive non-perturbative analytical tools in predicting the properties of heavy hadrons  \cite{Aliev:2009jt,Aliev:2010uy,Aliev:2012ru,Agaev:2016mjb,Azizi:2016dhy}, and associated predictions are confirmed with the worldwide accelerator experiments for many of hadrons. So far, the theoretical analysis has desired to concentrate on the masses and residues of the ground state triply heavy baryons  more than the excited states. On the other hand, the predicted  masses have been mainly reported with higher uncertainties giving large mass range for each member. Consequently, more comparisons of theoretical predictions are required in the mass and residue spectra of triply heavy baryons, which stimulates us to calculate them not only in the ground but also in the orbitally and radially exited states; and both in pole and $\overline{MS} $ schemes for heavy quarks.
	With the aim of achieving higher accuracies in the calculations, we consider the non-perturbative operators up to eight mass dimensions.
	In light of this, the present manuscript is organized as follows: In Sec. \ref{sec:two}, the formulation of the QCD sum rules was utilized to determine the masses and residues of the triply heavy 1S, 1P and 2S baryons. Sec. \ref{sec:three}, is dedicated to our numerical analysis, along with a comparison to other theoretical predictions in the literature, and Sec. \ref{sec:four}, pertains to a review and our conclusions. We move some lengthy expressions obtained from the calculations to the Appendix.
	
	\section {Spectroscopic parameters of the triply heavy Spin--1/2 Baryons }\label{sec:two}
	
	The spectroscopic parameters, mass and residue, can be elicited via the QCD sum rule method \cite{Shifman:1978bx,Shifman:1978by,Ioffe81}. To achieve these quantities, one must utilize an appropriate correlation function. In this regard, we use the following two-point correlation function:
	\begin{equation}
		\Pi(q)=i\int d^{4}xe^{iqx}\langle 0|\mathcal{T}\{\eta(x)\bar{\eta}(0)\}|0\rangle,  \label{eq:CF1}
	\end{equation}
	where $\eta (x)$ is the interpolating current for the baryons subjected to evaluations, which is the mathematical demonstration of particles. The symbol $\mathcal{T}$ denotes the time ordering of two $\eta (x)$  and $\bar{\eta}(0)$ interpolating currents and $q$ is the
	four-momentum of the relating triply heavy baryons. In the present work, the most general form of the interpolation current is considered:
	\begin{equation}\label{cur1}
		\eta(x)=2\varepsilon^{abc}\Big\{\Big(Q^{aT}(x)CQ^{'b}(x)\Big)\gamma_{5}Q^c(x)+
		\beta\Big(Q^{aT}(x)C\gamma_{5}Q^{'b}(x)\Big)Q^c (x)\Big\},
	\end{equation}
	where $Q$ and $Q^{'}$ illustrate the heavy quark $b$ or $c$ ($Q\neq Q^{'} $ for the triply heavy spin-1/2 baryons), $a$, $b$ and $c$ are color indices, $C$ is the charge conjugation operator and  $\beta$ is an arbitrary auxiliary parameter which for the Ioffe current is expressed as $\beta = -1$. We will fix the working region of $\beta$ when performing numerical analyses. The members of  triply heavy spin-1/2 baryons anticipated by the quark model are seen in Table~\ref{tab1}. 
	
	\begin{table}[htb]
		\begin{center}
			\begin{tabular}{|c|c|c|}\hline\hline
				Baryon        &$Q$ &  $Q'$    \\ \hline
				$\Omega_{bbc}$ &  $b$           &$c$    \\ \hline
				$\Omega_{ccb}$ &  $c$           &$b$     \\ 	\hline\hline
			\end{tabular}
		\end{center}
		\caption{The members of the triply heavy spin-1/2 baryons. }
		\label{tab1}
	\end{table}
	
	In the QCD sum rule approach, one initially has to  determine the above correlation function in two distinctive sides: i) hadronic side, which is calculated by including the hadronic parameters in the time-like region. The obtained results of this procedure include the mass and residue of the states under investigation. ii) QCD side, which is calculated by including the quark and gluon degrees of freedom in the space-like region. Extracted results of this side contain the gluon condensates of various dimensions representing the interaction of gluons with QCD vacuum, QCD coupling constant, the masses of the quarks and other corresponding parameters. By relating the results of these two representations, via  dispersion integrals and by using the quark-hadron duality  assumption, one can calculate the masses and residues in terms of other parameters. We apply  Borel transformation and continuum subtraction techniques to suppress the contributions of the higher states and continuum. The QCD sum rules for the physical quantities are obtained by matching the coefficients of the Lorentz structures entering the calculations.

	The initial point  to obtain the correlation function in terms of hadronic parameters is to insert relevant complete  sets, which have the same quantum numbers as the interpolating current into the adequate locations. As we aim to consider first three resonances, we use the” ground state $+$ orbitally excited state $+$ radially excited state $+$ continuum” scheme. After fulfilling this duty and performing the integrals over four-$x$, we can write the hadronic or phenomenological representation of the correlation function as:
	\begin{eqnarray}
		\Pi^{\mathrm{Had}}(q)&=&\frac{\langle0|\eta(0)|B_{QQQ^{'}}(q,s)\rangle\langle B_{QQQ^{'}}(q,s)|\bar{\eta}(0)|0\rangle}{m^2-q^2}
		+\frac{\langle0|\eta(0)|\tilde{B}_{QQQ^{'}}(q,s)\rangle\langle\tilde{B}_{QQQ{'}}(q,s)|\bar{\eta}(0)|0\rangle}{\tilde{m}^2-q^2}+\nonumber\\
		&+&\frac{\langle0|\eta(0)|B_{QQQ{'}}'(q,s)\rangle\langle B_{QQQ{'}}'(q,s)|\bar{\eta}(0)|0\rangle}{m'{}^2-q^2}+\cdots.
		\label{Eq:cor:Phys}
	\end{eqnarray}
	The $|B_{QQQ^{'}}(q,s)\rangle$, $|\tilde{B}_{QQQ^{'}}(q,s)\rangle$ and $|B_{QQQ^{'}}'(q,s)\rangle$ states are utilized to represent the various baryonic one-particle states: The ground $(1S)$, the first orbital excitation  $(1P)$ and the first radial excitation $(2S)$, respectively. Here, symbols $m$, $\tilde{m}$ and $m'$ are  the corresponding masses and dots  indicate an abbreviation of  the contribution of
	the higher states and continuum. The matrix elements of the interpolating current between the vacuum and the baryonic states under study in Eq.~(\ref{Eq:cor:Phys})  are determined as follows:
	\begin{eqnarray}
		\langle 0|\eta(0)|B_{QQQ^{'}}(q,s)\rangle&=&\lambda u(q,s),\nonumber\\
		\langle 0|\eta(0)|\tilde{B}_{QQQ^{'}}(q,s)\rangle&=&\tilde{\lambda}\gamma_5 \tilde{u}(q,s),\nonumber\\
		\langle 0|\eta(0)|B_{QQQ^{'}}'(q,s)\rangle&=&\lambda' u'(q,s),
		\label{Eq:Matrixelm}
	\end{eqnarray}  
	where $\lambda$, $\tilde{\lambda}$ and $\lambda'$ are residues of the considered states;  and  $u(q,s)$, $\tilde{u}(q,s)$ and  $u'(q,s)$ represent the corresponding Dirac spinors with spin~$s$,  satisfying  the following
	identity:
	\begin{eqnarray}
		\sum_{s}u(q,s)\bar{u}(q,s)=(\not\!q+m).
		\label{Eq:sumspin}
	\end{eqnarray}  
	By inserting  Eq.~\eqref{Eq:Matrixelm} into Eq.~\eqref{Eq:cor:Phys} and performing summations over the spins of $ B_{QQQ^{'}}$
	, the hadronic  representation of the correlation function in momentum-space is found:
	\begin{eqnarray}
		\Pi^{\mathrm{Had}}(q)=\frac{\lambda^2(\not\!q+m)}{m^2-q^2}+\frac{\tilde{\lambda}^2(\not\!q-\tilde{m})}{\tilde{m}^2-q^2}+\frac{\lambda'^2(\not\!q+m')}{m'{}^2-q^2}+\cdots.
		\label{Eq:cor:Phys1}
	\end{eqnarray}
	Eventually, in order to unveil the final configuration of  the hadronic correlation function, one  uses the Borel transformation to elevate the contribution of the three first resonances  and  suppress  the contributions of higher states and continuum. To this end, we apply
	\begin{equation}
		\label{eq14}
		\Pi^{\mathrm{Had}}(M^2)=\lim_{\stackrel{Q^2,n \to \infty}{Q^2/n=M^2}}
		\frac{(-q^2)^{n}}{(n-1)!}\left( \frac{d}{dq^2}\right)^{n-1} \Pi^{\mathrm{Had}}(q),
	\end{equation}
where $M^2$ is the Borel parameter and $ Q^2=-q^2 $. This leads to
	\begin{eqnarray}
		\Pi^{\mathrm{Had}}(M^2)=\lambda^2(\not\!q+m)e^{-\frac{m^2}{M^2}}+\tilde{\lambda}^2(\not\!q-\tilde{m})e^{-\frac{\tilde{m}^2}{M^2}}+\lambda'^2(\not\!q+m')e^{-\frac{m'{}^2}{M^2}}+\cdots. 
		\label{Eq:cor:Fin}
	\end{eqnarray}
Due to the presence of the  exponential function in the above equation,  the higher the value of the mass of the resonance, the lower is its contribution.  We will also apply the continuum subtraction procedure supplied by the quark-hadron duality assumption in next steps that further suppresses the contributions of the higher resonances and continuum. Hence, we keep only the  first three resonances  and, in the numerical analyses, we will show that the main contribution in the correlation function belongs to these first three resonances.
As is seen, we  have only two independent Lorentz structures:  $\not\!q$ and the unit matrix  $I$, which are used to calculate the masses and residues of the relevant  states.
	
\begin{figure}[!h]
		\centering
		\subfigure[{\,dim-0 (perturbative)}]{\label{a1}
			\includegraphics*[width=.12\textwidth]{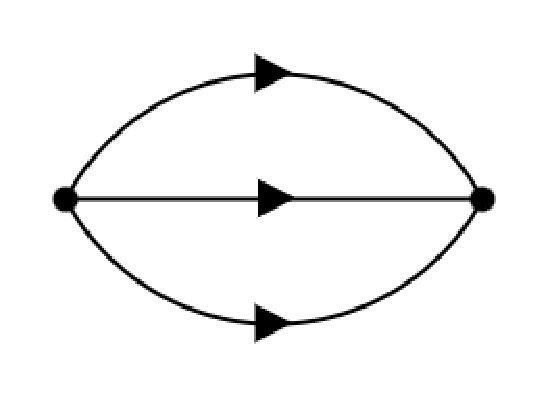}}
		\hspace{8mm}
		\subfigure[{\,dim-4 }]{\label{a2}
			\includegraphics*[width=.12\textwidth]{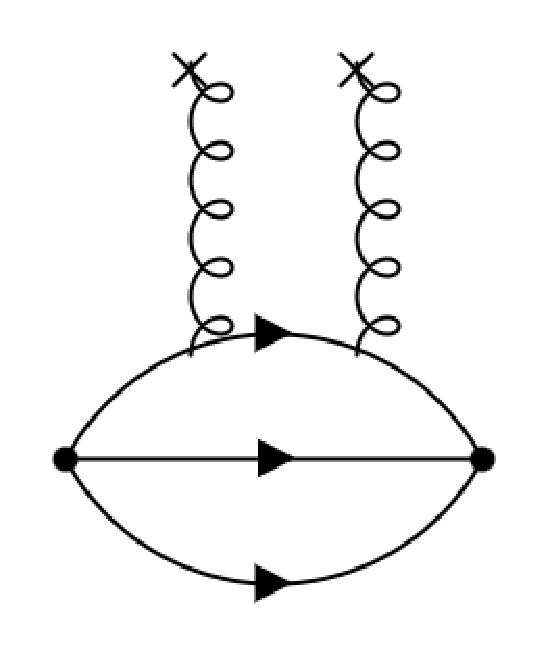}}
		\hspace{8mm}
		\subfigure[{\,dim-4}]{\label{a3}
			\includegraphics*[width=.12\textwidth]{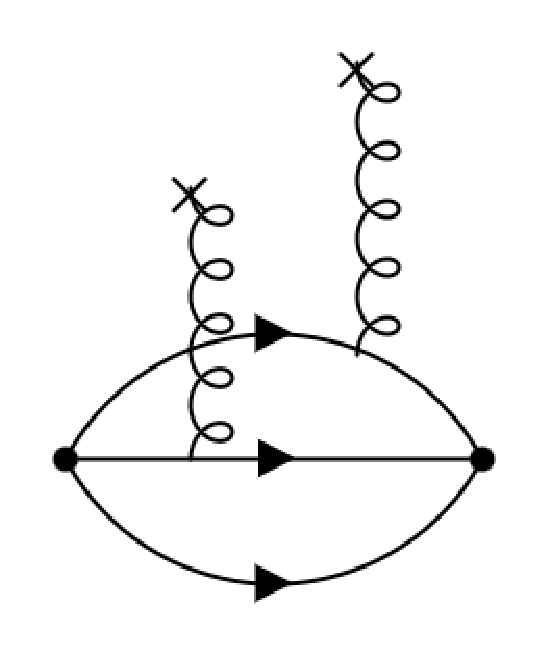}}
		\hspace{8mm}
		\hfill \\
		\subfigure[{\,dim-6 }]{\label{a4}
			\includegraphics*[width=.12\textwidth]{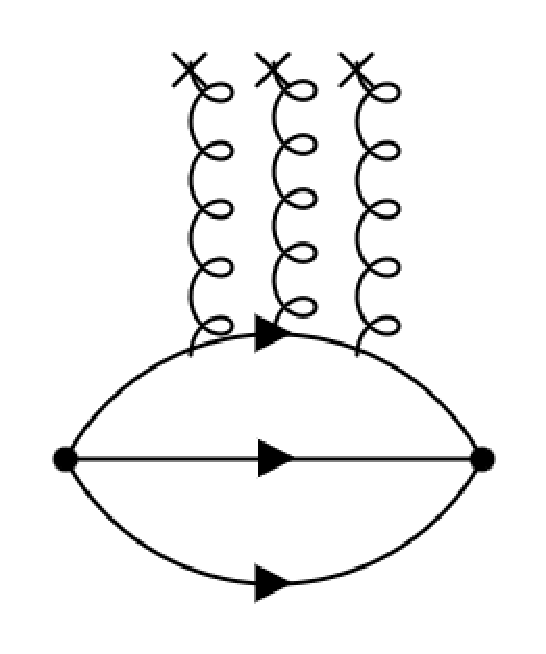}}
		\hspace{10mm}
		\subfigure[{\,dim-6 }]{\label{a5}
			\includegraphics*[width=.12\textwidth]{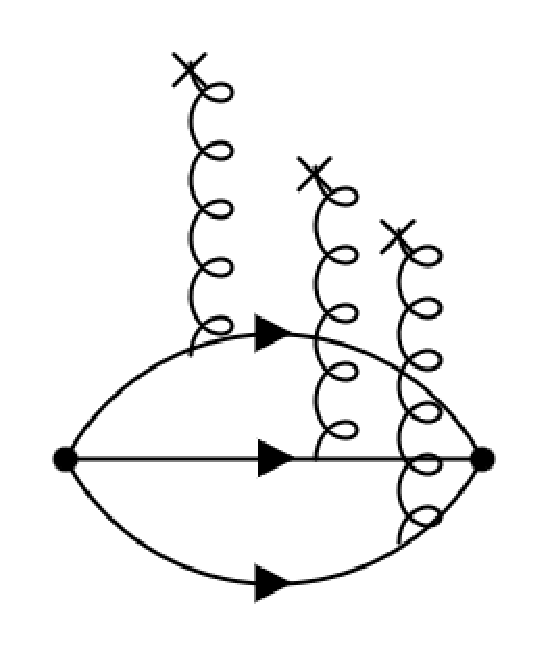}}
		\hspace{10mm}
		\subfigure[{\,dim-8 }]{\label{a6}
			\includegraphics*[width=.12\textwidth]{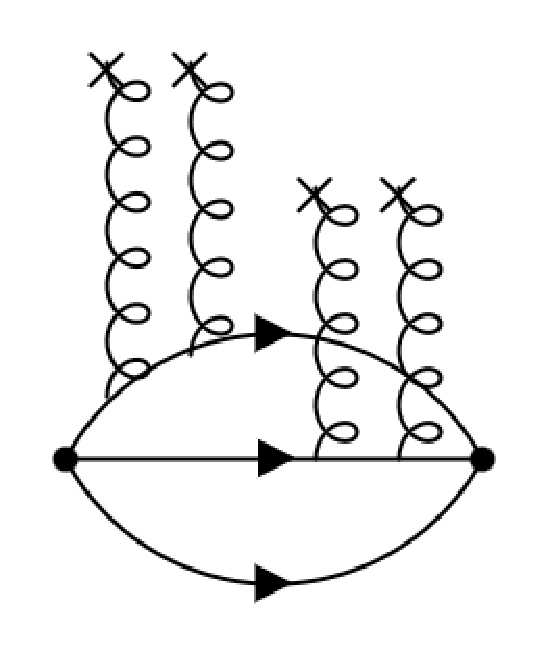}}
		\hspace{10mm}
		\subfigure[{\,dim-8}]{\label{a7}
			\includegraphics*[width=.12\textwidth]{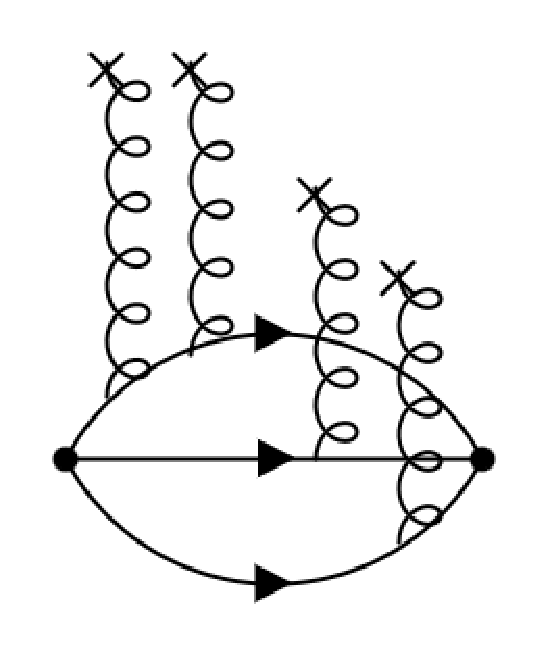}}
		\caption{Sample diagrams considered in the present study }
		{\label{bb}}
	\end{figure}

	After obtaining the hadronic correlation function in the time-like region, the subsequent phase is to evaluate the  QCD side of the correlation function in the deep Euclidean space-like region, where $q^2 \rightarrow -\infty$, by applying the operator product expansion (OPE). For this purpose, one must insert  the foregoing  interpolating current of $\Omega_{QQQ^{'}}$ given in Eq.~\eqref{cur1}
	within Eq.~\eqref{eq:CF1} and  carry out  all the possible contractions of the heavy
	quark-antiquark fields via  the Wick theorem. Accordingly, one can find  a perspicuous expression consisting of heavy quark propagators:
	\begin{equation}\label{eqQCD1}
		\begin{split}
			\Pi(q)&=4i\epsilon^{abc}\epsilon^{a^{'}b^{'}c^{'}}\int d^4x e^{i q x} \Big\{-\gamma_{5}
			S^{c'b}_{Q}(x)S'^{b'a}_{Q'}(x)S^{a'c}_{Q}(x)\gamma_{5}+
			\gamma_{5}S^{c'c}_{Q}(x)\gamma_{5}Tr\Big[S^{a'b}_{Q}(x)S'^{b'a}_{Q'}(x)\Big]\\
			&+ \beta\Big( -\gamma_{5}S^{c'b}_{Q}(x)\gamma_{5}S'^{b'a}_{Q'}(x)S^{a'c}_{Q}(x)
			-S^{c'b}_{Q}(x)S'^{b'a}_{Q'}(x)\gamma_{5}S^{a'c}_{Q}(x)\gamma_{5}
			+\gamma_{5}S^{c'c}_{Q}(x)Tr\Big[S^{a'b}_{Q}(x)\gamma_{5}S'^{b'a}_{Q'}(x)\Big]\\
			&+ S^{c'c}_{Q}(x)
			\gamma_{5}Tr\Big[S^{a'b}_{Q}(x)S'^{b'a}_{Q'}(x)\gamma_{5}\Big]\Big)+
			\beta^2\Big( -S^{c'b}_{Q}(x)\gamma_{5}S'^{b'a}_{Q'}(x)\gamma_{5}S^{a'c}_{Q}(x)+S^{c'c}_{Q}(x)
			Tr\Big[S^{b'a}_{Q'}(x)\gamma_{5}S'^{a'b}_{Q}(x)\gamma_{5}\Big]
			\Big)
			\Big\},
		\end{split}
	\end{equation}
	where $S$ is the full heavy quark propagator and $S'=CS^TC$. For the heavy quark propagator, the following explicit formula is utilized in coordinate space \cite{Agaev:2020zad}:
	\begin{eqnarray}\label{eqQCD2}
		&&S_{Q}^{ab}(x)=i\int \frac{d^{4}k}{(2\pi )^{4}}e^{-ikx}\Bigg \{\frac{\delta
			_{ab}\left( {\slashed k}+m_{Q}\right) }{k^{2}-m_{Q}^{2}}-\frac{%
			g_{s}G_{ab}^{\alpha \beta }}{4}\frac{\sigma _{\alpha \beta }\left( {\slashed %
				k}+m_{Q}\right) +\left( {\slashed k}+m_{Q}\right) \sigma _{\alpha \beta }}{%
			(k^{2}-m_{Q}^{2})^{2}}  \notag  \label{eq:A2} \\
		&&+\frac{g_{s}^{2}G^{2}}{12}\delta _{ab}m_{Q}\frac{k^{2}+m_{Q}{\slashed k}}{%
			(k^{2}-m_{Q}^{2})^{4}}+\frac{g_{s}^{3}G^{3}}{48}\delta _{ab}\frac{\left( {%
				\slashed k}+m_{Q}\right) }{(k^{2}-m_{Q}^{2})^{6}}\left[ {\slashed k}\left(
		k^{2}-3m_{Q}^{2}\right) +2m_{Q}\left( 2k^{2}-m_{Q}^{2}\right) \right] \left(
		{\slashed k}+m_{Q}\right) +\cdots \Bigg \},  \notag \\
		&&
	\end{eqnarray} 
	where $k$ is the four-momentum of the heavy quark and $m_Q$ is its mass.
	In Eq.~\eqref{eqQCD2}  we have:
	\begin{equation}\label{eqQCD3}
		G_{ab}^{\mu \nu }=G_{A}^{\mu \nu }\lambda_{ab}^{A}/2,\,\,~~G^{2}=G_{\mu
			\nu }^{A}G_{A }^{\mu \nu},\ \ G^{3}=\,\,f^{ABC}G^{\mu \nu }_{A}G^{\nu
			\delta }_{B}G^{\delta \mu }_{C},
	\end{equation}
	where   $ f^{ABC}$ denote the structure constants of the color
	group $SU_c(3)$; $A,B,C=1,\,2\,\ldots 8$;  $\lambda ^{A}$ are the Gell-Mann matrices; and  $\mu$, $\nu$ and $\delta$ represent the Lorentz indices. The first term in Eq.~\eqref{eqQCD2} represents the perturbative contribution (free heavy propagator) and the others are  non-perturbative contributions (the gluonic terms) which include emission of one  gluon, and two and three gluon condensates. Placing the heavy quark propagator, Eq.~\eqref{eqQCD2}, into the  QCD side of the correlation function,  Eq.~\eqref{eqQCD1}, results in various terms representing different contributions that are equivalent to  some  Feynman diagrams corresponding to both the  perturbative and non-perturbative contributions. In the current study, the calculation of the QCD correlation function involves non-perturbative terms up to the eight mass dimensions. As depicted in Fig.~\ref{bb}, the diagram \ref{a1} represents the perturbative contribution and the sample diagrams (\ref{a2}, \ref{a3}, \ref{a4}, \ref{a5}, \ref{a6},  \ref{a7}) stand for gluon condensates of different dimensions. For the two-gluon condensate, $ \langle 0 |G^n_{\alpha \beta}(x)G^m_{\alpha' \beta'}(0)|0 \rangle $, which leads to the four dimension non-perturbative contribution (diagrams \ref{a2} and \ref{a3}), we consider the first term of the Taylor expansion for the gluon field at $x=0$. We utilize \cite{Barsbay:2022gtu}:
		\begin{equation}
			\label{eq:2condensate}
				\langle 0 |G^A_{\alpha \beta}(0)G^B_{\alpha' \beta'}(0)|0 \rangle =
			\frac{\langle G^2\rangle }{96}\delta^{AB} [g_{\alpha \alpha'} g_{\beta \beta'}
			-g_{\alpha \beta'} g_{\alpha'\beta }],
			\end{equation}
	and
		\begin{equation}
			\begin{split}
				\label{eq:2condensate2}
				\mathrm t^{a b} \mathrm t^{a'b'}=\frac{1}{2}\left(\delta^{ab'}\delta^{a'b}-\frac{1}{3}\delta^{ab}\delta^{a'b'}\right),
			\end{split}
		\end{equation}
	where $t=\lambda^A/2$.	The six mass dimension non-perturbative contribution is calculated through the three-gluon condensate (diagrams \ref{a4} and \ref{a5}). We decompose  $ \langle 0 |G^A_{\alpha \beta}G^B_{\alpha' \beta'}G^C_{\alpha'' \beta''}|0 \rangle $ in terms of three-gluon condensate, $\langle G^3 \rangle$,  and other parameters as:
\begin{equation}
	\label{eq:3condensate}
	\begin{split}
	& \langle 0 |G^A_{\alpha \beta}G^B_{\alpha' \beta'}G^C_{\alpha'' \beta''}|0 \rangle =
	\frac{\langle G^3\rangle }{576}f^{ABC} [g_{\alpha' \beta} g_{\alpha'' \beta'} g_{\beta'' \alpha}
	-g_{\beta'' \alpha} g_{\beta \beta' }g_{\alpha' \alpha''}-g_{\alpha \alpha''} g_{\alpha' \beta }g_{\beta' \beta''}+g_{\alpha \alpha''} g_{\beta \beta' }g_{\alpha' \beta''}\\
	&+g_{\alpha \alpha'} g_{\beta \alpha'' }g_{\beta' \beta''} -g_{\alpha \alpha'} g_{\beta \beta'' }g_{\alpha'' \beta'} -g_{\alpha \beta'} g_{\beta \alpha'' }g_{\beta'' \alpha'}+g_{\alpha \beta'} g_{\beta \beta'' }g_{\alpha' \alpha''}].
	\end{split}
\end{equation} 
We also find:
\begin{equation}
	\label{eq:3condensate1}
	\begin{split}
	& f^{ABC}\mathrm t_{A}^{a\,b} \mathrm t_{B}^{a^{\prime }\,b^{\prime }}\mathrm t_{C}^{a^{\prime\prime }\,b^{\prime\prime }}=\frac{18}{56}i\bigg[\delta^{ba'}\delta^{b'a''}\delta^{b''a}-\frac{1}{3}\Big(\delta^{ba'}\delta^{ab'}\delta^{a''b''}+\delta^{a'b''}\delta^{b'a''}\delta^{ab}+\delta^{ab''}\delta^{ba''}\delta^{a'b'}\Big)\\
	&+\frac{2}{9}\delta^{ab}\delta^{a'b'}\delta^{a''b''}\bigg].
\end{split}
\end{equation}
The eight dimension diagrams \ref{a6} and \ref{a7} are  written in terms of two-gluon condensate, $\langle G^2 \rangle^2$.  Such contributions are obtained by multiplication of the  third term in Eq. (\ref{eq:A2} ) for two propagators with the perturbative part of the third quark propagator. 
	For all mass dimensions in the non-perturbative part of the correlation function, all possible permutations have been taken into account. 
	
After insertion of the expression of the heavy quark propagator in $x$-space into Eq.~\eqref{eqQCD1}, we perform the resultant Fourier integrals. In the calculations, various types of integrals  appear. For instance, integrals of the following types arise in the perturbative part, the diagram \ref{a1}:
\begin{equation}\label{int}
	\begin{split}
		&\int d^4k_1\,\int d^4k_2\,\int d^4k_3 \,\int d^4x\,e^{i(q-k_1-k_2-k_3)x}\frac{f(k_1,\,k_2,\,k_3)}{(k_1^2-m^2_Q)^{n_1}(k_2^2-m^2_{Q'})^{n_2}(k_3^2-m^2_Q)^{n_3}},
	\end{split}
\end{equation} 
where $n_1$, $n_2$ and $n_3$ are  the natural numbers. In the first step, we perform the Fourier integrals over four-$x$:
\begin{equation}
	\int d^4x e^{i(q-k_1-k_2-k_3)x}=(2\pi)^4\delta^4(q-k_1-k_2-k_3).
\end{equation}
The presence of Dirac's delta function facilitates the four-integration over $k_3$. The remaining four-integrals over $k_2$ and $k_1$ are performed using the Feynman parametrization:
\begin{equation}
	\label{feynman }
	\frac{1}{A_1^{n_1}A_2^{n_2}A_3^{n_3}}=\frac{\Gamma(n_1+n_2+n_3)}{\Gamma({n_1})\Gamma({n_2})\Gamma({n_3})}\int_0^1\int_0^{1-r}dzdr\frac{r^{n_1-1} \ z^{n_2-1}(1-r-z)^{n_3-1}}{[rA_1 \ +zA_2 \ +(1-r-z)A_3]^{n_1+n_2+n_3}}.
\end{equation} 
 To perform the four-integrals over $k_2$ and $k_1$, we utilize \cite{Azizi:2017ubq}:
\begin{equation}
	\label{eq318 }
	\int d^4 \ell\frac{(\ell^2)^m}{(\ell^2+\Delta)^n}=\frac{i\pi^2 (-1)^{m-n} \Gamma[m+2]\Gamma[n-m-2]}{\Gamma[2]\Gamma[n] (-\Delta)^{n-m-2}},
\end{equation} 
where $\Delta$ is a function of quark masses,  Feynman parameters and other related parameters, but does not depend on four-$\ell$. By using the following relation, we extract the imaginary parts of obtained results:
\begin{equation}
	\label{imag }
	\Gamma\Big[\frac{D}{2}-n\Big]\Big(-\frac{1}{\Delta}\Big)^{D/2-n}=\frac{(-1)^{n-1}}{(n-2)!}(-\Delta)^{n-2}ln[-\Delta].
\end{equation} 
This procedure is applied also to calculate the two-gluon condensate contributions, diagrams \ref{a2} and \ref{a3}. To calculate the contributions of higher dimensional non-perturbative operators, we follow the standard procedures of method and find their contributions in momentum-space.	Eventually, by applying the Borel transformation as well as continuum subtraction, the final form of the QCD  correlation function is:
	\begin{eqnarray}
		\Pi_i^{\mathrm{QCD}}(s_0,M^2)=\int_{(2m_Q+m_{Q'})^2}^{s_0}ds\,e^{-\frac{s}{M^2}}\rho_i(s)+\Gamma_i(M^2),~~~~~~~~~i= \not\!q~\mbox{or}~I.
		\label{Eq:finalCor:QCD}
	\end{eqnarray}
	Here $s_0$ represents the continuum threshold,  $\rho_i(s)=\frac{1}{\pi}\mathrm{Im}[\Pi_i^{\mathrm{QCD}}]$ are  spectral densities that include a perturbative and four-dimensional non-perturbative contributions.  $\Gamma_i(M^2)$ denote the results of the non-perturbative operators for the mass dimensions six and eight in the Borel scheme. The explicit mathematical expressions for $\rho_i(s)$ and parts of $ \Gamma_i(M^2)$ are provided in the Appendix.
	As mentioned earlier, both the QCD and hadronic sides are comprised of two independent  Lorentz structures, labeled as $\not\!q$ and  the unit matrix $I$. By matching the coefficients of these   Lorentz structures from both the QCD and hadronic sides, the desired sum rules are obtained:
	\begin{eqnarray}
		\lambda^2 e^{-\frac{m^2}{M^2}}+\tilde{\lambda}^2e^{-\frac{\tilde{m}^2}{M^2}}+\lambda'^2e^{-\frac{m'{}^2}{M^2}}=\Pi^{\mathrm{QCD}}_{\not\!q}(s_0,M^2),
		\label{Eq:cor:match1}
	\end{eqnarray}
	and
	\begin{eqnarray}
		\lambda^2 m e^{-\frac{m^2}{M^2}}-\tilde{\lambda}^2\tilde{m}e^{-\frac{\tilde{m}^2}{M^2}}+\lambda'^2m'e^{-\frac{m'{}^2}{M^2}}=\Pi^{\mathrm{QCD}}_{I}(s_0,M^2).
		\label{Eq:cor:match2}
	\end{eqnarray}
	These sum rules contain six unknowns (three masses and three residues). To solve them, we need five more equations for each structure, which would be found by applying successive derivatives with the respect to $-\frac{1}{M^2} $ to both sides of the above sum rules. This may impose higher uncertainties to the results, hence, we follow a three-step procedure to find the physical quantities. First we choose the continuum threshold, $s_0$, such that the sum rules contain only the ground state (first terms in Eqs. \eqref{Eq:cor:match1} and \eqref{Eq:cor:match2}). After some manipulations, the mass and residue for the ground state are found. As an example for the $\not\!q$ structure, we have:
	\begin{eqnarray}
		m^2=\frac{\frac{d}{d(-\frac{1}{M^2})}\Pi^{\mathrm{QCD}}_{\not\!q}(s_0,M^2)}{\Pi^{\mathrm{QCD}}_{\not\!q}(s_0,M^2)},
		\label{Eq:mass:Groundstates1}
	\end{eqnarray}
and
	\begin{eqnarray}
		\lambda^2=e^{\frac{m^2}{M^2}}\Pi^{\mathrm{QCD}}_{\not\!q}(s_0,M^2).
		\label{Eq:residumass:Groundstates1}
	\end{eqnarray}

Having calculated the mass and residue of the ground state,	we try to calculate the parameters of the first orbital excitation (1P). To this end, we follow the ground $+$ first orbitally excited state $+$ continuum scheme and by adjusting $s_0$ we put the radial excitation  (2S) inside the continuum. Now, using the ground state parameters, we can calculate $\tilde{m}$ and $\tilde{\lambda}$ as the physical quantities related to 1P state using similar procedure mentioned above. Finally, we increase the continuum threshold, $s_0$, and use 1S state $+$ 1P state $+$ 2S $+$ continuum scheme to calculate the mass and residue of the 2S state by considering  the parameters of the 1S and 1P states as inputs.

\section {NUMERICAL ANALYSES}\label{sec:three}	
\begin{figure}[h!]
	\begin{center}
		\includegraphics[width=.43\textwidth]{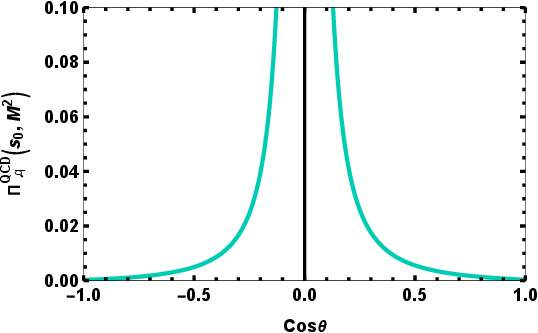}
	\end{center}
	\caption{ $ \Pi^{\mathrm{QCD}}_{\not\!q}(s_0,M^2) $ (in units of $\mathrm{GeV}^{6}$)  as a function of $\cos\theta$ at the medial values of $M^2$ and $s_0$.}
	\label{gr:costheta}
\end{figure}

	In this section, in order to perform numerical analyses of the expressions related to the mass and residue for  triply heavy spin-1/2 baryons in their ground  and first  orbital and radial excited states, we need a set of input parameters such as the quark masses in two  $\overline{MS}$ and pole schemes,  two-gluon  and three-gluon condensates  which are presented in  Table~\ref{tab:Parameter}.  
	\begin{table}[htb]
		\begin{tabular}{|c|c|}
			\hline\hline
			Parameters & Values \\ \hline\hline
			$\bar{m}_{c}(\bar{m}_{c})$                                     & $1.27\pm 0.02~\mathrm{GeV}$ \cite{ParticleDataGroup:2022pth}\\
			$\bar{m}_{b}(\bar{m}_{b})$                                     & $4.18^{+0.03}_{-0.02}~\mathrm{GeV}$ \cite{ParticleDataGroup:2022pth}\\
			$m_{c}$                                     & $1.67\pm 0.07~\mathrm{GeV}$ \cite{ParticleDataGroup:2022pth}\\
			$m_{b}$                                     & $4.78\pm0.06~\mathrm{GeV}$ \cite{ParticleDataGroup:2022pth}\\
					$\langle \frac{\alpha_s}{\pi} G^2 \rangle $ & $(0.012\pm0.004)$ $~\mathrm{GeV}^4 $\cite{Belyaev:1982cd}\\
			$\langle g_s^3 G^3 \rangle $                & $ (0.57\pm0.29)$ $~\mathrm{GeV}^6 $\cite{Narison:2015nxh}\\
			\hline\hline
		\end{tabular}%
		\caption{Numerical values utilized in the analyses.}
		\label{tab:Parameter}
	\end{table}
 An important task in our calculations is to find  working windows for three auxiliary parameters that are entered in the QCD sum rules for the physical quantities, namely  Borel parameter $M^2$,  threshold parameter $s_0$ and arbitrary mixing factor $\beta$.  They are set by analyzing the results using the standard requirements of the method  like pole dominance and convergence of the OPE,  in such a way that the physical quantities show relatively weak dependence on them.  The first auxiliary parameter is  $\beta$ to be fixed. As we previously  mentioned, this parameter appears in the interpolating currents to help all the possible configurations of the quark fields to be taken into account.  Since this parameter can take the values from $- \infty $ to $+\infty$, we introduce a more convenient variable $\theta$  by defining $\beta=\tan\theta$. Then, by examining the dependence of the results on $\cos\theta$ in the range of $[-1,1]$, we can encompass all desired values for $\beta$. In order to find the working regions of  $\cos\theta$, as an example, we plot the function  $\Pi^{QCD}_{\not\!q}(s_0,M^2)$   in term of $\cos\theta$ in  Fig.~\ref{gr:costheta}.   The  $\beta$ or $\cos\theta$ is  mathematical object, so in principle it  should not  affect the physical quantities. But in practice, we see some dependencies of the results on these helping parameters. We should select the regions that show relatively small dependence of the physical quantities on these parameters as the working windows.  As it is seen from Fig.~\ref{gr:costheta}, the following intervals show the slightest variations  in terms of  $\cos\theta$:
\begin{eqnarray}\label{beta fun}
	-1.0 \le\ \cos\theta \le\ -0.5~~~ \mathrm{and}~~~0.5 \le\ \cos\theta \le\ 1.0.
\end{eqnarray}

\begin{figure}[h!]
	\begin{center}
		\includegraphics[width=.43\textwidth]{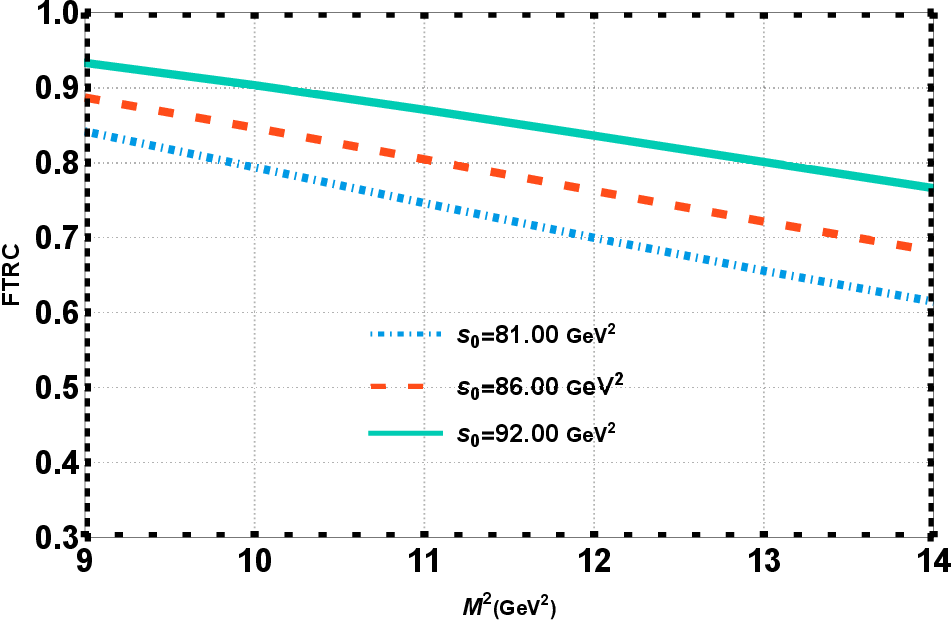}
	\end{center}
	\caption{ FTRC versus  $M^2$  at three fixed values of $s_0$ for  $\Omega_{ccb}$ channel. }
	\label{FTRC}
\end{figure}

The next auxiliary parameter that needs to be fixed is $M^2$. The upper limit for  this parameter is set based on the dominance of the pole contribution over the higher states  and continuum. In technique language, we demand 
\begin{eqnarray}
		\frac{\Pi(s_0,M^2,\beta)}{\Pi(\infty,M^2,\beta)}\geq 0.5.
	\end{eqnarray}
Its lower limit is obtained from the convergence of the OPE,  implying that the perturbative contribution should exceed  the non-perturbative one and the higher dimensional operators have relatively lower contributions. For this, we require  that the last non-perturbative operator contribution (eight-mass dimension)  does not surpass  0.05 of the total perturbative $+$ non-perturbative contributions. That is
 \begin{equation}
	 \frac{\Pi ^{\mathrm{Dim8}}(s_0,M^2,\beta)}{\Pi (s_0,M^2,\beta)}\le\ 0.05.
	  	\label{eq:Convergence}
	  \end{equation} 
 The obtained intervals  for $M^2$ for all the members in all schemes are shown in Table~\ref{results}.  The final auxiliary parameter to be set is the continuum threshold $s_0$. The values of $s_0$ are not entirely arbitrary and depend on the energy of the next excited state and differ for the ground  and the first  orbital and radial excited states. The selection of thresholds are such that the higher states do not contribute to the calculations for each considered state. The   working regions for  $s_0$ for all under study channels are also displayed in Table~\ref{results}.  It is instructive to check the pole dominance and OPE convergence by using the determined  working intervals of the auxiliary parameters. To this end, we depict Fig.~\ref{FTRC}, showing the variation of  first three resonances' contribution (FTRC)  with respect to $M^2$  at three fixed values of $s_0$ for  $\Omega_{ccb}$ channel as an example. From this figure we see that the pole dominance for the considered resonances is  nicely satisfied in the working windows of auxiliary parameters. In the average values of all the auxiliary parameters,  the higher dimensional term contributes with maximally 1\% satisfying the requirements of the method.  We shall note that the dimension six operators constitute maximally 
5\%  of the total contribution referring to the nice convergence of OPE. 

\begin{figure}[!htb]
	\centering
	\subfigure[]{\label{a11}
		\includegraphics*[width=.43\textwidth]{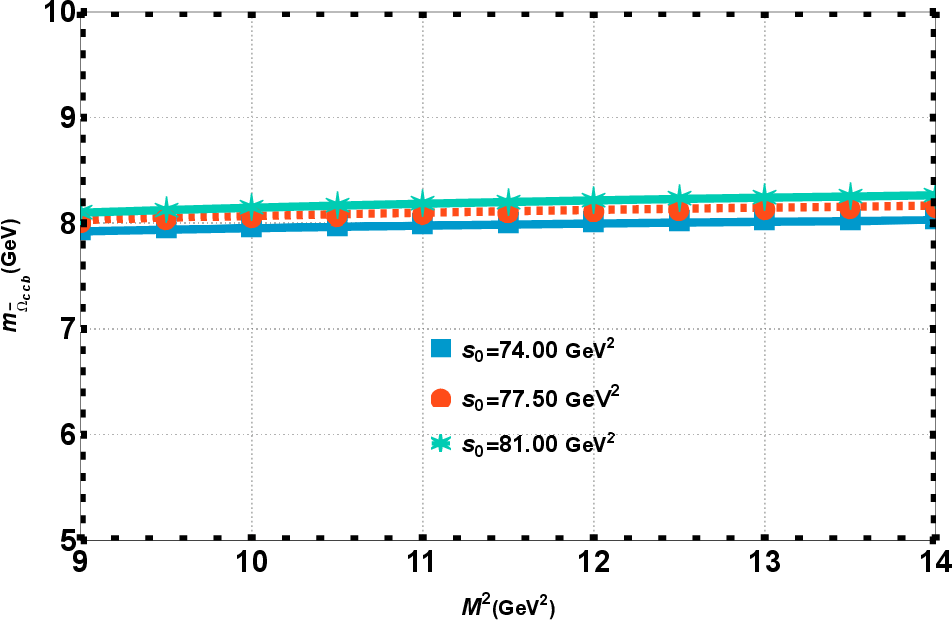}}
	\hspace{3mm}
	\subfigure[ ]{\label{a22}
		\includegraphics*[width=.43\textwidth]{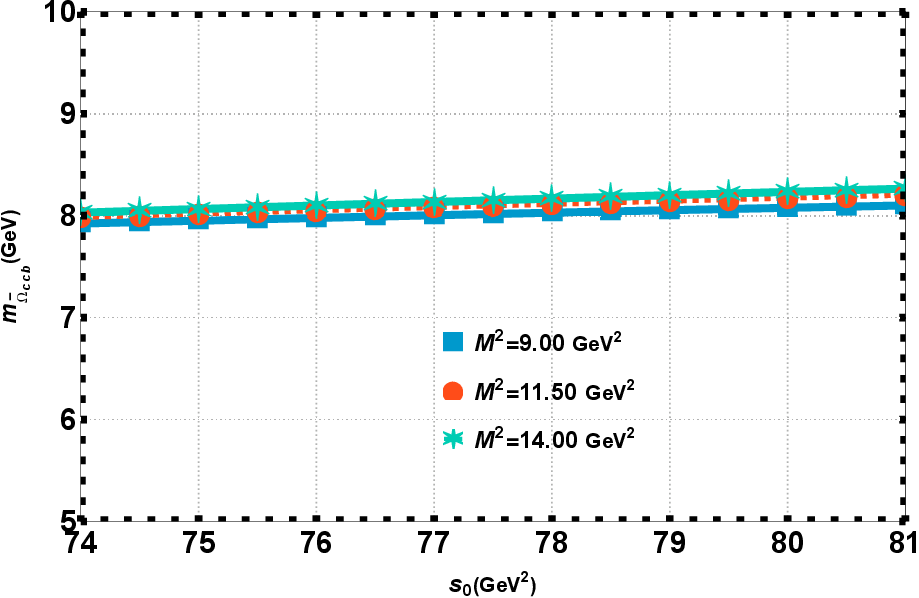}}
		\hfill \\
	\caption{(a) The mass of $\overline{\Omega}_{ccb}$ in the ground state  with respect to $M^2$ at three fixed values for the $s_0$ and at  $\cos\theta=-0.71$. 
	(b) The mass of $\overline{\Omega}_{ccb}$ in the ground state  with  respect to  $s_0$ at three fixed values for the $M^2$ and at  $\cos\theta=-0.71$. }
		{\label{aaa}}
\end{figure}

\begin{figure}[h!]

	\centering
	\subfigure[]{\label{a33}
		\includegraphics*[width=.43\textwidth]{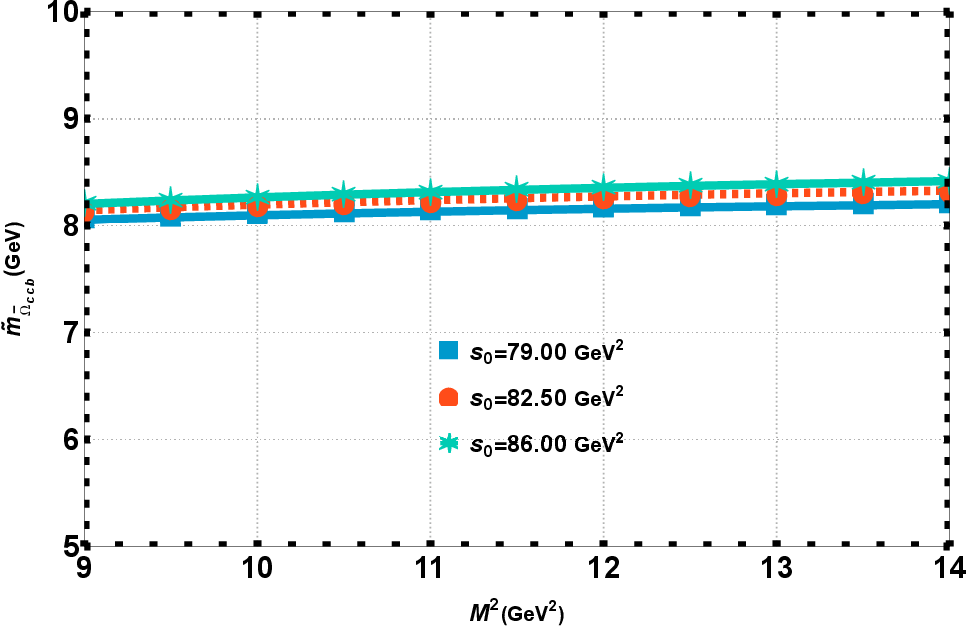}}
	\hspace{3mm}
	\subfigure[ ]{\label{a44}
		\includegraphics*[width=.43\textwidth]{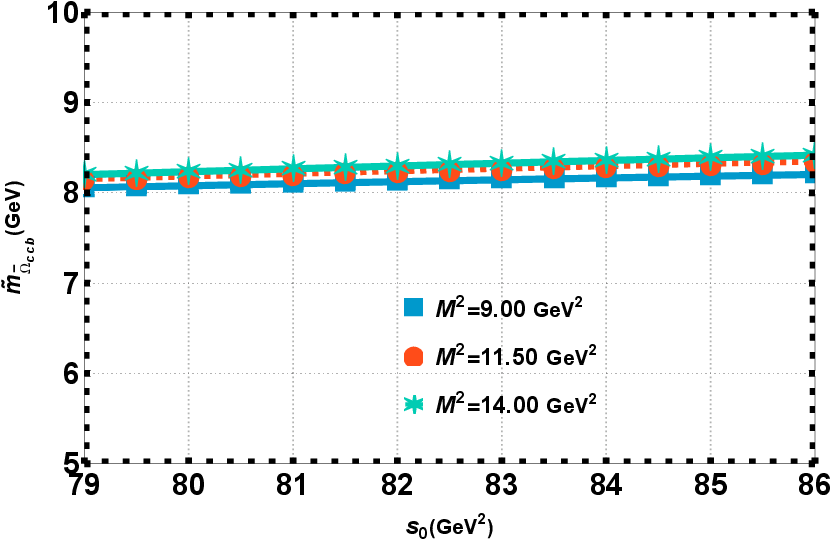}}
		\hfill \\
	\caption{(a) The mass of $\overline{\Omega}_{ccb}$ in the orbital exited state  with respect to $M^2$ at three fixed values for the $s_0$ and at $\cos\theta=-0.71$. (b) The mass of $\overline{\Omega}_{ccb}$ in the orbital exited state with respect to $s_0$ at three fixed values for the $M^2$ and at  $\cos\theta=-0.71$. }
		{\label{aa2}}
\end{figure}

\begin{figure}[h!]

	\centering
	\subfigure[]{\label{a55}
		\includegraphics*[width=.43\textwidth]{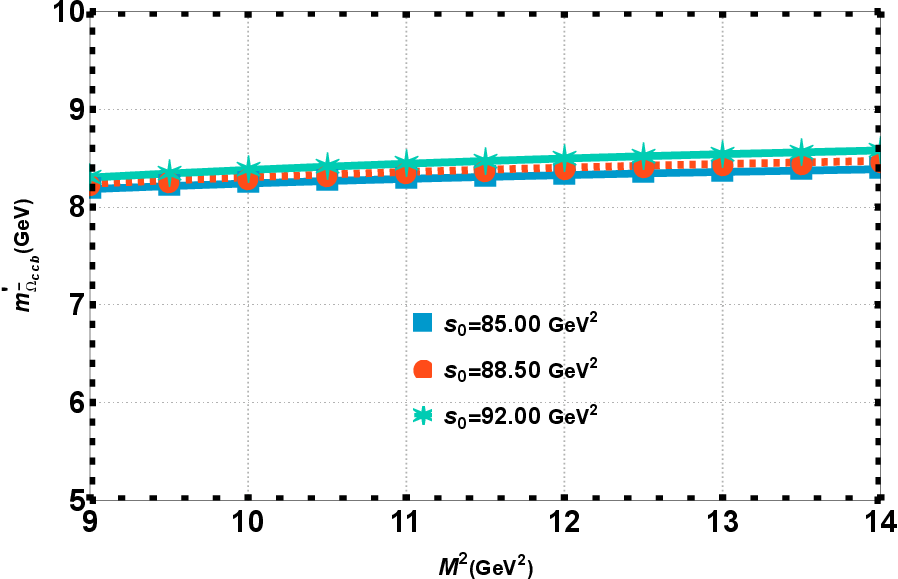}}
	\hspace{3mm}
	\subfigure[ ]{\label{a66}
		\includegraphics*[width=.43\textwidth]{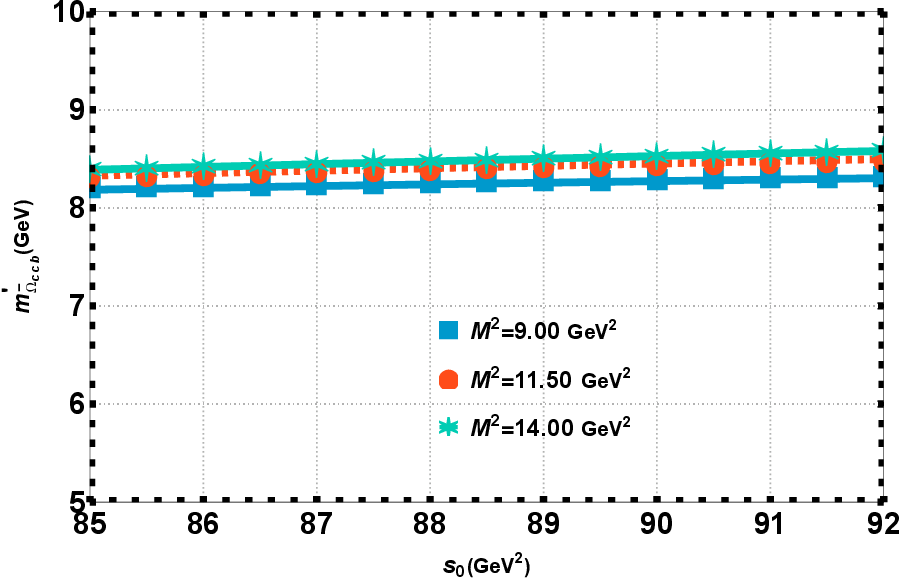}}
		\hfill \\
	\caption{(a) The mass of $\overline{\Omega}_{ccb}$ in the radial exited state  with respect to $M^2$ at three fixed values for the $s_0$ and at  $\cos\theta=-0.71$. (b) The mass of $\overline{\Omega}_{ccb}$ in the radial exited state  with respect to  $s_0$ at three fixed values for the $M^2$ and at  $\cos\theta=-0.71$. }
		{\label{aa3}}
\end{figure}

\begin{figure}[!h]
	\centering
	\subfigure[]{\label{Ra1}
		\includegraphics*[width=.43\textwidth]{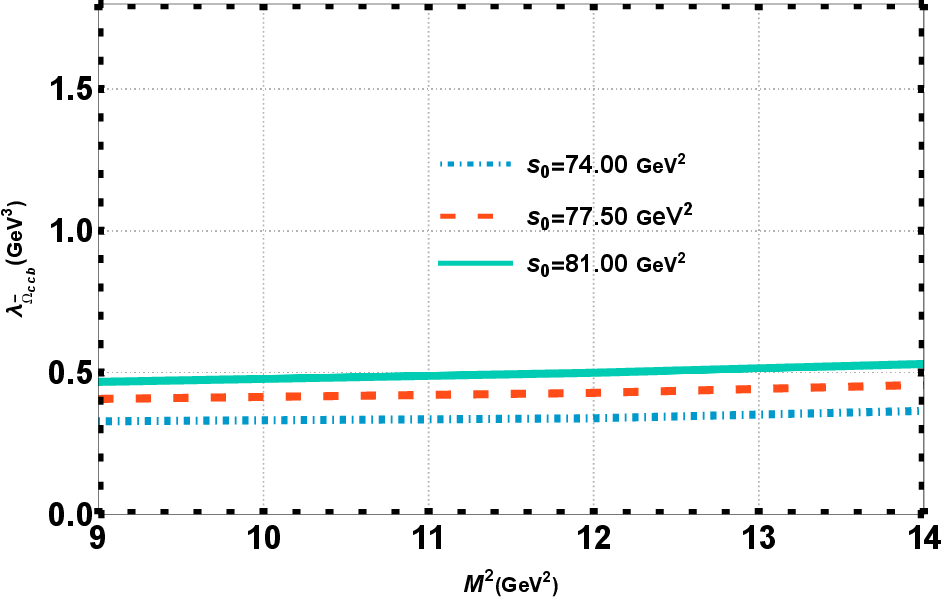}}
		\hfill \\
	\subfigure[]{\label{Ra2}
		\includegraphics*[width=.43\textwidth]{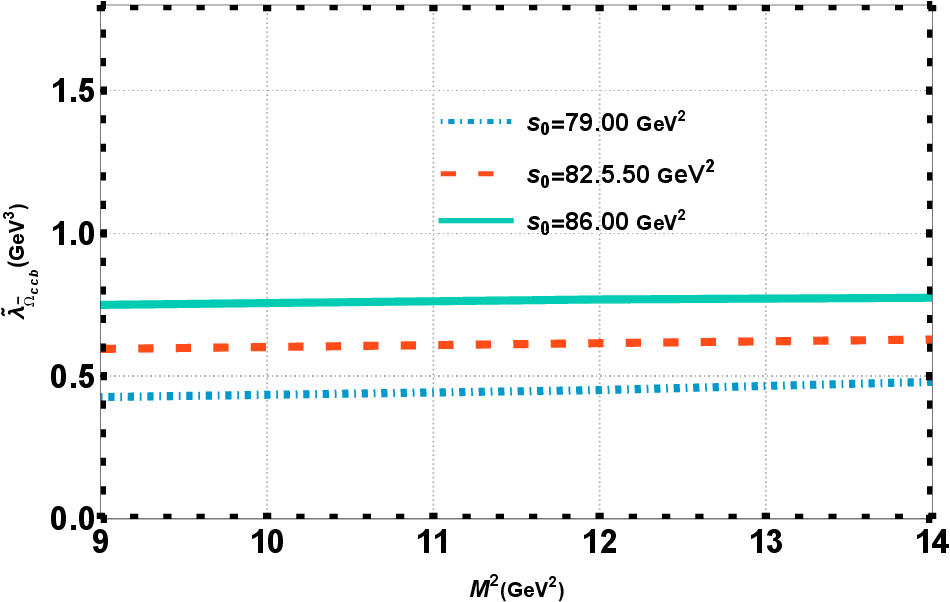}}
	\hspace{3mm}
	\subfigure[]{\label{ra3}
		\includegraphics*[width=.43\textwidth]{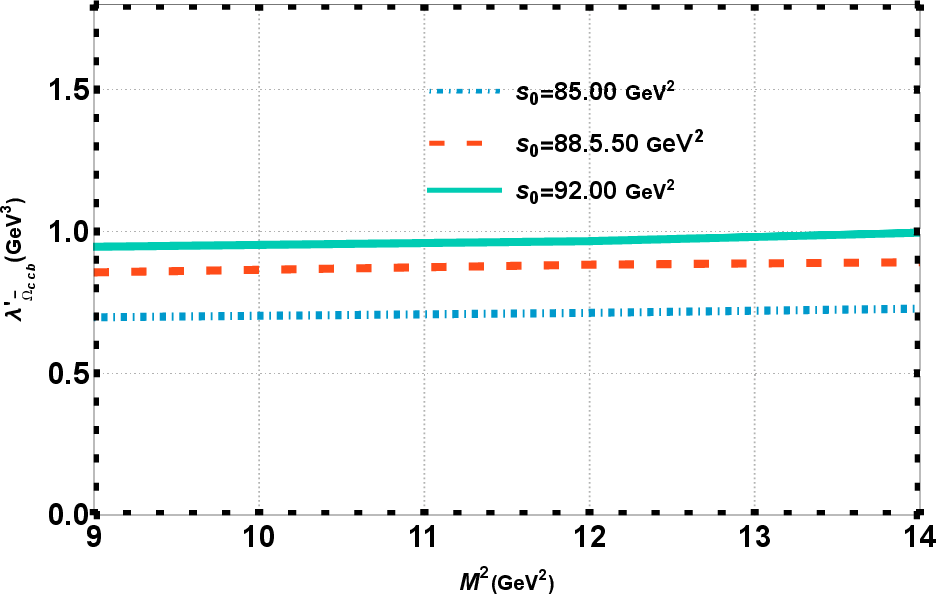}}
	\caption{(a) The residue of $\overline{\Omega}_{ccb}$ in the ground state  with respect to $M^2$ at three fixed values for the $s_0$ and at  $\cos\theta=-0.71$. (b)  The residue of $\overline{\Omega}_{ccb}$ in the orbital exited state  with respect to $M^2$ at three fixed values for the $s_0$ and at  $\cos\theta=-0.71$. (c) The residue of $\overline{\Omega}_{ccb}$ in the radial exited state  with respect to $M^2$ at three fixed values for the $s_0$ and at  $\cos\theta=-0.71$.}
	{\label{rrr}}
\end{figure}

 The stability diagrams of physical  quantities (masses and residues of $\overline{\Omega}_{ccb}$ in the ground and excited states as examples)  with respect to the variations of the  auxiliary parameters are depicted in  Figs.~\ref{aaa},~\ref{aa2},~\ref{aa3} and \ref{rrr}. In Fig.~\ref{a11}, the mass of $\overline{\Omega}_{ccb}$ in the ground state is drawn with respect to $M^2$ at three fixed values for threshold parameter and at,   $\cos\theta=-0.71$. Note that $\cos\theta=-0.71$ corresponds to the Ioffe current,  $\beta=-1$.  In Fig.~\ref{a22}, the mass of $\overline\Omega_{ccb}$ in the ground state is plotted in terms of  $s_0$ at three fixed values for the  Borel parameter and   at $\cos\theta=-0.71$. In Figs.~\ref{aa2} and \ref{aa3}, the masses of orbital and radial excited states for  $\overline{\Omega}_{ccb}$ are shown, respectively. As it is seen from these figures, the masses of ground, first orbitally excited and first radially excited states for $\overline{\Omega}_{ccb}$ show very weak dependence on the parameters $M^2$ and $s_0$. In Fig.~\ref{Ra1} the residue of $\overline{\Omega}_{ccb}$ in the ground state is plotted with respect to $M^2$ at three fixed values for threshold parameter and at   $\cos\theta=-0.71$.  The residues of orbital and radial excited states for  $\overline{\Omega}_{ccb}$ are drawn in Figs.~\ref{Ra2} and \ref{ra3}, respectively. The residual dependence appear as the uncertainties in the obtained results. Such weak dependence of the results on the auxiliary parameters in their working intervals are the case for the masses of $\Omega_{ccb}$, $\overline{\Omega}_{bbc}$ and $\Omega_{bbc}$ as well as all the residues. 

Having calculated the working windows for the auxiliary parameters, we proceed  to find  the numerical values of the pole and $\overline{MS}$ masses and residues for the states under consideration. To this end, we follow a three-step procedure mentioned previously. For calculation of  the mass and residue of the ground state, we set $s_0$ such that the first orbital and radial excitations remain inside the continuum. In other words, we choose a ground state $+$ continuum scheme. The obtained results for the ground state masses and residues for all the channels in this step are shown in Table~\ref{results}. Now, we increase the value of $s_0$ and use the obtained values for the ground state as inputs to calculate the parameters of the first orbital excited states applying  the ground state $+$ orbitally excited state $+$ continuum scheme. And finally, for the first radial excitation, we consider the ground state $+$ first orbitally excited state $+$ first radially excited state $+$ continuum, and choose an appropriate threshold parameter to find the values of the physical quantities for the 2S states. We collect the values obtained for the properties of the 1P and 2S states in Table~\ref{results}, as well. The presented uncertainties  arise from the  errors in the input parameters and uncertainties coming from the calculations of the working regions of  the auxiliary parameters. We shall note that the order of uncertainties in the values of the masses are very low compared to those of the residues. This is because of the fact that the mass is obtained from the ratio of two sum rules killing the errors of each other, while the residue is found only from one sum rule as are clear from  Eqs. (\ref{Eq:mass:Groundstates1}) and (\ref{Eq:residumass:Groundstates1}). As is seen from Table ~\ref{results}, the mass difference between the ground and first orbitally exited state (first radially excited state) in structure $\not\!q$ for $\overline{\Omega}_{ccb}$,  $\Omega_{ccb}$, $\overline{\Omega}_{bbc}$ and $\Omega_{bbc}$ are 0.15 (0.30) $\mathrm{GeV}$, 0.19 (0.39) $\mathrm{GeV}$, 0.14 (0.25) $\mathrm{GeV}$ and 0.23 (0.42) $\mathrm{GeV}$, whereas for structure $I$, these variances are  0.15 (0.29) $\mathrm{GeV}$, 0.18 (0.37) $\mathrm{GeV}$, 0.19 (0.23) $\mathrm{GeV}$ and 0.20 (0.38) $\mathrm{GeV}$, respectively. 
 \begin{table}[!h]
	\begin{tabular}{|c|c|c|c|c|c|c|c|}
			\hline\hline
		Particle  & State &$M^2~(\mathrm{GeV^2})$&$s_0~(\mathrm{GeV^2})$  & m $(\not\!q)~(\mathrm{GeV})$&m $ (I)~(\mathrm{GeV})$ & $\lambda(\not\!q)~(\mathrm{GeV^3})$& $\lambda(I)~(\mathrm{GeV^3})$ \\ \hline\hline
		\multirow{3}{*}{} &$\overline{\Omega}_{ccb}(\frac{1}{2}^+)(1S)$ &$9.0-14.0$& $74-81$& $8.15{}^{+0.27}_{-0.23}$& $8.11{}^{+0.27}_{-0.23}$& $0.44{}^{+0.23}_{-0.25}$ & $0.42{}^{+0.09}_{-0.10}$  \\ \cline{2-8} 
		$\overline{\Omega}_{ccb}$    &$\overline{\Omega}_{ccb}(\frac{1}{2}^-)(1P)$ &$9.0-14.0$& $79-86$& $8.30{}^{+0.20}_{-0.25}$& $8.26{}^{+0.28}_{-0.27}$& $0.59{}^{+0.15}_{-0.18}$ & $0.55{}^{+0.18}_{-0.12}$ \\ \cline{2-8} 
		&$\overline{\Omega}_{ccb}(\frac{1}{2}^+)(2S)$ &$9.0-14.0$& $85-92$& $8.45{}^{+0.31}_{-0.27}$& $8.40{}^{+0.32}_{-0.33}$& $0.85{}^{+0.15}_{-0.10}$ & $0.69{}^{+0.12}_{-0.11}$  \\ \hline\hline
			\multirow{3}{*}{} &$\Omega_{ccb}(\frac{1}{2}^+)(1S)$ &$9.0-14.0$& $74-81$& $8.63{}^{+0.12}_{-0.13}$& $8.64{}^{+0.18}_{-0.12}$& $0.35{}^{+0.15}_{-0.15}$ & $0.26{}^{+0.08}_{-0.11}$  \\ \cline{2-8} 
		$\Omega_{ccb}$    &$\Omega_{ccb}(\frac{1}{2}^-)(1P)$ &$9.0-14.0$& $79-86$& $8.82{}^{+0.17}_{-0.15}$& $8.82{}^{+0.18}_{-0.16}$& $0.53{}^{+0.12}_{-0.11}$ & $0.39{}^{+0.11}_{-0.11}$ \\ \cline{2-8} 
		&$\Omega_{ccb}(\frac{1}{2}^+)(2S)$ &$9.0-14.0$& $85-92$& $9.02{}^{+0.20}_{-0.15}$& $9.01{}^{+0.32}_{-0.33}$&  $0.67{}^{+0.15}_{-0.12}$& $0.57{}^{+0.13}_{-0.11}$\\ \hline\hline
			\multirow{3}{*}{} &$\overline{\Omega}_{bbc}(\frac{1}{2}^+)(1S)$ &$12.0-16.0$& $140-148$& $11.13{}^{+0.23}_{-0.19}$& $11.10{}^{+0.28}_{-0.23}$& $0.68{}^{+0.24}_{-0.23}$ & $0.52{}^{+0.14}_{-0.22}$ \\ \cline{2-8} 
		$\overline{\Omega}_{bbc}$     &$\overline{\Omega}_{bbc}(\frac{1}{2}^-)(1P)$ &$12.0-16.0$& $148-155$& $11.27{}^{+0.24}_{-0.17}$& $11.29{}^{+0.23}_{-0.21}$& $0.86{}^{+0.23}_{-0.19}$ & $0.85{}^{+0.10}_{-0.17}$  \\ \cline{2-8} 
		&$\overline{\Omega}_{bbc}(\frac{1}{2}^+)(2S)$ &$12.0-16.0$&	 $155-163$& $11.38{}^{+0.28}_{-0.23}$& $11.33{}^{+0.35}_{-0.30}$& $0.98{}^{+0.33}_{-0.32}$ & $0.92{}^{+0.09}_{-0.09}$ \\ \hline\hline
		\multirow{3}{*}{} &$\Omega_{bbc}(\frac{1}{2}^+)(1S)$ &$12.0-16.0$& $140-148$& $11.74{}^{+0.13}_{-0.13}$& $11.78{}^{+0.16}_{-0.18}$& $0.66{}^{+0.19}_{-0.15}$ & $0.50{}^{+0.13}_{-0.13}$ \\ \cline{2-8} 
		$\Omega_{bbc}$     &$\Omega_{bbc}(\frac{1}{2}^-)(1P)$ &$12.0-16.0$& $148-155$& $11.97{}^{+0.16}_{-0.11}$& $11.98{}^{+0.12}_{-0.17}$& $0.76{}^{+0.28}_{-0.33}$ & $0.73{}^{+0.11}_{-0.10}$  \\ \cline{2-8} 
		&$\Omega_{bbc}(\frac{1}{2}^+)(2S)$ &$12.0-16.0$&	 $155-163$& $12.16{}^{+0.18}_{-0.14}$& $12.16{}^{+0.17}_{-0.15}$& $0.95{}^{+0.14}_{-0.14}$ & $0.91{}^{+0.16}_{-0.14}$\\ 	\hline\hline
	\end{tabular}
	\caption{Working windows for the auxiliary parameters and calculated mass and residue results. For the  baryons with over-line and baryons without over-line, the $\overline{MS}$ and pole values of the quark masses 
				are used, respectively.}
	\label{results}
\end{table}
 Since there is no experimental information available for the triply heavy baryons, Tables \ref{comper results3} and \ref{comper results2} only include comparisons among the predictions of present study and other existing theoretical approaches. As we previously mentioned,  with the aim of achieving higher accuracies, we carry out  the calculations  by considering the non-perturbative operators up to eight mass dimensions and for the first three resonances, while, in the previous study \cite{Aliev:2012tt}, the mass and residue  were only computed for the ground state and non-perturbative part up to four mass dimensions. We also include the existing predictions of other studies in Tables \ref{comper results3} and \ref{comper results2}. We see that some of them report only the masses of the ground, but others include the parameters of the excited states as well. We shall remark that most of these studies do not present the scheme of the quark masses, hence, we compare them with our values obtained using the $\overline{MS}$  scheme for the quark masses.   As shown in  tables \ref{comper results3} and \ref{comper results2}, our obtained mass for ground state of $\overline{\Omega}_{ccb}$   aligns well  with the predictions of various methods such as the non- relativistic quark  \cite{Roberts:2007ni, Shah:2018bnr}, the relativistic quark model \cite{Yang:2019lsg,Faustov:2021qqf}, QCD sum rules \cite{Wang:2020avt}, the Regge trajectories \cite{Oudichhya:2023pkg} and  effective Hamiltonian \cite{Serafin:2018aih}  within the indicated uncertainties, whiles it is slightly different with predictions of some methods like  Faddive equation \cite{Qin:2019hgk} and QCD sum rules \cite{Wang:2011ae,Aliev:2012tt}, which contain the non-perturbative operators up to dimension four and less numbers of resonances.  The small differences with the predictions of Refs.  \cite{Wang:2011ae,Aliev:2012tt} can be attributed to the dimension six and eight operators that are considered in the present study and contribute respectively with 5\% and 1\% to the total integral as previously mentioned.  Our prediction for the mass of ground state of $\overline{\Omega}_{bbc}$ is in good consistency with the various theoretical predictions, including  the non- relativistic quark  \cite{ Shah:2018div}, the relativistic quark model \cite{Yang:2019lsg,Faustov:2021qqf}, QCD sum rules \cite{Wang:2020avt}, Faddive equation \cite{Qin:2019hgk}  and  effective Hamiltonian \cite{Serafin:2018aih}  within the errors and there is small difference with predictions of some approaches, containing the Regge trajectories \cite{Oudichhya:2023pkg}, the non- relativistic quark  \cite{Roberts:2007ni} and  QCD sum rules \cite{Wang:2011ae,Aliev:2012tt}. Our results for 1P and 2S of $\overline{\Omega}_{ccb}$   are in agreement, within the
 errors, with the other mentioned theoretical predictions \cite{Roberts:2007ni,Shah:2018bnr, Yang:2019lsg,Faustov:2021qqf, Oudichhya:2023pkg, Qin:2019hgk,Serafin:2018aih}. For 1P of  $\overline{\Omega}_{ccb}$, our result  is a little bit
 higher than the predictions of QCD sum rule approach \cite{Wang:2011ae}. Within the presented errors of 1P and 2S, our results of  $\overline{\Omega}_{bbc}$, except some predictions \cite{Shah:2018div,Roberts:2007ni, Oudichhya:2023pkg,Wang:2011ae} which show little differences, exhibit good consistency with the theoretical predictions  \cite{ Yang:2019lsg,Faustov:2021qqf, Qin:2019hgk,Serafin:2018aih}. Our obtained  masses for ground states of  $\Omega_{ccb}$ and $\Omega_{bbc}$ are in good agreement with predictions of QCD sum rule method \cite{Aliev:2012tt}  within the errors. Our results for 1S and 1P of  $\Omega_{ccb}$ and $\Omega_{bbc}$ are slightly
 higher than findings of  Ref.~\cite{Wang:2011ae} using QCD sum rules.
\begin{table}[h!]
	\begin{tabular}{|c|c|c|c|c|c|c|}
			\hline\hline
	\multicolumn{1}{|c|}{particle}&  \multicolumn{3}{c|}{$\overline{\Omega}_{ccb}$} & \multicolumn{3}{c|}{$\Omega_{ccb}$}  \\ \hline
	state&  $1S$ & $1P$ & $2S$ & $1S$ & $1P$ & $2S$ \\ \hline\hline
	Present\,Work&$8.15{}^{+0.27}_{-0.23}$&$8.30{}^{+0.20}_{-0.25}$&$8.45{}^{+0.31}_{-0.27}$&$8.63{}^{+0.12}_{-0.13}$& $8.82{}^{+0.17}_{-0.15}$&$9.02{}^{+0.20}_{-0.15}$\\ \hline
	Ref.\cite{Aliev:2012tt}&$7.61{}^{+0.13}_{-0.13}$ &-&-&$8.50{}^{+012}_{-0.12}$	&-&- \\ \hline
	Ref.\cite{Wang:2011ae}&$7.61{}^{+0.13}_{-0.13}$  &$7.74{}^{+0.13}_{-0.13}$&-& $8.23{}^{+013}_{-0.13}$& $8.36{}^{+0.13}_{-0.13}$&-\\ \hline
	Ref.\cite{Roberts:2007ni}&$8.24$&$8.42$&$8.53$&-&-&-\\ \hline
	Ref.\cite{Shah:2018bnr}&$8.00$&$8.38$&$8.61$&-&-&-\\ 	\hline
	Ref.\cite{Yang:2019lsg}&$8.00$&$8.30$&$8.45$&-&-&-\\ \hline
	Ref.\cite{Faustov:2021qqf}&$7.98$&$8.25$&$8.36$&-&-&-\\ \hline
	Ref.\cite{Wang:2020avt}&$8.02{}^{+0.08}_{-0.08}$ &-&-	&-&-&-\\ \hline
	Ref.\cite{Oudichhya:2023pkg} &$8.19$&$8.49$&$8.62$&-&-&-\\ \hline
	Ref.\cite{Qin:2019hgk}&$7.86$&$8.16$&$8.33$&-&-&-\\ \hline
Ref.\cite{Serafin:2018aih}&$8.30$&$8.49$&$8.64$&-&-&-\\ \hline
\hline
	\end{tabular} 
	\caption{The mass spectra (in units of $\mathrm{GeV}$) of the $\overline{\Omega}_{ccb}$ and  $\Omega_{ccb}$ baryons compared to other theoretical predictions.}
	\label{comper results3}
\end{table}
\begin{table}[h!]
	\begin{tabular}{|c|c|c|c|c|c|c|}
			\hline\hline
		\multicolumn{1}{|c|}{particle}&  \multicolumn{3}{c|}{$\overline{\Omega}_{bbc}$} & \multicolumn{3}{c|}{$\Omega_{bbc}$}  \\ \hline
		state&  $1S$ & $1P$ & $2S$ & $1S$ & $1P$ & $2S$ \\ \hline\hline
		Present\,Work&$11.13{}^{+0.23}_{-0.19}$&$11.27{}^{+0.24}_{-0.17}$&$11.38{}^{+0.28}_{-0.23}$& $11.74{}^{+0.13}_{-0.13}$&  $11.97{}^{+0.16}_{-0.11}$&$12.16{}^{+0.18}_{-0.14}$\\ \hline
		Ref.\cite{Aliev:2012tt}&$10.59{}^{+0.14}_{-0.14}$ &-&-&$11.73{}^{+0.16}_{-0.16}$	&-&- \\ \hline
		Ref.\cite{Wang:2011ae}&$10.47{}^{+0.12}_{-0.12}$  &$10.60{}^{+0.12}_{-0.12}$&-&$11.50{}^{+0.11}_{-0.11}$& $11.62{}^{+0.11}_{-0.11}$&-\\ \hline
			Ref.\cite{Roberts:2007ni}&$11.53$&$11.71$&$11.78$&-&-&-\\ \hline
		Ref.\cite{Shah:2018div}&$11.23$&$11.57$&$11.75$&-&-&-\\ \hline
			Ref.\cite{Yang:2019lsg}&$11.20$&$11.48$&$11.61$&-&-&-\\ \hline
			Ref.\cite{Faustov:2021qqf}&$11.19$&$11.41$&$11.50$&-&-&-\\ \hline
		Ref.\cite{Wang:2020avt}&$11.22{}^{+0.08}_{-0.08}$ &-&-&-&-&-	\\ \hline
			Ref.\cite{Oudichhya:2023pkg} &$11.52$&$11.88$&$11.75$&-&-&-\\ \hline
				Ref.\cite{Qin:2019hgk}& $11.07$&$11.41$&$11.60$&-&-&-\\ \hline
		Ref.\cite{Serafin:2018aih}&$ 11.21  $&$11.43$&$11.58$&-&-&-\\ \hline
	\hline
	\end{tabular} 
	\caption{The mass spectra (in units of $\mathrm{GeV}$) of the $\overline{\Omega}_{bbc}$ and $\Omega_{bbc}$ baryons compared to other theoretical predictions.}
	\label{comper results2}
\end{table}
On the other hands, we examine the residues of the ground, first orbitally and first radially excited states of the triply
heavy baryons, and depicted our findings in Table~\ref{copm r}. As previously mentioned, our residues' findings include the impacts of non-perturbative operators  up to eight mass dimensions as well. So far, there have been few studies conducted on the residues of triply heavy spin-1/2 baryons in the literature. Table~\ref{copm r} summarizes the predictions that are available. In the case of the residues, there are also  good agreements, within the uncertainties, presented between existing results of 1S and 1P in the literature  \cite{Aliev:2012tt,Wang:2011ae} and our findings. The results  attained for
the residues can be utilized as inputs  to  analyze various  decays of the
considered triply heavy spin-1/2  baryons. Our results may help experimental groups in their ongoing search for the triply heavy baryons at different hadron colliders.
\begin{table}[th!]
	\begin{tabular}{|c|c|c|c|c|c|c|}
		\hline\hline
		Particle  & State &Present\,Work $(\not\!q)$ &Present\,Work$(I)$  & Ref.\cite{Aliev:2012tt}($\not\!q$)& Ref.\cite{Aliev:2012tt}(I) & Ref.\cite{Wang:2011ae} \\ \hline\hline
		\multirow{3}{*}{} &$\overline{\Omega}_{ccb}(\frac{1}{2}^+)(1S)$ & $0.44{}^{+0.23}_{-0.25}$ & $0.42{}^{+0.09}_{-0.10}$& $0.56{}^{+0.18}_{-0.18}$&$0.38{}^{+0.13}_{-0.13}$&$0.47{}^{+0.10}_{-0.10}$ \\ \cline{2-7} 
		$\overline{\Omega}_{ccb}$    &$\overline{\Omega}_{ccb}(\frac{1}{2}^-)(1P)$ & $0.59{}^{+0.15}_{-0.18}$ & $0.55{}^{+0.18}_{-0.12}$&-&-& $0.57{}^{+0.11}_{-0.11}$ \\ \cline{2-7} 
		&$\overline{\Omega}_{ccb}(\frac{1}{2}^+)(2S)$ & $0.85{}^{+0.15}_{-0.10}$ & $0.69{}^{+0.12}_{-0.11}$&-&-&-  \\ \hline\hline
		\multirow{3}{*}{} &$\Omega_{ccb}(\frac{1}{2}^+)(1S)$ & $0.35{}^{+0.15}_{-0.15}$ & $0.26{}^{+0.08}_{-0.11}$&$0.38{}^{+0.13}_{-0.13}$&$0.30{}^{+0.10}_{-0.10}$ &$0.47{}^{+0.10}_{-0.10}$ \\ \cline{2-7} 
		$\Omega_{ccb}$    &$\Omega_{ccb}(\frac{1}{2}^-)(1P)$ & $0.53{}^{+0.12}_{-0.11}$ & $0.39{}^{+0.11}_{-0.11}$&-&-& $0.57{}^{+0.11}_{-0.11}$ \\ \cline{2-7} 
		&$\Omega_{ccb}(\frac{1}{2}^+)(2S)$ &  $0.67{}^{+0.15}_{-0.12}$& $0.57{}^{+0.13}_{-0.11}$&-&-&-\\ \hline\hline
		\multirow{3}{*}{} &$\overline{\Omega}_{bbc}(\frac{1}{2}^+)(1S)$ & $0.68{}^{+0.24}_{-0.23}$ & $0.52{}^{+0.14}_{-0.22}$&$0.85{}^{+0.28}_{-0.28}$&$0.65{}^{+0.22}_{-0.22}$&$0.68{}^{+0.15}_{-0.15}$ \\ \cline{2-7} 
		$\overline{\Omega}_{bbc}$     &$\overline{\Omega}_{bbc}(\frac{1}{2}^-)(1P)$ & $0.86{}^{+0.23}_{-0.19}$ & $0.85{}^{+0.10}_{-0.17}$&-&-&$0.84{}^{+0.17}_{-0.17}$  \\ \cline{2-7} 
		&$\overline{\Omega}_{bbc}(\frac{1}{2}^+)(2S)$ & $0.98{}^{+0.33}_{-0.32}$ & $0.92{}^{+0.09}_{-0.09}$&-&-&- \\ \hline\hline
		\multirow{3}{*}{} &$\Omega_{bbc}(\frac{1}{2}^+)(1S)$ & $0.66{}^{+0.19}_{-0.15}$ & $0.50{}^{+0.13}_{-0.13}$&$0.53{}^{+0.17}_{-0.17}$&$0.45{}^{+0.15}_{-0.15}$ &$0.68{}^{+0.15}_{-0.15}$\\ \cline{2-7} 
		$\Omega_{bbc}$     &$\Omega_{bbc}(\frac{1}{2}^-)(1P)$ & $0.76{}^{+0.28}_{-0.33}$ & $0.73{}^{+0.11}_{-0.10}$&-&-&$0.86{}^{+0.17}_{-0.17}$  \\ \cline{2-7} 
		&$\Omega_{bbc}(\frac{1}{2}^+)(2S)$ & $0.95{}^{+0.14}_{-0.14}$ & $0.91{}^{+0.16}_{-0.14}$&-&-&-\\ 	\hline\hline
	\end{tabular}
\caption{The residue spectra (in units of $\mathrm{GeV^3}$) of the $\overline{\Omega}_{ccb}$, $\Omega_{ccb}$,  $\overline{\Omega}_{bbc}$ and $\Omega_{bbc}$ baryons  compared to other theoretical predictions.}
\label{copm r}
\end{table}
\section {CONCLUSION}\label{sec:four}
	
Calculation of mass and residue is crucial as they are among the most fundamental properties of  particles. Their values can be  used as inputs in various analyses of the  interactions and decays of particles. The mass and residue spectra of the triply heavy  spin-1/2  baryons  have been calculated using the QCD sum rule approach in the current work.  One of the advantages of this approach is its independence of arbitrary parameters, leading to a final result that is not affected by such choices. With the intent of improving accuracy in mass and residue calculations for the ground, the first orbital   and the first radial excited states, we extended the non-perturbative contribution to include the operators up to eight mass dimensions. Various predictions for the spectroscopic parameters of these states exist in the literature, but we need more information on the interactions/decays of these particles with/to other known states. Our results obtained with higher accuracies can be used in future related analyses. The obtained results for the parameters of the triply heavy baryons in their ground and excited states can also shed light on the search of different experimental groups for these states at various hadron colliders. Their identification in the experiment will be another impressive success in the colliders and comparison of the future data with theoretical predictions will provide a good insight into the non-perturbative nature of QCD as the successful theory of strong interaction. Such possible progresses will also  put the successes of the quark model to the top point as this model has predicted the ground and excited triply heavy baryons decades ago.   
	
	\section*{ACKNOWLEDGEMENTS}
 K. Azizi is thankful to Iran National Science Foundation (INSF)
	for the  partial financial support provided under the elites Grant No.  4025036.	
	
\section*{APPENDIX: Some expressions obtained in QCD side of the calculations}

Here, we bring forward the explicit forms of different components of the spectral densities $\rho_i(s)$  and parts of  $\Gamma_i(M^2)$ attained from calculations for both the structures:
\begin{equation}
	\begin{split}
		\rho^{pert}_{\not\!q}(s)&=\frac{3}{64 \pi ^4}
		\int^{1}_{0} dz \int^{1-z}_{0} dr \ \frac{D\,\Theta(D)}{U_1^4\, U_3^3} \Bigg\{\bigg[2\, m_Q^2\, U_1^2\, U_2\, U_3^2\, z + 2 \,m_Q\, m_{Q'}\, U_1^2 \,U_3 \,(U_3 + 4\, z) \,(U_1 + U_2\, z) \\
		&+ 
		z \,(10\, s\, U_2\, (U_3 - z) \,z \,(U_1 + U_2\, z)^2 - 
		D\, U_1^2\, (U_1\, (U_2 + 14\, U_3 - 14\, z) + 15\, U_2\, (U_3 - z)\, z))\bigg]\\
		&+2 \beta\bigg[-2\, m_Q^2\, U_1^2\, U_2\, U_3^2 + 2\, s\, U_2 \,(U_3 - z) \,z\, (U_1 + U_2\, z)^2 + 
		D\, U_1^2\, (3 \,U_2\, z \,(-U_3 + z)\\
		& + U_1\, (U_2 - 4\, U_3 + 4\, z))\bigg]	+	\beta^2\bigg[-2\, m_Q^2\, U_1^2\, U_2\, U_3^2\, z + 2\, m_Q \,m_{Q'}\, U_1^2\, U_3\, (U_3 + 4\, z) (U_1 + U_2\, z)\\
		& + 
		z\, (-10\, s\, U_2 (U_3 - z)\, z \,(U_1 + U_2 \,z)^2 + 
		D \,U_1^2 (U_1\, (U_2 + 14\, U_3 - 14\, z) + 15\, U_2\, (U_3 - z)\, z)) \bigg]
		\Bigg\},
	\end{split}
\end{equation}
	\begin{equation}
		\begin{split}
			\rho_I^{Pert}(s)&=\frac{3}{32 \pi ^4}
			\int^{1}_{0} dz \int^{1-z}_{0} dr \ \frac{D\, \Theta (D)}{U_1^3\, U_3^3} \Bigg\{\bigg[ -5 \,m_Q^2\, m_{Q'}\, U_1^2\, U_3^2 - m_Q \,s \,U_2 \,U_3\, (5 \,U_3 - 4\, z)\, z\, (U_1 + U_2 \,z) \\
			&- 
			m_{Q'}\, s\, (U_3 - z) \,z \,(U_1 + U_2\, z)^2 + 
			D \,U_1^2\, (m_Q \,U_3 \,(5\, U_3 - 4\, z) + m_{Q'}\, (U_1 + (U_3 - z)\, z))
			\bigg]\\
			&-2 \beta\,m_{Q'} \bigg[ m_Q^2\, U_1^2 \,U_3^2 - s\, (U_3 - z) \,z\, (U_1 + U_2\, z)^2 + 
			D\, U_1^2\, (U_1 + (U_3 - z)\, z)\bigg]\\
			&+
			\beta^2\bigg[-5\, m_Q^2\, m_{Q'}\, U_1^2\, U_3^2 + m_Q \,s\, U_2\, U_3\, (5\, U_3 - 4\, z)\, z \,(U_1 + U_2\, z)\\
			& - 
			m_{Q'}\, s\, (U_3 - z) \,z\, (U_1 + U_2\, z)^2 + 
			D\, U_1^2\, (m_Q\, U_3\, (-5\, U_3 + 4\, z)\\
			& + m_{Q'} (U_1 + (U_3 - z)\, z))\bigg]
			\Bigg\},
		\end{split}
	\end{equation}
	\begin{equation}
		\begin{split}
			\rho^{dim-4}_{\not\!q}(s)&=\frac{1}{192 \pi ^2}
			\int^{1}_{0} dz \int^{1-z}_{0} dr \ \Big\langle\frac{\alpha_{s}GG}{\pi}\Big\rangle\ \frac{\Theta(D)}{U_1^4\, U_3} \Bigg\{\bigg[8\,z\Big(-9 \,r^2\, (U_3 - z)\, (U_1 + U_2 \,z) + 
			3\, U_4\, z\, (3\, U_2\, z\, (-U_3 + z)\\
			& + U_1 \,(U_2 - 2\, U_3 + 2 \,z)) + 
			r \,(U_1\, U_3\, (9 + 68 \,z) - U_1 \,z\, (9 + 9 \,U_2 + 68\, z) + 
			U_2\, (U_3 - z) \,z \,(9 + 77\, z))\Big)\bigg]\\
			&+2\, \beta \bigg[3\,z\Big(U_4 \,z\, (27\, U_1\, U_2 + 50\, U_1\, (-U_3 + z) + 77 \,U_2\, z\, (-U_3 + z)) + 
			11\, r^2 \,(7\, U_2\, z\, (-U_3 + z)\\
			& + U_1\, (U_2 - 6\, U_3 + 6\, z)) + 
			r\, (-11\, U_1\, U_2 + 6\, U_1\, (U_3 - z) (11 + 12\, z) + U_2\, (U_3 - z)\, z \,(77 + 72\, z))\Big)\bigg]\\
			&	+	\beta^2\bigg[8\,z\Big(-9\, r^2\, (U_3 - z)\, (U_1 + U_2\, z) + 
			3\, U_4 \,z\, (3\, U_2\, z\, (-U_3 + z) + U_1 \,(U_2 - 2\, U_3 + 2\, z)) \\
			&+ 
			r\, (U_1 \,U_3\, (9 + 68\, z) - U_1\, z\, (9 + 9\, U_2 + 68 \,z) + 
			U_2\, (U_3 - z)\, z \,(9 + 77\, z))\Big) \bigg]
			\Bigg\},
		\end{split}
	\end{equation}
\begin{equation}
	\begin{split}
		\rho^{dim-4}_I(s)&=\frac{1}{192 \pi ^2}
		\int^{1}_{0} dz \int^{1-z}_{0} dr \ \Big\langle\frac{\alpha_{s}GG}{\pi}\Big\rangle\ \frac{\Theta(D)}{U_1^4\, U_3} \Bigg\{\bigg[m_Q \Big(6\, (U_3 - z)\, z^2\, (-3\, U_3^3 + 9\, U_3^2\, z - (9 - 9\, r + U_3)\, z^2 + 4\, z^3)\\
		& +
		U_1\, (-3\, U_3^3 \,(r + 4\, U_3) + 
		U_3^2\, (3 - 31\, r + 54\, U_3)\, z + (-36 + 64\, r - 57\, U_3)\, U_3\, z^2 + 
		68\, U_3\, z^3 - 24 \,z^4)\Big)\\
		& + 
		m_{Q'}\Big(-3\, r^2\, U_1 (U_1 + z\, (-U_3 + z)) - 3 \,U_1\, U_4\, z\, (U_1 + z\, (-U_3 + z)) + 
		3\, r\, U_1 \,(1 + z) \,(U_1 + z\, (-U_3 + z)) \\
		&+ 
		2\, r^3\, U_3\, (2\, U_1 + 3\, z\, (-U_3 + z))\Big)\bigg]
		+2\, \beta \bigg[m_{Q'}\,\Big(-3\, r^2\, U_1\, (U_1 + z\, (-U_3 + z)) - 3\, U_1\, U_4\, z\, (U_1 + z\, (-U_3 + z)) \\
		&+ 
		3\, r\, U_1 (1 + z) (U_1 + z\, (-U_3 + z)) + 2\, r^3 \,U_3\, (2\, U_1 + 3 \,z (-U_3 + z))\Big)\bigg]\\
		&	+	\beta^2\bigg[m_Q\,\Big(6\, (U_3 - z) \,z^2\, (5\, U_3^3 - 15\, U_3^2\, z + (15 - 15\, r + U_3)\, z^2 - 
		6 \,z^3) + 
		U_1\, (-3\, (-1 + r)\, r\, U_3^2 + 20\, U_3^4\\
		& + 
		U_3^2\, (31\, r - 3\, (30\, U_3 + U_4))\, z + 
		2\, U_3\, (30 - 44\, r + 45\, U_3)\, z^2 - 112 \,U_3\, z^3 + 36\, z^4)\Big)\\
		& + 
		m_{Q'}\, \Big(-3 \,r^2 \,U_1\, (U_1 + z\, (-U_3 + z)) - 3\, U_1\, U_4\, z \,(U_1 + z\, (-U_3 + z)) + 
		3 \,r\, U_1 (1 + z)\, (U_1 + z (-U_3 + z))\\
		& + 
		2 \,r^3\, U_3 \,(2\, U_1 + 3\, z\, (-U_3 + z))\Big) \bigg]
		\Bigg\}.
	\end{split}
\end{equation}
$\Gamma_i(M^2) $ are displayed in  6 and 8 mass dimensions, as follows:
\begin{eqnarray}
	\Gamma_i^{dim-j}(M^2)=\Gamma_{i,1}^{dim-j}(M^2)+\,\beta \,\Gamma_{i,2}^{dim-j}(M^2)+\,\beta^2\,\Gamma_{i,3}^{dim-j}(M^2),~~~~~~~~~i= \not\!q~\mbox{or}~I \,~\mbox{and} \,~j= 6~\mbox{or}~8.
	\label{Eq:dim}
\end{eqnarray}
Due to the lengthy expression of $\Gamma_i^{dim-j}(M^2)$, we only explicitly write the coefficient of $ \Gamma_{\not\!q,2}^{dim-j}(M^2)	$
\begin{eqnarray}
	\Gamma^{dim-6}_{\not\!q,2}(M^2)&=&\frac{1}{30720 \, \pi ^4 \, M^6}
	\int^{1}_{0} dz \int^{1-z}_{0} dr \, \frac{\langle g_s^3 G^3 \rangle}{U_1^5\, U_2^6 \, U_3^8\,r^7 \,z^3}\, e^{-\frac{K}{M^2}}\Bigg\{2\,M^6\,
	 U_2^3\,r^3\,z^3\,V_1 \nonumber \\
	&+&
	2\,m_Q^2\,M^4\,U_1\,U_2^2\,U_3\,r^2\,z^2\,V_2
	+ 2\,m_{Q'}^2\,M^4\, U_1\,U_2^2\,r^3\,z^2\,V_3\nonumber \\ 
&	+&
	m_{Q}^4\,M^2\,  U_1^2\, U_2\, U_3^2\,r\, z\,V_4
	+2\,m_{Q'}^2\,m_Q^2\,M^2\,U_1^2\,U_2\,U_3\,r^2\,z\,V_5\nonumber \\
	&+&m_{Q'}^4\,M^2\, U_1^2\,U_2\,r^3\,z\,V_6+m_Q^6\,U_1^3 U_3^3\,V_7 \nonumber \\
&+&m_{Q'}^2\,m_Q^4\, U_1^3 \,U_2\, U_3^2\,r\,V_8+m_{Q'}^4\,m_{Q}^2\,U_1^3\,U_3\,r^2\,V_9	+ m_{Q'}^6 \, U_1^3\,r^3\,V_{10}\Bigg\},
\end{eqnarray}
\begin{eqnarray}
	\Gamma^{dim-8}_{\not\!q,2}(M^2)&=&\frac{1}{55296  \, M^8}
	\int^{1}_{0} dz \int^{1-z}_{0} dr \, \frac{\langle \frac{\alpha_s}{\pi} G^2 \rangle\,^2}{U_1^2\, U_2^6 \, U_3^4\,r^6 \,z^4}\, e^{-\frac{K}{M^2}}\Bigg\{
	m_Q^2\,M^4\,U_2^2\,r\,z^3\,S_1
	+ 3\,m_{Q'}^2\,M^4\, U_1\,U_2^4\,U_3^2\,r^5\,z^3\,S_2\nonumber \\ 
	&	-&
	m_{Q}^4\,M^2\,  U_1\, U_2\, U_3\, z^2\,S_3
	-\,m_{Q'}^2\,m_Q^2\,M^2\,U_1\,r\,z\,S_4
	-57\,m_{Q'}^4\,M^2\, U_1^2\,U_2^4\,U_3^4\,r^7\,z^2-32\,m_Q^6\,U_1^2\,U_2^3\, U_3^5\,r^2\,z^3\,S_5 \nonumber \\
	&+&32\,m_{Q'}^2\,m_Q^4\, U_1^2 \, U_3^4\,r^3\,S_6 +32\,m_{Q'}^4\,m_{Q}^2\,U_1^2\,U_3^4\,r^4\,S_7	\Bigg\},
\end{eqnarray}
and we need to define
\begin{eqnarray}
	V_1&=&6\, r^2\, U_1\, (r^5 + U_1^2 - r \,U_1^2)\, U_2^2\, (U_1 + 5\, r\, U_2)\, U_3^6\nonumber \\
	& + &
	2\, r\, U_2\, U_3^5\, \bigg[-3\, r^6\, U_1^2\, U_2 + 4\, U_1^5 \,U_3 - 
	r\, U_1^2\, \big[4\, (-1 + r)\, r^4 + 8\, U_1^2\, U_2 + 81\, r\, U_1\, U_2^2\big]\, U_3\nonumber \\
	& + &
	5\, r^6\, U2\, (4 \,U_1 + 3\, r\, U_2)\, U_3^2 + 
	3 \,r\, U_1^2\, U_2\, (2 \,U_1 + 3\, r\, U_2) \,U_3^3\bigg]\, z\nonumber \\
	& - &
	r\, U_3^5\, \bigg[U_1^5\, (43\, U_2 + 3\, U_3) + 2\, U_1^4\, U_2\, (7\, r\, U_2 - 16\, U_3^2) + 
	5\, r^5\, U_1 \,U_2 \,U_3\, (16\, r \,U_2 - 5\, U_3^2)\nonumber \\
	& + &
	30\, r^6 \,U_2^2\, U_3\, (r\, U_2 - U_3^2) + 
	3\, r \,U_1^2 \,U_3\, (r^3 + 2\, U_2^2 \,U_3)\, (8\, r\, U_2 - U_3^2) + 
	12\, r\, U_1^3\, U_2^2\, (-29\, r\, U_2 + U_3^2)\bigg]\, z^2\nonumber \\
	& +& 
	U_3^4\, \bigg[4 \,r\, U_1^3\, U_2\, (25 \,U_1^2 + 23\, r \,U_1\, U_2 - 93\, r^2\, U_2^2) + 
	U_1\, (15\, U_1^4 + 24\, r^6\, U_1\, U_2 + 40\, r^7\, U_2^2)\, U_3\nonumber \\
	& - &
	21\, r\, U_1^2\, U_2\, (9\, U_1^2 + 8\, r \,U_1 \,U_2 + 2\, r^2\, U_2^2) \,U_3^2 - 
	5\, r^5\, (3\, U_1^2 + 20\, r\, U_1\, U_2 + 18\, r^2\, U_2^2)\, U_3^3\nonumber \\
	& + &
	12\, r\, U_1\, U_2\, (4\, U_1^2 + 3\, r\, U_1\, U_2 + 4\, r^2\, U_2^2)\, U_3^4 + 
	3\, r^5\, (2 \,U_1 + 5\, r \,U_2) \,U_3^5\bigg] \,z^3\nonumber \\
	& +& 
	U_3^3\, \bigg[2\, r \,U_1^3 \,U_2 \,(-61\, U_1^2 - 51 \,r \,U_1\, U_2 + 93\, r^2\, U_2^2) + 
	30\, (-3 + 2\, r) \,U_1^5 \,U_3\nonumber \\
	& - &
	8\, r\, \Big(-4 + 5\, r\, \big[2 + (-2 + r)\, r\big]^2\Big)\, U_1^2\, U_2\, U_3 + 
	r\, U_1^2\, U_2\, (497\, U_1^2 + 612\, r\, U_1\, U_2 + 276\, r^2\, U_2^2)\, U_3^2\nonumber \\
	& + &
	30\, r^5\, (U_1^2 + 5\, r \,U_1\, U_2 + 3\, r^2\, U_2^2)\, U_3^3 - 
	3\, U_1\, (-15 \,U_1^3 + 98\, r\, U_1^2\, U_2 + 118\, r^2\, U_1\, U_2^2 + 
	86 \,r^3\, U_2^3) \,U_3^4\nonumber \\
	& -&
	3\, r^5\, (12\, U_1 + 25\, r\, U_2)\, U_3^5 + 
	32\, r^2 \,U_1 \,U_2^2\, U_3^6 + 3 \,r^5\, U_3^7\bigg] \,z^4\nonumber \\
	& + &
	U_3^2\, \bigg[2 \,r U_1^4 \,U_2\, (38\, U_1 + 17\, r\, U_2) + 15\, (15 - 13 \,r)\, U_1^5\, U_3 - 
	6 \,r\, U_1^2\, U_2\, (123\, U_1^2 + 146\, r \,U_1 \,U_2 + 51\, r^2\, U_2^2)\, U_3^2\nonumber \\
	& - &
	2\, r\, \bigg(15\, r^4 \,U_1^2 + 
	2\, \Big[-2 + r\, \Big(10 + r\, \big[-20 + r\,\big (20 + r\, (-10 + 27\, r)\big)\big]\Big)\Big]\, U_1\, U_2 + 
	15\, r^6\, U_2^2\bigg)\, U_3^3\nonumber \\
	& + &
	3\, U_1\, (-105\, U_1^3 + 268 \,r\, U_1^2\, U_2 + 366\, r^2 \,U_1\, U_2^2 + 
	200\, r^3\, U2^3)\, U_3^4 + 30\, r^5\, (3 \,U_1 + 5\, r\, U_2)\, U_3^5\nonumber \\
	& +& 
	2\, (30 \,U_1^3 - 89\, r\, U_1^2\, U_2 - 88\, r^2\, U_1\, U_2^2 + 15\, r^3\, U_2^3)\, U_3^6 - 
	21\, r^5\, U_3^7\bigg]\, z^5 \nonumber \\
	&+ &
	U_3\, \bigg[-19\, r\, U_1^5\, U_2 + 15\, (-20 + 19\, r) \,U_1^5\, U_3 + 
	6\, r\, U_1^2\, U_2\, (107\, U_1^2 + 96\, r\, U_1\, U_2 + 17\, r^2\, U_2^2)\, U_3^2\nonumber \\
	& +& 
	5\, U_1\, \bigg(\Big[9 - 3\, r\,\Big (15 + r\, \big[-30 + r\, \big(30 + r\, (-15 + 2\, r)\big)\big]\Big)\Big]\, U_1 \nonumber \\
	&+ &
	r\, \Big[-3 + r\, \Big(15 + r\, \big[-30 + r \,\big(30 + r\, (-15 + 8\, r)\big)\big]\Big)\Big]\, U_2\bigg)\, U_3^3\nonumber \\
	& - &
	3 U_1\, (-315\, U_1^3 + 430 \,r\, U_1^2\, U_2 + 592\, r^2\, U_1\, U_2^2 \nonumber \\
	&+ &
	244\, r^3\, U_2^3) \,U_3^4 - 30\, r^5\, (4\, U_1 + 5\, r \,U_2)\, U_3^5\nonumber \\
	& +& 
	2\, (-240 \,U_1^3 + 191\, r\, U_1^2\, U_2 + 208\, r^2\, U_1\, U_2^2 - 
	75\, r^3\, U_2^3) \,U_3^6 + 63\, r^5 \,U_3^7 + 30\, r^2\, U_2^2\, U_3^8\bigg]\, z^6 \nonumber \\
	&+ &
	U_3\, \bigg[3\, (75 - 74\, r) \,U_1^5 - r\, U_1^3\, U_2\, (305\, U_1 + 144\, r\, U_2)\, U_3\nonumber \\
	& + &
	3\, \Big[-135 + 
	r\, \Big(675 + r\, \big[-1350 + r\, \big(1350 + r\, (-675 + 134\, r)\big)\big]\Big)\Big]\, U_1^2 \,U_3^2 \nonumber \\
	&+ &
	3 \,U_1\, (-525\, U_1^3 + 440\, r\, U_1^2\, U_2 + 538 \,r^2 \,U_1\,U_2^2 + 
	152\, r^3\, U_2^3)\, U_3^3\nonumber \\
	& + &
	3\, \Big[1 + r\,\Big( -5 + r\, \big[10 + r\, \big(-10 + r\, (5 + 4\, r)\big)\big]\Big)\Big]\, (6\, U_1 + 
	5\, r\, U_2) \,U_3^4\nonumber \\
	& + &
	2 \,(840 \,U_1^3 - 173\, r\, U_1^2\, U_2 - 280\, r^2\, U_1\, U_2^2 + 
	150 \,r^3\, U_2^3) \,U_3^5 - 105\, r^5\, U_3^6 - 
	10\, r\, U_2\, (16\, U_1 + 21\, r\, U_2)\, U_3^7\bigg]\, z^7 \nonumber \\
	&+& 
	U3 \,\bigg[U_1^4\, (-90\, U_1 + 61\, r \,U_2) - 
	3\, U_1\, (-525\, U_1^3 + 286\, r\, U_1^2\, U_2 + 260\, r^2\, U_1\, U_2^2 + 
	38\, r^3\, U_2^3)\, U_3^2 \nonumber \\
	&+& 
	3 \,\bigg(12\, \Big[-5 + r\, \big(5 + (-5 + r)\, r\big)\,\big (5 + r\, (-5 + 4\, r)\big)\Big]\, U_1 \nonumber \\
	&+ &
	5\, r\,\Big[-9 + 
	r\, \Big(45 + r\, \big[-90 + r\, \big(90 + r\, (-45 + 8\, r)\big)\big]\Big)\Big]\, U_2\bigg) \,U_3^3\nonumber \\
	& - &
	10 \,(336\, U_1^3 + r \,U_1^2 \,U_2 - 48\, r^2 \,U_1 \,U_2^2 + 30\, r^3\, U_2^3) \,U_3^4\nonumber \\
	& +& 
	3\,\Big[1 + r\, \Big(-5 + r\, \big[10 + r\, \big(-10 + r\, (5 + 34\, r)\big)\big]\Big)\Big]\, U_3^5
	+ 
	2\, (810\, U_1^2 + 434\, r\, U_1\, U_2 + 315\, r^2\, U_2^2)\, U_3^6\bigg]\, z^8\nonumber \\
	& +&
	\bigg[-33 + 
	15\, r^9 + 30\, r^10 + r^7\, (3330 - 6462 \,U_1) + r^8\, (-855 + 804\, U_1) - 
	9\, r^6\, (735 - 2518\, U_1 + 420\, U_1^2)\nonumber \\
	& +& 
	3\, r^5 \,(2751 + 2 \,U_1\, (-7559 + 3780\, U_1)) + 
	210 \,r^4\, \Big(-33 + 10 \,U_1\, \big[27 + U_1 \,(-27 + 2\, U_1)\big]\Big)\nonumber \\
	& + &
	15\, U_1 \,|big(54 + U_1\, \big[-252 + U_1\, (280 + (-63 + U_1)\, U_1)\big]\Big)\nonumber \\
	& -& 
	30\, r^3\, \Big(-132 + 56 \,U_1\, \big[27 + 5 \,U_1\, (-9 + 2\, U_1)\big] - 5\, U_2^3\, U_3^4\Big)\nonumber \\
	& - &
	r^2 \,\Big[45 \,\Big(33 + 7 \,U_1\, \big[-72 + U_1\, \big(180 + U_1\, (-80 + 3\, U_1)\big)\big]\Big)\nonumber \\
	& - &
	156 \,U_1^2\, U_2^2 \,U_3^2 + 272\, U_1\, U_2^2\, U_3^4 + 1050\, U_2^2\, U_3^6\Big]\nonumber \\
	& +& 
	2\, r\, \big[165 + 945 \,U_1^4 + 270\, U_2 \,U3^8 + 6\, U_1^3\, (-1400 + 27\, U_2\, U_3^2)\nonumber \\
	& + &
	U_1^2 \,(11340 + 149\, U_2\, U3^4) - 72\, U_1\, (45 + 14\, U_2\, U_3^6)\big]\bigg] \,z^9\nonumber \\
	& + &
	U_3\, \bigg[9 \,U_1^3 \,(35\, U_1 - 6\, r\, U_2) - 
	2\, (1680\, U_1^3 + 131\, r\, U_1^2 \,U_2 - 48\, r^2\, U_1\, U_2^2 + 
	15\, r^3\, U_2^3)\, U_3^2\nonumber \\
	& - &
	3 \,\Big[-55 + r\, \Big(275 + r\, \big[-550 + r\, \big(550 + r\, (-275 + 48\, r)\big)\big]\Big)\Big]\, U_3^3
	+
	14\, 405\, U_1^2 + 193\, r\, U_1\, U_2 + 75\, r^2 \,U_2^2\, U_3^4\nonumber \\
	& - &
	180\, (12 \,U_1 + 7\, r\, U_2)\, U_3^6\bigg]\, z^{10} + \bigg[-45\, U_1^4 + 
	2\, U_1 \,(840\, U_1^2 + 49\, r\, U_1\, U2 - 8 \,r^2\, U_2^2)\, U_3^2\nonumber \\
	& + &
	3\, \Big[-165 + 
	r\, \Big(825 + r\, \big[-1650 + r\, \big(1650 + r\, (-825 + 164\, r)\big)\big]\Big)\Big]\, U_3^3\nonumber \\
	& -& 
	70\, (81\, U_1^2 + 32 \,r \,U_1\, U_2 + 9\, r^2 \,U_2^2) \,U_3^4 + 
	1890\, (2\, U_1 + r\, U_2)\, U_3^6\bigg] \,z^{11} \nonumber \\
	&+& 
	2\, U\,3 \bigg[-U_1^2\, (240\, U_1 + 7\, r\, U_2) + 
	15 \,(126\, U_1^2 + 38 \,r \,U_1\, U_2 + 7\, r^2\, U_2^2)\, U_3^2\nonumber \\
	& - &
	189\, (12\, U_1 + 5 \,r\, U_2)\, U_3^4 + 495\, U_3^6\bigg]\, z^{12} + 
	2\, \bigg[30\, U_1^3 - (810\, U_1^2 + 164\, r\, U_1\, U_2 + 15 \,r^2\, U_2^2)\, U_3^2\nonumber \\
	& +& 
	630 \,(3\, U_1 + r\, U_2) \,U_3^4 - 693\, U_3^6\bigg] \,z^{13} + 
	U3\, \bigg[U_1\, (405 \,U_1 + 41\, r\, U_2) - 540\, (4\, U_1 + r\, U_2)\, U_3^2\nonumber \\
	& + &
	1386\, U_3^4\bigg] \,z^{14} - 
	45\, \bigg[U_1^2 - 3\, (6 \,U_1 + r\, U_2)\, U_3^2 + 22\, U_3^4\bigg]\, z^{15} - 
	15\, U_3\, \bigg[12\, U_1 + r\, U_2 - 33\, U_3^2\bigg] \,z^{16}\nonumber \\
	& +& 3\, \bigg[6 U_1 - 55\, U_3^2\bigg]\, z^{17} + 
	33\, U_3\, z^{18} - 3\, z^{19},
\end{eqnarray}
\begin{eqnarray}
	V_2&=&9\, r^{15}\, z^4 
	+
	9 U_4^7\, z^4 \,(U_1 + z - z^2) \,(10\, U_1^3 + 10\, U_1^2\, U_4\, z + 5\, U_1\, U_4^2 \,z^2 + 
	U_4^3\, z^3)\nonumber \\
	& + &
	r\, U_4^4\, z^2\, \bigg[-6\, U_1^4\, U_2\, U_3 - 
	12\, U_1^3 \,U_4\, \big[U_2\, (U_3 - 11\, U_1\, U_3) + 3\, U_1\, U_4\big]\, z\nonumber \\
	& - &
	2\, U_1^2\, \big[4\, U_2\, U_3 + 9\, U_1\, (35\, U_1 - 18\, U_2\, U_3)\big]\, U_4^2\, z^2 - 
	2\, U_1\, (720\, U_1^2 + U_2\, U_3 - 158\, U_1\, U_2\, U_3)\, U_4^3\, z^3\nonumber \\
	& + &
	U_1\, (-1215 \,U_1 + 154\, U_2 \,U_3)\, U_4^4\, z^4 + 
	30\, (-18\, U_1 + U_2\, U_3) \,U_4^5\, z^5 - 99 \,U_4^6 \,z^6\bigg]\nonumber \\
	& + &
	3\, r^{14}\, z^3\, \bigg[-6\, U_1 + 10\, U_2\, U_3 + 3\, z\, (-10 + 7 \,z)\bigg] + 
	r^{13}\, z^2\, \bigg[9\, U_1^2 + 30\, U_2^2\, U_3^2 + 30\, U_2 \,U_3 \,z (-8 + 5\, z)\nonumber \\
	& + &
	U_1\, (-53\, U_2\, U_3 + 54\, (3 - 2\, z)\, z) + 27 \,z^2\, (15 + 7\, (-3 + z)\, z)\bigg]
	+ 
	r^2\, U_4\, \bigg[15\, U_4^6 \,z^6\, (2\, U_2^2\, U_3^2 - 18\, U_2\, U_3\, U_4 \,z + 33\, U_4^2\, z^2)\nonumber \\
	& + &
	2\, U_1\, U_4^4\, z^4\, \big[1215\, U_4^3\, z^3 + 8 \,U_2^2\, U_3^2 (-1 + 11 \,U_4\, z) + 
	U_2\, U_3\, U_4\, z\, (9 - 618\, z + 616\, z^2)\big]\nonumber \\
	& +& 
	U_1^4\, \big[-2\, U_2^2\, U_3^2\, (-1 + 5\, U_4\, z)\, (-6 + 13\, U_4\, z) + 
	9\, U_4^4 \,z^2\, (1 + 28 \,z + 206 \,z^2)\nonumber \\
	& +& 
	2\, U_2\, U_3 \,U_4^2 \,z\, (1 + z - 319\, z^2 + 330\, z^3)\big] + 
	12\, U_1^3\, U_4 \,z\, \Big(420\, U_4^4\, z^4 + 
	U_2\, U_3\, U_4^2\, z^2\, \big[7 + 2\, z\, (-82 + 81\, z)\big]\nonumber \\
	& +& 
	U_2^2\, U_3^2\, \big[-2 + U_4\, z\, (13 - 12\, U_4\, z)\big]\Big) + 
	2\, U_1^2\, U_4^2\, z^2\, \Big(2430\, U_4^4 \,z^4 + 
	2\, U_2\, U_3 \,U_4^2\, z^2 \,\big[16 + z \,(-557 + 553\, z)\big]\nonumber \\
	& + &
	3 \,U_2^2\, U_3^2\, \big[-2 + U_4\, z\, (9 + 22\, U_4\, z)\big]\Big)\bigg] + 
	r^{12}\, z\, \bigg[U_1^2\, \big[16\, U_2\, U_3 + 9\, z\, (-8 + 5\, z)\big] \nonumber \\
	&-& 
	U_1\, \big[48\, U_2^2\, U_3^2 + 53\, U_2\, U_3\, z\, (-7 + 4 \,z) + 
	54\, (-2 + z) \,z^2 \,(-6 + 5\,z)\big]\nonumber \\
	& +& 
	3\, z\,\Big (30\, U_2^2\, U_3^2 \,(-2 + z) + 10\, U_2\, U_3\, z\, \big[28 + 5\, z\, (-7 + 2 \,z)\big] + 
	3\, z^2 \,\big[-120 + 7\, z\, \big(36 + z\, (-24 + 5\, z)\big)\big]\Big)\bigg]\nonumber \\
	& +& 
	r^{10}\, \bigg[3\, U_1^2\, \Big(2 \,U_2^2 \,U_3^2 \,(-4 + z) + 16\, U_2\, U_3\, z\, \big[5 +\, (-5 + z)\, z\big] + 
	3\, z^2\, \big[-56 + 5 \,z \,(21 + 2\, (-6 + z) \,z)\big]\Big)\nonumber \\
	& +& 
	U_1\, \Big[18\, U_2^3\, U_3^3 - 48\,U_2^2\, U_3^2\, z\, \big[10 + (-8 + z)\, z\big] - 
	5\, U_2\, U_3\, z^2\,\Big(-371 + 6\, z\, \big[106 + z\, (-53 + 7\, z)\big]\Big)\nonumber \\
	& - &
	54\, z^3\, \Big(42 + z\, \big[-112 + z\, \big(105 + 4\, (-10 + z)\, z\big)\big]\Big)\Big] + 
	3\, z^2\, (10\, U_2^2\, U_3^2\, (-2 + z)\, (10 + (-10 + z)\, z)\nonumber \\
	& + &
	10\, U_2\, U_3\, z\, (70 + z\, (-175 + 2\, (-5 + z)\, z\, (-15 + 2\, z)))\nonumber \\
	& - &
	3\, U_4\, z^2\, (252 + z\, (-630 + z\, (546 + z\, (-189 + 10 \,z)))))\bigg] + 
	r^5\, \bigg[2\, U_1\, U_4\, z^3\, \Big(9\, U_4^4\, (U_4 - 756\, z^4)\nonumber \\
	& + &
	40\, U_2^2\, U_3^2\, U_4\, z \,\big[7 + 5\, z\, (-10 + 9\, z)\big] - 
	18\, U_2^3\, U_3^3\, \big[10 + z\, (-32 + 25\, z)\big]\nonumber \\
	& - &
	7\, U_2\, U_3\, U_4^2\, z^2\, \big[18 + z\, (-786 + 773\, z)\big]\Big) + 
	U_1^4\, \Big[-45\, U4^2 \,z^2\, \big[7 + z\, (22 + 27\, z)\big]\nonumber \\
	& +& 
	2\, U_2^2\, U_3^2\, \big[60 + z\, (-196 + 151\, z)\big] + 
	2\, U_2\, U_3\, z\, \Big(-20 + z\, \big[155 + (106 - 245\, z) \,z\big]\Big)\Big]\nonumber \\
	& + &
	6\, U_1^3\, \Big[-1680\, U_4^3\, z^5 - 9\, U_2^3\, U_3^3\, \big[3 + z\, (-9 + 7\, z)\big] - 
	10\, U_2\, U_3\, U_4\, z^3\, \big[7 + z\, (-89 + 83\, z)\big]\nonumber \\
	& +& 
	2\, U_2^2\, U_3^2 \,z\, \Big(40 + z\, \big[-170 + (212 - 87\, z)\, z\big]\Big)\Big]\nonumber \\
	& +& 
	3\, U_4^4\, z^4\, \Big(3 - 
	z\, \big[9 + z \,\big(350\, U_2^2\, U_3^2 - 1260\, U_2\, U_3\, U_4\, z + 
	3\,(-3 + z + 462\, U_4^2 \,z^2)\big)\big]\Big)\nonumber \\
	& + &
	U_1^2\, z \,\Big(9\, U_4^4\, z \,(U_4 - 1890\, z^4) - 
	20\, U_2\, U_3\, U_4^2\, z^3\, \big[28 + z\, (-581 + 559\, z)\big]\nonumber \\
	& +& 
	18\, U_2^3\, U_3^3\,\big(-15 + z\, \big[57 + 8\, z\, (-9 + 4\, z)\big]\Big) - 
	12\, U_2^2\, U_3^2 \,z\, \Big[-35 + 
	2\, z\, \big(75 + 2 \,z\, \big[-25 + z\, (-27 + 23 \,z)\big]\big)\Big]\Big)\bigg] \nonumber \\
	&+& 
	3\, r^{11}\, \bigg[U_1^2\, \Big(2\, U_2^2\, U_3^2\, + 16\, U_2\, U_3 (-2 + z)\, z + 
	3\, z^2\, \big[28 + 5\, z\, (-7 + 2\, z)\big]\Big)\nonumber \\
	& +& 
	U_1\, \Big(-2 \,U_2^3\, U_3^3 + 16\, U_2^2\, U_3^2\, (5 - 2\, z)\, z - 
	53\, U_2\, U_3 \,z^2\, \big[7 + 2\, (-4 + z)\, z\big]1 + 
	6\, z^3\, \big[84 + z \,(-168 + 5\, (21 - 4\, z)\, z)\big]\Big)\nonumber \\
	& + &
	z^2\, \Big(30\, U_2^2 \,U_3^2\, \big[5 + (-5 + z) \,z\big] + 
	10\, U_2 \,U_3\, z\, \big[-56 + 5\, z\, (21 + 2\, (-6 + z)\, z)\big]\nonumber \\
	& +& 
	3\, z^2\, \Big[210 + z\,\big (-588 + z\, \big[588 + z\, (-245 + 34\, z)\big]\big)\Big]\Big)\bigg]\nonumber \\
	& + &
	r^9\, \bigg[-9 \,U_1^4 \,z^2 + 
	U_1^2\, \Big[6\, U_2^2\, U_3^2\, \big[6 + z\, (-3 + 2\, z)\big]\nonumber \\
	& + &
	8\, U_2 \,U_3\, z\, \big[-40 + z \,(60 + (-24 + z)\, z)\big] - 
	45\, z^2\, \Big(-14 + z\, \big[35 + 2 \,z\, \big(-15 + z\, (5 + z)\big)\big]\Big)\Big]\nonumber \\
	& +& 
	U_1\, \Big[-18\, U_2^3\, U_3^3 + 16\, U_2^2 \,U_3^2\, z \big[30 + (-3 + z)\, z\, (12 + z)\big] - 
	54\, U_4\, z^3\, \Big(-42 + z\, \big[98 + z\, \big(-77 + z\, (23 + 8\, z)\big)\big]\Big) \nonumber \\
	&+& 
	5\, U_2\, U_3\, z^2\, \Big(-371 + z\, \big[848 + z\, \big(-636 + z\, (166 + 23\, z)\big)\big]\Big)\Big]\nonumber \\
	& + &
	3\, z^2\, \Big(-10\, U_2^2\, U_3^2\, \Big(-15 + z\, \big[30 + (-3 + z)\, z\, (6 + z)\big]\Big)\nonumber \\
	& + &
	10\, U_2\, U_3\, U4\, z\, \Big(-56 + z\, \big[119 + z \,\big(-81 + z\, (19 + 8 \,z)\big)\big]\Big)\nonumber \\
	& - &
	3\, U_4^2\, z^2\, \big[-210 + z\, \big(462 + z \,\big[-336 + z\, (91 + 48\, z)\big]\big)\big]\Big)\bigg] \nonumber \\
	&+ &
	r^3\, \bigg[-15\, U_4^6\, z^6\, (14\, U_2^2\, U_3^2 - 72\, U_2\, U_3\, U_4\, z + 99\, U_4^2\, z^2) + 
	2\, U_1\, U4^3\, z^3\, \Big(-18\, U_2^3\, U_3^3 \,(1 - 2 z)^2 - 3240\, U_4^4\, z^4\nonumber \\
	& +& 
	8\, U_2^2\, U_3^2\, U4\, z\, (7 - 68 \,z + 66\, z^2) - 
	3\, U_2\, U_3\, U_4^2\, z^2 \,\big[12 + z \,(-724 + 719\, z)\big]\Big)\nonumber \\
	& + &
	3\, U_1^3\, \Big[-3360 \,U_4^5\, z^5 + 9 \,U_2^3 \,U_3^3\, (1 - 2 z)^2\, (-1 + 2 \,U_4\, z) - 
	4\, U_2\, U_3\, U_4^3\, z^3\, \big[21 + z\, (-417 + 406 \,z)\big]\nonumber \\
	& +& 
	4\, U_2^2\, U_3^2\, U_4\, z\, \Big(12 + 
	z\, \big[-73 + z\, \big(139 + 2 \,z \,(-61 + 24\, z)\big)\big]\Big)\Big]\nonumber \\
	& + &
	U_1^2\, U_4\, z\, \Big[-11340 \,U_4^5\, z^5 + 
	9\, U_2^3\, U_3^3\, (1 - 2\, z)^2\,(-3 + 4\, U4\, z)\nonumber \\
	& -& 
	4\, U_2\, U_3\, U_4^3\, z^3 \,\big[56 + z\, (-1687 + 1661\, z)\big] - 
	12\, U_2^2\, U_3^2\, U_4\, z\, \Big(-7 + z\, \big[31 + z\, \big(19 + z\, (-101 + 55 \,z)\big)\big]\Big)\Big]\nonumber \\
	& + &
	U_1^4\, \Big(-9 \,U_4^4\, z^2\, (7 + 82\, z + 326 \,z^2) + 
	2\, U_2\, U_3\, U_4^2 \,z\, \big[-6 + z\, (43 + 588\, z - 646\, z^2)\big]\nonumber \\
	& +& 
	2 \,U_2^2\, U_3^2\, \Big[30 + z\, \big(-196 + z\, \big[453 + z\, (-476 + 195\, z)\big]\big)\Big]\Big)\bigg]\nonumber \\
	& + &
	r^4\, \bigg[90\, U_4^5\, z^6\, (7\, U_2^2\, U_3^2 - 28\, U_2\, U_3 \,U_4\, z + 33\, U_4^2\, z^2)\nonumber \\
	& + &
	2\, U_1\, U_4^2\, z^3\, \Big(5670 U_4^4\, z^4 + 
	18\, U_2^3\, U_3^3 \,(-1 + 2 \,z)\, (-5 + 8\, z) - 
	8\, U_2^2 \,U_3^2\, U_4 \,z\, \big[21 + z\, (-177 + 166\, z)\big]\nonumber \\
	& +& 
	7 \,U_2\, U3 \,U_4^2 \,z^2 \,\big[12 + z\, (-624 + 617\, z)\big]\Big) + 
	U_1^4\, \Big(9\, U_4^3\, z^2\, \big[21 + z\, (128 + 291\, z)\big]\nonumber \\
	& -& 
	2\, U_2\, U_3\, U_4\, z\, \big[-15 + U_4\, z\, (125 + 604\, z)\big] + 
	2\, U_2^2\, U_3^2\, \big[-60 + z\,\big (294 + z\, (-453 + 238\, z)\big)\big]\Big)\nonumber \\
	& +& 
	6\, U_1^3\, \Big[2100\, U_4^4\, z^5 - 
	9\, U_2^3\, U_3^3\, (-1 + 2 \,z)\, \big[2 + z\, (-5 + 4\, z)\big] + 
	10\, U_2\, U_3\, U_4^2\, z^3\, \big[7 + z\, (-114 + 109\, z)\big]\nonumber \\
	& -& 
	2\, U_2^2\, U_3^2\, z\,\Big(30 + z\, \big[-170 + z\, \big(318 + z\, (-261 + 85\, z)\big)\big]\Big)\Big]\nonumber \\
	& + &
	U_1^2\, z\, \Big(17010 \,U_4^5\, z^5 + 
	4\, U_2\, U_3\, U_4^3\, z^3\, \big[112 + z\, (-2849 + 2777\, z)\big]\nonumber \\
	& + &
	9\, U_2^3\, U_3^3\, (-1 + 2 \,z)\, \Big(-15 + 2\, z\, \big[23 + 2\, z\, (-13 + 6\, z)\big]\Big) \nonumber \\
	&+& 
	6\, U_2^2\, U_3^2\, U_4\, z\, \big[-42 + 
	z\, \big(183 + z \,\big[-17 + z \,(-341 + 211 \,z)\big]\big)\Big]\Big)\bigg]\nonumber \\
	& + &
	r^7\, \bigg[-3 \,U_1^3\, \big[9\, U_2^3\, U_3^3 + 4\, U_2\, U_3\, (7 - 32\, z)\, z^3 + 480\, U_4\, z^5 + 
	4\, U_2^2\, U_3^2\, z\, (-12 + 17\, z)\big]\nonumber \\
	& + &
	U_1^4\, \big[12\, U_2^2\, U_3^2 + 2\, U_2\, U_3\, z\, (-6 + 17\, z) + 
	9\, z^2\, (-21 + 2\, z + 4\,z^2)\big] + 
	3\, U_4^2\, z^2\, \Big[10\, U_2^2\, U_3^2\, (U_4 - 21\, z^4)\nonumber \\
	& + &
	10\, U_2\, U_3\, U_4\, z\, \big[-8 + z\, (11 - 3\, z + 84\, z^3)\big] - 
	9\, U_4^2\, z^2\, \Big(-15 + z \,\big[24 + z\, (-10 + z + 110\, z^2)\big]\Big)\Big]\nonumber \\
	& +& 
	U_1\, z\, \Big[36\, U_2^3\, U_3^3\, z^2\, (-5 + 7 \,z) - 
	54\, U_4^3\, z^2\,\Big (-12 + z\, \big[20 + z\, (-9 + z + 120\, z^2)\big]\Big)\nonumber \\
	& +& 
	16\, U_2^2\, U_3^2\, \Big(3 + z\, \big[-6 + z\,\big (3 + z\, (21 - 96\, z + 76\, z^2)\big)\big]\Big)\nonumber \\
	& -& 
	U_2\, U_3\, U_4\, z\, \Big(371 + 
	z \,\big[-901 + z\, \big(689 + z\, [9 + 14 z\, (-324 + 313\, z)]\big)\big]\Big)\Big]\nonumber \\
	& + &
	U_1^2\, \Big(9 \,U_2^3\, U_3^3 \,z\, (-15 + 19\, z) + 
	6\, U_2^2\, U_3^2\, \big[1 + z\, (-1 + 2 z)\, (1 + 10\, (-4 + z)\, z)\big]\nonumber \\
	& -& 
	9\, U_4^2\, z^2\, \Big(-28 + z\, \big[49 + 3\, z\, (-8 + z + 180\, z^2)\big]\Big)\nonumber \\
	& - &
	4\, U_2\,U_3\, z\, \Big[24 + 
	z\, \big(-60 + z\, \big[48 + z\, (44 + z\, (-637 + 583\, z))\big]\big)\Big]\Big)\bigg] \nonumber \\
	&+& 
	r^8\, \bigg[U_1^4\, z\, \big[2\, U_2 \,U_3 - 9\, (-7 + z)\, z\big] + 
	12\, U_1^3\, z\, (-2\, U_2^2\, U_3^2 + U_2\, U_3\, z^2 + 15\, z^4)\nonumber \\
	& + &
	U_1^2\, \Big[27\, U_2^3\, U_3^3 \,z - 6\, U_2^2 \,U_3^2\, U_4\, (4 + z + 15 \,z^2) + 
	4\, U_2\, U_3\, z\, \Big(60 + z \,\big[-120 + z \,(72 + (4 - 91 \,z) \,z)\big]\Big)\nonumber \\
	& +& 
	9\, U_4 \,z^2\, \Big(-56 + z\, \big[119 + z\, \big(-81 + z\, (19 + 134\, z)\big)\big]\Big)\Big] + 
	3\, U_4\, z^2\, \Big[10\, U_2^2\, U_3^2\, \big[-6 + z\, (9 - 3\, z + 7\, z^3)\big]\nonumber \\
	& - &
	10\, U_2\, U_3\, U_4\, z\,\Big (-28 + z\, \big[49 + 3\, z\, (-8 + z + 12\, z^2)\big]\Big)\nonumber \\
	& + &
	3\, U_4^2\, z^2\, \Big(-120 + z\,\big[228 + z\,\big (-132 + z\, (25 + 164\, z)\big)\big]\Big)\Big] \nonumber \\
	&+& 
	U_1\, \Big(6\, U_2^3\, U_3^3\, (1 + 6\, z^3) - 
	16\, U_2^2\, U_3^2\, z\, \Big(15 + z\, \big[-24 + U_4\, z\, (9 + 16\, z)\big]\Big)\nonumber \\
	& +& 
	18\, U_4^2\, z^3\, \Big[-84 + z \,\Big(168 + z\, \big[-105 + 2\, z\, (11 + 67\, z)\big]\Big)\Big] \nonumber \\
	&+ &
	U_2\, U_3\, z^2\, \Big[1113 + 
	z\, \big(-3180 + z \,\big[3180 + z\, (-1200 + z\, (-1185 + 1274\, z))\big]\big)\Big]\Big)\bigg] \nonumber \\
	&+ &
	r^6\, \bigg[U_1^4\, \Big(2\, U_2^2\, U_3^2\, (-30 + 49\, z) + 
	9\, U_4 \,z^2\, \big[35 + 4\, z\, (11 + 5\, z)\big] - 
	2\, U_2\, U_3\, z\, \big[-15 + z\, (82 + 7 \,z)\big]\Big)\nonumber \\
	& -& 
	6\, U_1^3\, \Big(-840\, U_4^2\, z^5 + 9\, U_2^3\, U_3^3\, (-2 + 3\, z) - 
	2\, U_2\, U_3\, z^3\, \big[21 + 4\, z\, (-48 + 43\, z)\big]\nonumber \\
	& +& 
	2 \,U_2^2\, U_3^2\, z\, \big[30 + z\, (-85 + 53\, z)\big]\Big) + 
	3\, U_4^3\, z^3\, \Big(350\, U_2^2\, U_3^2\, z^3 + 10\, U_2\, U_3\, U_4 \,(U_4 - 126\, z^4) \nonumber \\
	&+ &
	3\, U_4^2\, z\, \big[-10 + z \,(13 - 3\, z + 462\, z^3)\big]\Big) + 
	U_1\, z^2\, \Big[36\, U2^3\, U3^3\, z \,\big[10 + z\, (-28 + 19\, z)\big]\nonumber \\
	& - &
	80\, U_2^2\, U_3^2\, U_4\, z^2\, \big[7 + z\, (-41 + 35\, z)\big] + 
	54\, U_4^4\, z\, \big[-3 + z\, (4 - z + 210\, z^3)\big]\nonumber \\
	& +& 
	U_2\, U_3\, U_4^2\, \Big(53 + 
	z\, \big[-106 + z \,\big(53 + 42 \,z\, (6 + z\, (-212 + 207\, z))\big)\big]\Big)\Big]+ 
	U_1^2\, z\,\Big (18\, U_2^3\, U_3^3\, (-3 + 4 \,z)\, (-5 + 6\, z) \nonumber \\
	&+ &
	9\, U_4^3\, z\, \big[-8 + z\, (11 - 3 \,z + 1260\, z^3)\big]
	 - 
	6\, U_2^2\, U_3^2 \,z\, \Big(70 + z\, \big[-225 + 2\, z\, (50 + 27\, z)\big]\Big)\nonumber \\
	& +& 
	4\, U_2\, U_3\, U_4\, \Big[4 + 
	z\,\big (-8 + z\, \big[4 + z\, (112 + z\, (-1799 + 1699\, z))\big]\big)\Big]\Big)\bigg],
\end{eqnarray}
\begin{eqnarray}
	V_3&=& 3\, r^2\, U_1^2\, U_2^2\, U_3^5\, \bigg[-6\, r^4 \,U_1 + 2\, U_1^2\, U_3^2 + 
	r^5\, (-15\, U_2 + 2\, U_3)\bigg]  \nonumber \\
	&+ &
	2\, r^2\, U_1\, U_2 \,U_3^4\, \bigg[-8 \,U_1^3\, U_3^2\, (U_2 + U_3) + 
	2\, r^5\, U_2\, U_3^2\, (-9\, U_2 + 10\, U_3) + 
	r^4\, U_1\, U_3\, \big[(-39 + 36\, r)\, U_2 + 8\, U_3^2\big] \nonumber \\ &+& 
	3\, U_1^2\, \Big(\Big[2 + r\, \big(-8 + r \,\big[12 + r\, (-8 + 5 r)\big]\big)\Big]\, U_2 - 
	3\, r^3\, U_3^2\Big)\bigg]\, z + 
	r\, U_3^5\, \bigg[2\, r \,U_1\, U_2\, \Big(27\, r^3\, U_1^2 \nonumber \\
	& +& \Big[8\, r^4\, (-3 + r + 2\, r^2) + 
	3\, \Big(1 + r\, \big[-4 + r\, \big(6 + r\, (-4 + 27\, r)\big)\big]\Big)\, U_1 - 7 \,U_1^3\Big] \,U_2 + 
	18\, r^5\, U_2^2\Big)\nonumber \\
	& + &
	2 \,U_1^2\, \big[-6\, r^4\, (3 + r) + (24 + 19\, r) \,U_1^2\big]\, U_2\, U_3 + 
	3\, r\, U_1^3\, (3\, U_1 - 4\, U_2^2)\, U_3^2\nonumber \\
	& +& 
	r^4\, (9 \,U_1^2 + 50\, r \,U_1\, U_2 + 30\, r^2\, U_2^2) \,U_3^3\bigg] \,z^2 
	- 
	r\, U_3^4\, \bigg[2\, U_1^2 \,U_2 \Big(-3 \,\big[4 + (-8 + r)\, r\big]\, \big[4 + r\, (-8 + 7\, r)\big]\, U_1\nonumber \\
	& + &
	3 \,r\, \Big[-6 + r\, \big(24 + r\, \big[-36 + r \,(24 + 7\, r)\big]\big)\Big]\, U2 - 
	46\, r \,U_1^2\, U_2\Big) + 
	4\, U_1\, U_2\, \big[-36\, r^4\, U_1 + 69\, U_1^3 - 19 \,r \,U_1^3\nonumber \\
	& + &
	12\, r^5\, (2\, U_1 - 3\, U_2) + 26\, r^6\, U_2\big] \,U_3 + 
	3\, U_1^3\, \big[3 \,(4 + r)\, U_1 + 56\, r \,U_2^2\big]\, U_3^2 \nonumber \\
	&+& 
	r^4 \,\big[45\, U_1^2 + 2\, (9 + 91\, r) \,U_1\, U_2 + 90\, r^2\, U_2^2\big]\, U_3^3 - 
	6\, r^4\, (3\, U_1 + 5\, r\, U_2)\, U_3^5\bigg] \,z^3\nonumber \\ &+& 
	U_3^3\, \bigg[-6\, r\, U_1^3\, U_2\, \big[98 + 
	r\, (-392 + r \,(588 + r\, (-392 + 95\, r)) + 17\, U_1\, U_2)\big]\nonumber \\ 
	&+& 
	4\, r\, U_1^2 \,\big[16 + 177\, U_1^2 + 
	2\, r\, (-40 + r\, (80 + r\, (-80 + r\, (13 + 17\, r))) - 
	58\, U_1^2)\big]\, U_2\, U_3\nonumber \\
	& +& 
	2\, U_1\, \Big[45\, U_1^3 + 
	2\, r^2\, \Big(8 - 4 \,r\, \big[8 + r \,(2 + r) \,(-6 + 7\, r)\big] + 
	153\, U_1^2\Big)\, U_2^2\Big] \,U_3^2 \nonumber \\
	&+ &
	6\, r \,\big[6\, U_1^4 + 35\, r^5\, U_1\, U_2 + 15 \,r^6\, U_2^2 + 
	15\, r^4\, U_1 \,(U_1 + U_2)\big] U_3^3 - 354\, r^2\, U_1^2\, U_2^2\, U_3^4\nonumber \\
	& - &
	6\, r^5\, (18\, U_1 + 25\, r\, U_2)\, U_3^5 + 9 \,r^5\, U_3^7\bigg]\, z^4\nonumber \\
	& -& 
	U_3^2\, \bigg[-34\, r^2\, U_1^4 \,U_2^2 + 8\, (129 - 110\, r) \,r\, U_1^4 \,U_2\, U_3 \nonumber \\
	&+ &
	2\, U_1\, \bigg(45 \,\big[7 + 2\, (-4 + r)\, r\big]\, U_1^3 + 
	2 \,r \,\Big(89 + r\, \big[-356 + r \,\big(534 + r\, (-356 + 53\, r)\big)\big]\Big)\, U_1\, U_2 \nonumber \\
	&+& 
	2\, r^2\, \Big[44 + 8\, r\, \big(-22 + r\, (33 - 22\, r + 4\, r^2)\big) + 
	219\, U_1^2\Big]\, U_2^2\bigg)\, U_3^2\nonumber \\
	&+& 
	2\, r\, \Big[45\, r^4\, U_1^2 + 
	2 \,\Big(-4 + r \,\big[20 + r\, \big(-40 + r\, [40 + r\, (25 + 9\, r)]\big)\big]\Big)\, U_1\, U_2 + 
	15\, r^6\, U_2^2\big]\, U_3^3 \nonumber \\
	&-& 6\, r\, U_1^2 \,U_2\, (268\, U_1 + 183\, r\, U_2)\, U_3^4 - 
	30 \,r^5\, (9 \,U_1 + 10\, r \,U_2)\, U_3^5 - 180\, U_1^3\, U_3^6 + 63\, r^5\, U_3^7\bigg]\, z^5 \nonumber \\
	&+&
	U_3^2\, \bigg[2\, (444 - 425\, r) \,r\, U_1^4\, U_2 + 
	U_1^2\, \Big[45\, \big[42 + r\, (-68 + 27\, r)\big]\, U_1^2\nonumber \\
	& + &
	4\, r\, \Big(191 + 2\, r\, \big[-382 + r\, \big(573 + r\, (-382 + 91\, r)\big)\big]\Big)\, U_2 + 
	576\, r^2\, U_1\, U_2^2\big]\, U_3 \nonumber \\
	&-& 
	5\, U_1\, \Big[9\, \Big(-3 + r\, \big[15 + r\, \big(-30 + r\, [30 + r \,(-15 + 2\, r)]\big)\big]\Big)\, U_1\nonumber \\
	& + &
	2\, r\, \Big(3 + 
	r \,\Big[-15 + r \,\big(30 + r\, \big[-30 + r\, (-3 + 10\, r)\big]\big)\Big]\Big)\, U_2\Big]\, U_3^2\nonumber \\
	& - &
	12\, r\, U_1^2\, U_2\, (215 \,U_1 + 148\, r\, U_2)\, U_3^3 - 
	60\, r^5 \,(6 \,U_1 + 5 \,r \,U_2)\, U_3^4\nonumber \\
	& + &
	32\, U_1\, (-45\, U_1^2 + 13 \,r^2\, U_2^2)\, U_3^5 + 189 \,r^5\, U_3^6 + 
	30\, r^2\, U_2^2\, U_3^7\bigg]\, z^6 \nonumber \\
	&+& 
	U_3^2\, \bigg[-3 \,U_1^3\, \Big(48 \,r^2\, U_2^2 + 
	U_1\, \big[1050 + 873\, r^2 + 20\, r\, (-96 + 7\, U_2)\big]\Big)\nonumber \\
	& +& 
	9\, \Big(-135 + 
	r\, \Big[675 + r\, \big(-1350 + r\, \big[1350 + r\, (-675 + 134\, r)\big]\big)\Big]\Big)\, U_1^2\, U_3 \nonumber \\
	&+ &
	2 \,r\, U_1\, U_2\, \Big[40 (-4 + 33\, U_1^2) + 
	r\, \Big(640 - 5\, r\, \big[192 + r \,(-128 + 41 \,r)\big] + 807\, U_1\, U_2\Big)\Big]\, U_3^2 \nonumber \\
	&+ &
	6 \,\Big[1 + r \,\Big(-5 + r\, \big[10 + r\, \big(-10 + r\, (5 + 4\, r)\big)\big]\Big)\Big]\, (9 \,U_1 + 
	5\, r\, U_2)\, U_3^3\nonumber \\ 
	&+ &
	4\, U_1\, (1260\, U_1^2 - 173\, r\, U_1\, U2 - 140\, r^2\, U_2^2) \,U_3^4 - 
	315\, r^5\, U_3^5 - 210\, r^2\, U_2^2\, U_3^6\bigg]\, z^7 \nonumber \\
	&+ &
	U_3\, \bigg[84\, r\, U_1^4\, U_2 + 18\, (175 - 163\, r)\, U_1^4\, U_3 + 
	2\, r\, U_1\, U_2\, \Big(868 - 858\, U_1^2 + 
	r\, \big[-3472 + r\, \big(5208 + r\, (-3472 + 877\, r)\big)\nonumber \\
	& -& 390\, U_1\, U_2\big]\Big)\, U_3^2 + 
	6\, \Big[18\, \Big(-5 + r \,\big[5 + (-5 + r)\, r\big]\, \big[5 + r\, (-5 + 4\, r)\big]\Big) \,U_1\nonumber \\
	& +& 
	5\, r\, \Big(-9 + 
	r\, \big[45 + r\, \big(-90 + r \,[90 + r\, (-45 + 8 \,r)]\big)\big]\Big)\, U_2\big]\, U_3^3
	 - 
	20 \,U_1 \,(504\, U_1^2 + r\, U_1\, U_2 - 24 \,r^2\, U_2^2)\, U_3^4\nonumber \\
	& + &
	9\, \Big(1 + r \,\big[-5 + r\, \big(10 + r \,[-10 + r (5 + 34\, r)]\big)\big]\Big)\, U_3^5 \nonumber \\
	&+ &
	90 \,(54\, U_1^2 + 7\, r^2\, U_2^2)\, U_3^6\bigg]\, z^8 + 
	U_3\, \bigg[18\, (-105 + 103\, r)\, U_1^4 + 12\, r\, U_1^2\, U_2\, (54\, U_1 + 13\, r\, U_2)\, U_3\nonumber \\
	& - &
	18\, \Big(-135 + 
	r \,\big[675 + r\, \big(-1350 + r\, [1350 + r\, (-675 + 134\, r)]\big)\big]\Big)\, U_1\, U_3^2 \nonumber \\
	&+ &
	4\, U_1\, (3150\, U_1^2 + 149\, r\, U_1\, U_2 - 68\, r^2\, U_2^2)\, U_3^3 - 
	9 \,\Big(11 + 5\, r\, \big[-11 + r \,\big(22 + r\, [-22 + r\, (11 + 2\, r)]\big)\big]\Big)\, U_3^4\nonumber \\
	& - &
	42 \,(270 \,U_1^2 + 96 \,r \,U_1\, U_2 + 25\, r^2\, U_2^2)\, U_3^5 + 
	1080\, r\, U_2\, U_3^7\bigg] \,z^9 \nonumber \\
	&+ &
	U_3\, \bigg[18\, U_1^3\, (35\, U_1 - 6\, r\, U_2) - 
	4\, U_1\, (40\, U_1 - 3 \,r\, U_2) \,(63\, U_1 + 8\, r\, U_2)\, U_3^2\nonumber \\
	& - &
	9\, \Big(-55 + r\, \big[275 + r\, \big(-550 + r\, [550 + r\, (-275 + 48\, r)]\big)\big]\Big)\, U_3^3 + 
	14\, (1215\, U_1^2 + 386\, r\, U_1\, U_2 + 75\, r^2\, U_2^2)\, U_3^4 \nonumber \\
	&- &
	360\, (18\, U_1 + 7\, r\, U_2)\, U_3^6\bigg]\, z^{10} + \bigg[-90\, U_1^4 + 
	4\, U_1\, (1260\, U_1^2 + 49 \,r\, U_1\, U_2 - 4\, r^2\, U_2^2)\, U_3^2\nonumber \\
	& + &
	9\, \Big(-165 + 
	r\, \big[825 + r \,\big(-1650 + r\, [1650 + r\, (-825 + 164\, r)]\big)\big]\Big)\, U_3^3 - 
	70\, (243 \,U_1^2 + 64 \,r\, U_1 \,U_2 + 9\, r^2 \,U_2^2)\, U_3^4\nonumber \\
	& +& 
	3780\, (3\, U_1 + r\, U_2) \,U_3^6\bigg] \,z^{11} + 
	2\, U_3\, \bigg[-2\, U_1^2\, (360\, U_1 + 7\, r \,U_2) + 
	15\, (378\, U_1^2 + 76\, r\, U_1\, U_2 + 7\, r^2\, U_2^2)\, U_3^2\nonumber \\
	& - &
	378\, (18\, U_1 + 5\, r\, U2)\, U_3^4 + 1485\, U_3^6\bigg]\, z^{12} + 
	2\, \bigg[90\, U_1^3 - (2430\, U_1^2 + 328\, r\, U_1\, U_2 + 15\, r^2\, U_2^2)\, U_3^2\nonumber \\
	& +& 
	630\, (9\, U_1 + 2\, r\, U_2)\, U_3^4 - 2079\, U_3^6\bigg]\, z^{13} + 
	U_3\, \bigg[U_1\, (1215\, U_1 + 82\, r\, U_2) - 1080 (6\, U_1 + r \,U_2)\, U_3^2 + 
	4158\, U_3^4\bigg] \,z^{14}\nonumber \\
	& -& 
	135\, \bigg[U_1^2 - 2 \,(9\, U_1 + r \,U_2)\, U_3^2 + 22\, U_3^4\bigg] \,z^{15} + 
	15\, U_3\, \bigg[-36\, U_1 - 2\, r\, U_2 + 99\, U_3^2\bigg]\, z^{16} \nonumber \\
	&+ &9 \,\bigg[6\, U_1 - 55\, U_3^2\bigg]\, z^{17} + 
	99\, U_3\, z^{18} - 9\, z^{19},
\end{eqnarray}
\begin{eqnarray}
	V_4&=&3\, r^3\, U_1^3\, U_2^2\, U_3^7 \,(-9\, U_2 + 13\, U_3) + 
	r^2\, U_1\, U_2\, U_3^6\, \bigg[8\, r^4\, U_1\, U_3 + 8\, r^5\, U_2\, U_3 \nonumber \\
	&- &
	39\, U_1^2 \,U_2\, U_3 + 
	r\, U_1\, (139\, U_1 \,U_2^2 - 213\, U_1\, U_2\, U_3 - 7\, (4\, U_1 + 3 \,U_2^2) \,U_3^2)\bigg]\, z \nonumber \\
	&+ &
	r^2 \,U_1\, U_3^5\, \bigg[-16\, r^5\, U_2^2\, U_3 + 9 \,r^3\, U_1\, U_3^3 + 
	r^4 \,U_2 \,U_3\, (-24\, U_1 + 31\, U_3^2)\nonumber \\
	& + &
	r\, U_1\, \big[-286\, U_1\, U_2^3 + 420 \,U_1\, U_2^2\, U_3 + 
	3\, U_2\, (51\, U_1 + 46\, U_2^2)\, U_3^2 - 9\, U_1\, U_3^3\big]\nonumber \\
	& +& 
	U_1\, U_2 \,U_3\, (243\, U_1\, U_2 + 56\, U_1\, U_3 - 39\, U_2\, U_3^3)\bigg]\, z^2 + 
	r\, U_3^4\, \bigg[294\, r^2\, U_1^3\, U_2^3\nonumber \\
	& + &
	r\, U_1\, U_2\, (24 \,r^4\, U_1 + 8\, r^5 \,U_2 - 579\, U_1^2\, U_2 - 345\, r U_1^2\, U_2)\, U_3\nonumber \\
	& -&
	U_1^2\, U_2\, \Big(\big[84 + r\, (321 + 179\, r)\big]\, U_1 + 375\, r^2\, U_2^2\Big) \,U_3^2 \nonumber \\
	&+ &
	r\, U_1\, \big[9\, U_1\, (-5\, r^3 + 3 \,U_1 + 2\, r\, U_1) - 124\, r^4 \,U_2\big] \,U_3^3 + 
	234\, r \,U_1^2\, U_2^2\, U_3^4 + 3\, r^4\, (6\, U_1 + 5\, r U_2)\, U_3^5\bigg]\, z^3\nonumber \\
	& + &
	r\, U_3^3\, \bigg[-147\, r^2\, U_1^3\, U_2^3 + 
	U_1^2\, U_2\, \Big(-112 + 8\, r \,\big[70 + r\, \big(-140 + r\, [140 + r\, (-70 + 13\, r)]\big)\big]\nonumber \\
	& + &
	9\, r\, (75 + 4 \,r)\, U1\, U_2\Big)\, U_3 + 
	U_1^2\, U_2\, \Big(\big[840 - r \,(19 + 43 \,r)\big]\, U_1 + 540 \,r^2\, U_2^2\Big)\, U_3^2\nonumber \\
	& + &
	186\, r^5\, U_1\, U_2\, U_3^3 - 438\, r\, U_1^2\, U_2^2 \,U_3^4 - 75\, r^5\, U_2\, U_3^5 - 
	48\, r \,U_1\, U_2^2\, U_3^6\nonumber \\
	& + &
	9\, U_3^3\, \Big(10\, r^4 \,U_1^2 + 2\, \big[-3 + (-3 + r)\, r\big]\, U_1^3 - 12\, r^4\, U_1\, U_3^2 +
	r^4\, U_3^4\Big)\bigg]\, z^4 \nonumber \\
	&+ &
	U_3^3\, \bigg[3\, r^2\, (-145 + 66\, r) \,U_1^3\, U_2^2 - 
	r\, U_1^2 \,U_2 \,\big[(3360 - 3846\, r + 986\, r^2) \,U_1 + 423\, r^2\, U_2^2\big]\, U_3\nonumber \\
	& - &
	2\, U_1\, \big[45\, (r^5 - U_1) \,U_1 + \big]	2\, r\, \Big(7 + 
	r\, \big[-35 + r\, \big(70 + r \,[-70 + r\, (35 + 24\, r)]\big)\big]\Big) \,U_2\Big] \,U_3^2\nonumber \\
	& +& 
	3\, r\, U_1^2\, \big[9\, (4 + r)\, U1 + 2\, r\, U_2^2\big]\, U_3^3 + 
	30\, r^5 \,(9\, U_1 + 5\, r\, U_2)\, U_3^4 + 
	2\, r\, U_1\, U_2\, (631\, U_1 + 216\, r \,U_2)\, U_3^5 - 63\, r^5\, U_3^6\bigg]\, z^5\nonumber \\
	& + &
	U_3^3\, \bigg[r \,U_1^2 \,U_2\,\Big(141\, r^2\, U_2^2 + 
	U1\, \big[7140 + r\, (-11574 + 4559\, r + 180 \,U_2)\big]\Big)\nonumber \\
	& +& 
	r\, \Big(381 - 
	5\, r\, \big[381 + r\, \big(-762 + r\, [762 + r \,(-381 + 70\, r)]\big)\big]\Big)\, U_1\, U_2\, U_3 + 
	1002\, r^2\, U1^2\, U_2^2\, U_3^2 - 150\, r^6\, U_2\, U_3^3\nonumber \\
	& - &
	2\, r \,U_1\, U_2\, (2905\, U_1 + 816\, r\, U_2)\, U_3^4 + 
	9 \,U_3\, \Big(-5\, \Big[-3 + 
	r\, \big(15 + r\, \big[-30 + r \,\big(30 + r \,(-15 + 2 r)\big)\big]\big)\big]\, U_1^2\nonumber \\
	& +&
	\Big(-80 + 
	3\, r\, \big[38 + r\, (-16 + 3 \,r)\big]\Big) \,U_1^3 - 40\, r^5\, U_1\, U_3^2 + 
	21\, r^5\, U_3^4\Big)\bigg]\, z^6\nonumber \\
	& + &
	U_3^2\, \bigg[-45\, r^2 \,U_1^3\, U_2^2 + 
	U_1^2 \,\Big(45\, (-27 + 56\, U_1) + 
	9\, r\, \big[675 - 630\, U_1 + 
	r\, \big(r\, (1350 + r \,(-675 + 134\, r) - 114 \,U_1)\nonumber \\
	& +& 
	15\, (-90 + 31\, U_1)\big)\big] + 5\, r\, (-1764 + 1499\, r) \,U_1\, U_2\Big)\, U_3 - 
	1353\, r^2\, U_1^2 \,U_2^2\, U_3^2 \nonumber \\
	&+ &
	3\, \Big(1 + r\, \big[-5 + r\, \big(10 + r\, [-10 + r \,(5 + 4\, r])\big)\big]\Big)\, (18\, U_1 + 
	5\, r\, U_2)\, U_3^3 \nonumber \\
	&+&
	70\, r\, U_1\, U_2\, (209\, U_1 + 48\, r\, U_2)\, U_3^4 - 
	315\, r^5\, U3^5 - 2068\, r\, U_1\, U_2\, U_3^6\bigg] \,z^7\nonumber \\
	& +& 
	U_3^2\, \bigg[(6384 - 6119\, r)\, r \,U_1^3\, U_2 + 
	3\, U_1^2\, \Big(-6 \,\big[280 + r\, (-455 + 184\,r)\big] \,U_1 + 245\, r^2\, U_2^2\Big)\, U_3 \nonumber \\
	&+& 
	3\, \Big[36\, \Big(-5 + r\, \big[5 + (-5 + r)\, r\big]\, \big[5 + r \,(-5 + 4\, r)\big]\Big)\, U_1\nonumber \\
	& + &
	5\, r\, \Big(-9 + 
	r\, \big[45 + r\, \big(-90 + r\, [90 + r\, (-45 + 8\, r)]\big)\big]\big)\, U_2\Big]\, U_3^2 - 
	10\, r\, U_1\, U_2\, (2233\, U_1 + 408\, r\, U_2) \,U_3^3\nonumber \\
	& + &
	9\, \Big(1 + r\, \big[-5 + r\, \big(10 + r\, [-10 + r\, (5 + 34\, r)]\big)\big]\Big)\, U_3^4 + 
	20\, U_1\, (243\, U_1 + 308\, r\, U_2) \,U_3^5\bigg]\, z^8\nonumber \\
	& + &
	U_3^2\, \bigg[3\, U_1^2\, \Big(-49 \,r^2\, U_2^2 + 
	3\, U_1 \,\big[700 + 577\, r^2 - 14\, r \,(91 + 20\, U_2)\big]\Big)\nonumber \\
	& - &
	18\, \Big(-135 + 
	r \,\big[675 + r\, \big(-1350 + r\, [1350 + r\, (-675 + 134\, r)]\big)\big]\Big)\, U_1\, U_3 + 
	2 \,r\, U_1\, U_2\, (10661\, U_1 + 1464\, r\, U_2)\, U_3^2\nonumber \\
	& -& 
	9\, \Big(11 + 5\, r\, \big[-11 + r\, \big(22 + r \,[-22 + r \,(11 + 2\, r)]\big)\big]\Big)\, U_3^3 \nonumber \\
	&-& 
	11340\, U_1\, (U_1 + r\, U_2) \,U_3^4 + 540 \,r\, U_2\, U_3^6\bigg]\, z^9\nonumber \\
	& + &
	U_3\, \bigg[420\, r\, U_1^3\, U_2 + 126\, (-40 + 37\, r)\, U_1^3\, U_3 - 
	2\, r\, U_1\, U_2\, (6251\, U_1 + 576\, r\, U_2) \,U_3^2\nonumber \\
	& -& 
	9\, \Big(-55 + r\, \big[275 + r\, \big(-550 + r\, [550 + r\, (-275 + 48\, r)]\big)\big]\Big)\, U_3^3 + 
	14\, U_1\, (1215 \,U_1 + 967\, r \,U_2)\, U_3^4 \nonumber \\
	&-& 
	180\, (36 \,U_1 + 7\, r \,U_2)\, U_3^6\bigg]\, z^{10} + 
	U3 \,\bigg[18 (140 - 137\, r)\, U_1^3 + 2\, r \,U_1 \,U_2\, (2065\, U_1\, + 96\, r\, U_2)\, U_3\nonumber \\
	& + &
	9\, \Big(-165 + 
	r\, \big[825 + r\, \big(-1650 + r\, [1650 + r \,(-825 + 164\, r)]\big)\big]\Big)\, U_3^2\nonumber \\
	& - &
	14\, U_1\, (1215 \,U_1 + 754\, r\, U_2) \,U_3^3 + 
	1890\, (6\, U_1 + r\, U_2)\, U_3^5\bigg]\, z^{11}\nonumber \\
	& +& 
	2\, U_3\, \bigg[-5\, U_1^2\, (72\, U_1 + 59\, r\, U_2) + 42\, U_1\, (135\, U_1 + 62\, r\, U_2)\, U_3^2\nonumber \\ &- &
	189\, (36\, U_1 + 5\, r\, U_2)\, U_3^4 + 1485\, U_3^6\bigg]\, z^{12} + 
	2 \,\bigg[45 \,U_1^3 - 10\, U_1\, (243\, U_1 + 74\, r \,U_2)\, U_3^2 \nonumber \\
	&+& 
	630 \,(9 \,U_1 + r\, U_2) \,U_3^4 - 2079\, U_3^6\bigg]\, z^{13} + 
	U3\, \bigg[5 \,U_1\, (243\, U_1 + 37 \,r\, U_2) - 540 \,(12\, U_1 + r\, U_2)\, U_3^2 + 
	4158\, U_3^4\bigg]\, z^{14}\nonumber \\
	& - &
	135\, \bigg[22 - 88\, r^3 + 22\, r^4 + (-18 + U_1) \,U_1 + 
	r\, (-88 + 36\, U_1 - U_2\, U_3) + r^2\, (132 - 18 \,U_1 + U_2\, U_3)\bigg]\, z^{15}\nonumber \\
	& - &
	15\, U_3\, \bigg[36\, U_1 + r\, U_2 - 99\, U_3^2\bigg]\, z^{16} + 9\, \bigg[6\, U_1 - 55\, U_3^2\bigg]\, z^{17} + 
	99\, U_3\, z^{18} - 9 \,z^{19},
\end{eqnarray}
\begin{eqnarray}
	V_5&=&-9 \,r^3\, U_1^2\, U_2^2\, U_3^5\, \bigg[r^4\, U_2 - (-1 + 2\, r)\, U_1\, (-1 + 
	r \,U_3)\bigg] \nonumber \\
	&+& 
	r^2 \,U_1\, U_2\, U_3^4\, \bigg[4\, r^5\, U_2\, U_3^3 + 8\, r^4\, U_1\, U_3^2\, (-6\, U_2 + U_3) + 
	U_1^2\, \Big(3\, \big[-3 + r\, \big(-7 + r\, [39 + r\, (-45 + 19\, r)]\big)\big]\, U_2\nonumber \\ 
	&-& 
	2\, r\, \big[5 + 2\, r\, (-5 + 7\, r)\big] \,U_3^2\Big)\bigg]\, z + 
	r^2\, U_1\, U_3^5\, \bigg[U_1\, U_2\, \Big(\big[20 + r\, \big(23 + r \,(-106 + 117\, r)\big)\big]\, U_1\nonumber \\
	& + &
	3 \,\big[-3 + r\, \big(12 + r\, [-18 + r\, (12 + 29\, r)]\big)\big]\, U_2\Big) + 
	2\, U_2 \,\big[6\, (-3 + r)\, r^3 U_1 + 4 \,r^4\, (-3 + 2\, r)\, U_2\nonumber \\
	& + &
	3\, (7 + 17\, r)\, U_1^2 \,U_2\big]\, U_3 + 
	r\, \big[9\, (r^2 - U_1) \,U_1 + 28 \,r^3\, U_2\big]\, U_3^3\bigg]\, z^2\nonumber \\
	& - &
	r \,U_3^4\, \bigg[U_1^2\, U2\, \Big(\big[30 + r\, \big(81 + r\, [-181 + r\, (-1 + 125\, r)]\big)\big]\, U_1 + 
	3\, r\, \big[-3 + r\, (6 + r)\big]\, \big[3 + r\, (-6 + 7\, r)\big]\, U_2\Big)\nonumber \\
	& + &
	r \,U_1\, U_2 \,\big[24\, r^3\, (-6 + 5\, r)\, U_1 + 4 \,r^4\, (-18 + 17\, r)\, U_2 + 
	3 \,(11 + 53\, r)\, U_1^2 \,U_2\big]\, U_3\nonumber \\
	& +& 
	r\, U_1\, \big[-9 \,U_1\, (-5\, r^3 + 3\, U_1 + 2\, r \,U_1) + 
	2\, r^3\, (9 + 47\, r) \,U_2\big]\, U_3^3 - 3\, r^4 \,(6\, U_1 + 5\, r\, U_2) \,U_3^5\bigg] \,z^3\nonumber \\
	& +&
	r\, U_3^3\, \bigg[\Big(354 + r\, \big[-691 + r\, (277 + (103 - 25\, r) \,r)\big]\Big)\, U_1^3\, U_2 + 
	8\, U_1^2\, U_2 \Big(-5 + 
	r\, \big[25 - 15\, U_1\, U_2\nonumber \\
	& + &
	r\, (-50 + r\, (50 + r\, (-52 + 31\, r)) + 33 \,U_1\, U_2)\big]\Big)\, U_3 - 
	24\, r\, (1 - 4\, r + 6\, r^2)\, U_1\, U_2^2\, U_3^2\nonumber \\
	& +& 
	6\, U_1 \,\Big(3 \,U_1\, \big[5\, r^4 + \big(-3 + (-3 + r)\, r\big) \,U_1\big] + 
	r^4\, (15 + 13\, r)\, U_2 + 16\, r^4\, U_2^2\Big)\, U_3^3 + 
	117\, r\, U_1^2\, U_2^2\, U_3^4\nonumber \\
	& -& 3\, r^4\, (36\, U_1 + 25\, r\, U_2)\, U_3^5 + 
	9\, r^4\, U_3^7\bigg]\, z^4 + 
	U_3^3\, \bigg[24\, (11 - 13\, r)\, r^2 \,U_1^3\, U_2^2\nonumber \\
	& + &
	2 \,r\, U_1\, U_2\, \Big(U_1\, \big[271 - 762\, U_1 + 
	r\, (-1084 + r\, (1626 + r\, (-1084 + 343\, r) - 223\, U_1) + 
	879\, U_1)\big]\nonumber \\
	& +& 
	12\, r\, \big[9 + 2\, r \,\big(-18 + r\, [27 + r\, (-18 + 5\, r)]\big)\big]\, U_2\Big)\, U_3 + 
	2\, U_1\, \Big[45\, U_1 \,(-r^5 + U_1)\nonumber \\
	& + &
	r\, \Big(-5 + r\, \big[25 + 
	r\, \big(-50 + r\, [50 + r\, (-115 + 39\, r)]\big)\big]\Big)\, U_2\Big] \,U_3^2 + 
	9 \,r\, U_1^2\, \big[3 \,(4 + r)\, U_1 - 79\, r \,U_2^2\big]\, U_3^3\nonumber \\
	& + &
	30\, r^5\, (9 \,U_1 + 5\, r\, U_2)\, U_3^4 - 63\, r^5\, U_3^6\bigg]\, z^5 + 
	U_3^3\, \bigg[r \,U_1^2\, U_2 \,\Big(-2714 + 3360\, U_1 + 
	r\, \big[10856 + 10856\, r^2 - 2750\, r^3 \nonumber \\
	&+& 59\, r \,(-276 + 37\, U_1) - 
	6\, U_1\, (915 + 32\, U_2)\big]\Big) + 
	U_1\, \Big[9\, U_1\, \Big(15 - 
	5\, r \,\big[15 + r\, \big(-30 + r\, [30 + r\, (-15 + 2\, r)]\big)\big]\nonumber \\
	& +& \big[-80 + 
	3\, r\, \big(38 + r\, (-16 + 3\, r)\big)\big] \,U_1\Big) + 
	r \,\Big(183 - 
	5 \,r \,\big[183 + 
	r\, \big(-366 + r\, [366 + r\, (-219 + 67 \,r)]\big)\big]\Big)\, U_2\Big]\, U_3 \nonumber \\
	&+& 
	1440\, r^2 \,U_1^2\, U_2^2 \,U_3^2 - 30 \,r^5\, (12\, U_1 + 5\, r \,U_2)\, U_3^3 - 
	816\, r^2\, U_1\, U_2^2\, U_3^4 + 189 \,r^5 \,U_3^5\bigg] \,z^6\nonumber \\
	& + &
	U_3^2 \,\bigg[48\, r^2\, U_1^3\, U_2^2 + 
	U_1^2\, \Big(45\, (-27 + 56\, U_1) + 
	9\, r\, \big[675 - 630\, U_1 + 
	r\, \big(r \,(1350 + r\, (-675 + 134 \,r) - 114\, U_1) \nonumber \\
	&+ &
	15\, (-90 + 31\, U_1)\big)\big] + 5\, r\, (-846 + 725\, r)\, U_1\, U_2\Big)\, U_3 - 
	2 \,r\, U_1\, U_2 \,\Big(557 + 
	2\, r \,\big[-1114 + r\, \big(1671 + r\, (-1114 + 301\, r)\big) \nonumber \\
	&+ &
	360\, U_1\, U_2\big]\Big) \,U_3^2 + 
	3 \Big(1 + r\, \big[-5 + r\, \big(10 + r\, [-10 + r\, (5 + 4\, r)]\big)\big]\Big) (18 \,U_1 + 
	5\, r \,U_2)\, U_3^3\nonumber \\
	& +& 2\, r \,U_1\, U_2\, (3571\, U_1 + 840\, r\, U_2) \,U_3^4 - 
	315\, r^5\, U_3^5\bigg]\, z^7 + 
	U_3^2\, \bigg[(3090 - 2969\, r) \,r\, U_1^3\, U_2\nonumber \\
	& + &
	2\, U_1\, \Big(-9\, \big[280 + r \,(-455 + 184\, r)\big]\, U_1^2 + 
	r\, \big[1757 + 2\, r\, \big(-3514 + r\, [5271 + r\, (-3514 + 883\, r)]\big)\big]\, U_2\nonumber \\
	& +& 
	360\, r^2\, U_1 \,U_2^2\Big)\, U_3 + 
	3\, \Big(36\, \big[-5 + r\, \big(5 + (-5 + r)\, r\big)\, \big(5 + r\, (-5 + 4\, r)\big)\big]\, U_1 \nonumber \\
	&+& 
	5\, r\, \big[-9 + 
	r\, \big(45 + r \,[-90 + r\, (90 + r\, (-45 + 8\, r))]\big)\big]\, U_2\Big)\, U_3^2 - 
	10\, r\, U_1\, U_2\, (1117\, U_1 + 204\, r\, U_2)\, U_3^3\nonumber \\
	& + &
	9\, \Big(1 + r\, \big[-5 + r\, \big(10 + r\, [-10 + r\, (5 + 34\, r)]\big)\big]\Big)\, U_3^4 + 
	4860\, U_1^2\, U_3^5\bigg]\, z^8\nonumber \\
	& +& 
	U_3^2\, \bigg[9\, U_1^2\, \Big(-16\, r^2\, U_2^2 + 
	U_1\, \big[700 + 577\, r^2 - 2 \,r (637 + 68\, U_2)\big]\Big)\nonumber \\
	& - &
	18\, \Big(-135 + 
	r\, \big[675 + r \,\big(-1350 + r \,[1350 + r\, (-675 + 134 \,r)]\big)\big]\Big)\, U_1\, U_3\nonumber \\
	& + &
	2\, r \,U_1\, U_2\, (5405\, U_1 + 732 \,r\, U_2)\, U_3^2 - 
	9\, \Big(11 + 5\, r\, \big[-11 + r\, \big(22 + r\, [-22 + r\, (11 + 2\, r)]\big)\big]\Big)\, U_3^3 \nonumber \\
	&- &
	126\, U_1 \,(90\, U_1 + 53\, r\, U_2)\, U_3^4 + 540\, r\, U_2\, U_3^6\bigg]\, z^9 \nonumber \\
	&+ &
	U_3\, \bigg[204\, r\, U_1^3\, U_2 + 126\, (-40 + 37\, r)\, U_1^3\, U_3 - 
	2\, r \,U_1\, U_2\, (3191\, U_1 + 288\, r\, U_2) \,U_3^2\nonumber \\
	& -& 
	9\, \Big(-55 + r\, \big[275 + r\, \big(-550 + r \,[550 + r\, (-275 + 48 \,r)]\big)\big]\Big) \,U_3^3 + 
	70\, U_1\, (243\, U_1 + 116\, r\, U_2) \,U_3^4\nonumber \\
	& - &
	180\, (36\, U_1 + 7\, r\, U_2)\, U_3^6\bigg]\, z^{10} + 
	U_3\, \bigg[18 \,(140 - 137\, r)\, U_1^3 + 2\, r\, U_1\, U_2\, (1057\, U_1 + 48\, r\, U_2)\, U_3\nonumber \\
	& + &
	9 \,\Big(-165 + 
	r \,\big[825 + r\, \big(-1650 + r\, [1650 + r\, (-825 + 164\, r)]\big)\big]\Big)\, U_3^2 - 
	14\, U_1\, (1215\, U_1 + 457 \,r\, U_2)\, U_3^3 \nonumber \\
	&+ &
	1890\, (6\, U_1 + r\, U_2)\, U_3^5\bigg]\, z^{11} + 
	2\, U_3\, \bigg[-U_1^2\, (360 \,U_1 + 151 \,r\, U_2) + 3\, U_1\, (1890\, U_1 + 529\, r\, U_2) \,U_3^2\nonumber \\
	& - &
	189\, (36\, U_1 + 5\, r \,U_2)\, U_3^4 + 1485\, U_3^6\bigg] z^{12} + 
	2\, \bigg[45\, U_1^3 - 2 \,U_1\, (1215\, U_1 + 226\, r\, U_2)\, U_3^2\nonumber \\
	& +& 
	630\, (9\, U_1 + r\, U_2)\, U_3^4 - 2079\, U_3^6\bigg] \,z^{13} + 
	U_3\, \bigg[U_1\, (1215\, U_1 + 113\, r\, U_2) - 540\, (12\, U_1 + r\, U_2)\, U_3^2 + 
	4158\, U_3^4\bigg]\, z^{14} \nonumber \\
	&-& 
	135\, \bigg[22 - 88 \,r^3 + 22\, r^4 + (-18 + U_1)\, U1 + 
	r\, (-88 + 36\, U_1 - U_2\, U_3) + r^2\, (132 - 18\, U_1 + U_2 \,U_3)\bigg]\, z^{15} \nonumber \\
	&- &
	15\, U_3\, \bigg[36\, U_1 + r\, U_2 - 99\, U_3^2\bigg] \,z^{16} + 9\, \bigg[6 \,U_1 - 55\, U_3^2\bigg]\, z^{17} + 
	99\, U_3 \,z^{18} - 9\, z^{19},
\end{eqnarray}
\begin{eqnarray}
	V_6&=&r^6\, U_1^3\, U_2^2 U_3^4\, (-35\, U_2 + 3\, U_3) \nonumber \\
	&+ &
	r^3\, U_1^2\, U_2\, U_3^4\, \bigg[3\, r^2\, (-7 + 6\, r)\, U_1\, U_2 - 21\, r^3\, U2^2\, U_3 - 
	2 \,\Big(2\, \big[-2 + r\, (4 + 7\, r)\big]\, U_1 + 39\, r^3\, U_2\Big)\, U_3^2 + 8\, r^3\, U_3^3\bigg]\, z\nonumber \\
	& + &
	r^2\, U_1\, U_3^4\, \bigg[21\, r^3\, U_1\, U_2^2\, (2\, U_1 + r\, U_2) + 
	U_1\, U_2\, \Big(\big[-16 + r\, \big(5 + r \,(38 + 81\, r)\big)\big]\, U_1\nonumber \\
	& + &
	3\, r^3\, (-7 + 59\, r)\, U_2\Big)\, U_3 + 24\, r^3\, (-3 + 2\, r)\, U_1\, U_2\, U_3^2 - 
	48\, r^4\, U_2^2\, U_3^3 + r\, \big[9\, (r^2 - U_1) \,U_1 + 25 \,r^3\, U_2\big]\, U_3^4\bigg]\, z^2\nonumber \\
	& + &
	r\, U_3^3\, \bigg[-21\, r^4\, U_1^3\, U_2^2 - 
	U_1^2\, U_2\, \Big(\big[-24 + r\, \big(9 + r\, [17 + r\, (35 + 71\, r)]\big)\big]\, U_1 + 
	3\, r^4\, (-21 + 47\, r)\, U_2\Big) \,U_3\nonumber \\
	& +&
	24\, (12 - 11\, r)\, r^4 U_1^2\, U_2\, U_3^2 + 
	144\, r^5\, U_1\, U_2^2\, U_3^3 + 
	r \,U_1\, \big[9\, U_1\, (-5\, r^3 + 3\, U_1 + 2\, r\, U_1)\nonumber \\
	& -& 
	4\, r^3 \,(9 + 16\, r)\, U_2\big]\, U_3^4 + 3\, r^4\, (6\, U_1 + 5\, r\, U_2)\, U_3^6\bigg]\, z^3\nonumber \\
	& - &
	r\, U_3^3\, \bigg[\Big(132 + r\, \big[-317 + r\, \big(281 + r \,(-139 + 7\, r)\big)\big]\Big)\, U_1^3\, U2\nonumber \\
	& + &
	U_1^2\, U_2\, \Big[-8 \,\Big(4 + r\, \big[-20 + r\, \big(40 + r \,[-40 + r\, (-34 + 49\, r)]\big)\big]\Big) + 
	63\, r^4\, U_2\Big]\, U_3 + 144 \,r^5\, U_1\, U_2^2\, U_3^2\nonumber \\
	& + &
	6\, U_1 \,\Big(-3\, U_1\, \big[5\, r^4 + (-3 + (-3 + r)\, r)\, U_1\big] + 
	5 \,(-6 + r)\, r^4\, U_2\Big)\, U_3^3 + 3 \,r^4\, (36\, U_1 + 25\, r\, U_2)\, U_3^5 - 
	9\, r^4\, U_3^7\bigg]\, z^4\nonumber \\
	& + &
	U_3^3\, \bigg[21\, r^5\, U_1^2 \,U_2^2 + 
	2\, r\, U_1 \,U_2\, \Big(U_1\, \big[-89 + 156\, U_1 + 
	r\, \big(356 - 165\, U_1 + r\, (-534 + r\, (356 + 55\, r) + 47\, U_1)\big)\big]\nonumber \\
	& + &
	24\, r^5\, U_2\Big)\, U_3 + 
	2\, U_1 \,\Big[45\, U_1\, (-r^5 + U_1) + 
	2\, r \,\Big(2 + 
	r\, \big[-10 + r\, \big(20 + r\, [-20 + r\, (-80 + 63\, r)]\big)\big]\Big)\, U_2\Big]\, U_3^2\nonumber \\
	& +& 
	27\, r\, (4 + r)\, U_1^3\, U_3^3 + 30\, r^5\, (9\, U_1 + 5 \,r\, U_2)\, U_3^4 - 
	63\, r^5\, U_3^6\bigg]\, z^5\nonumber \\
	& +& 
	U_3^3\, \bigg[r \,U_1^2 \,\Big[382 - 420 \,U_1 + 
	r \,\big(-1528 + r\, (2292 + 2\, r\, (-764 + 155 \,r) - 193 \,U_1) + 
	594\, U_1\big)\Big]\, U_2\nonumber \\
	& + &
	U_1\, \Big[9\, U_1\, \Big(15 - 
	5\, r\, \big[15 + r \,\big(-30 + r\, [30 + r\, (-15 + 2\, r)]\big)\big] + (-80 + 
	3\, r\, (38 + r\, (-16 + 3\, r)))\, U_1\Big)\nonumber \\
	& - &
	5\, r\, \Big(3 + 
	r\, \big[-15 + r\, \big(30 + r\, [-30 + r\, (-57 + 64\, r)]\big)\big]\Big)\, U_2\Big]\, U_3 - 
	30\, r^5\, (12\, U_1 + 5\, r\, U_2)\, U_3^3 + 189\, r^5\, U_3^5\bigg]\, z^6 \nonumber \\
	&+ &
	U_3^3\, \bigg[U_1^2 \,\Big(45 (-27 + 56\, U_1) + 
	9\, r\, \big[675 - 630\, U_1 + 
	r \,\big(r\, (1350 + r \,(-675 + 134\, r) - 114\, U_1) + 
	15\, (-90 + 31\, U_1)\big)\big] \nonumber \\
	&+&
	5\, (72 - 49\, r)\, r\, U_1\, U_2\Big) - 
	20\, r\, \Big(8 + r\, \big[-32 + r \,\big(48 + r \,(-32 + 17\, r)\big)\big]\Big)\, U_1\, U_2\, U_3\nonumber \\
	& + &
	3\, \Big(1 + r\, \big[-5 + r\, \big(10 + r\, [-10 + r\, (5 + 4 \,r)]\big)\big]\Big) (18\, U_1 + 
	5\, r\, U_2)\, U_3^2 - 346\, r\, U_1^2\, U_2\, U_3^3 - 315\, r^5\, U_3^4\bigg]\, z^7 \nonumber \\
	&+& 
	U_3^2\, \bigg[r \,(-204 + 181\, r)\, U_1^3\, U_2 + 
	2\, U_1\, \Big(-9\, \Big[280 + r\, (-455 + 184\, r)\big]\, U_1^2 \nonumber \\
	&+ &
	2 \,r\, \big[217 + 2\, r\,\big (-434 + r\, [651 + r\, (-434 + 113\, r)]\big)\big] \,U_2\Big)\, U_3 \nonumber \\
	&+ &
	3\, \Big[36\, \Big(-5 + r\, \big[5 + (-5 + r)\, r\big]\, \big[5 + r\, (-5 + 4\, r)\big]\Big) \,U_1 \nonumber \\
	&+& 
	5\, r\, \Big(-9 + 
	r \,\big[45 + r\, \big(-90 + r\, [90 + r\, (-45 + 8\, r)]\big)\big]\Big) \,U_2\Big]\, U_3^2 - 
	10\, r\, U_1^2\, U_2\, U_3^3\nonumber \\
	& + &
	9\, \Big(1 + r\, \big[-5 + r\, \big(10 + r\, [-10 + r\, (5 + 34\, r)]\big)\big]\Big)\, U_3^4 + 
	4860\, U_1^2\, U_3^5\bigg] \,z^8\nonumber \\
	& +&
	U_3^2\, \bigg[9\, U_1^3\, (700 + r\, (-1274 + 577\, r + 8 \,U_2))\nonumber \\
	& - &
	18\,\Big (-135 + 
	r\, \big[675 + r\, \big(-1350 + r \,[1350 + r \,(-675 + 134 \,r)]\big)\big]\Big)\, U_1\, U_3 + 
	298\, r\, U_1^2\, U_2\, U_3^2\nonumber \\
	& - &
	9\, \Big(11 + 5 \,r \,\big[-11 + r\, \big(22 + r\, [-22 + r\, (11 + 2\, r)]\big)\big]\Big)\, U_3^3 - 
	252\, U_1\, (45\, U_1 + 8\, r\, U_2)\, U_3^4 + 540 \,r\, U_2\, U_3^6\bigg]\, z^9\nonumber \\
	& - &
	U3\, \bigg[12\, r\, U_1^3\, U_2 + 126 \,(40 - 37\, r)\, U_1^3\, U_3 + 262\, r\, U_1^2\, U_2\, U_3^2 \nonumber \\
	&+ &
	9\, \Big(-55 + r\, \big[275 + r\, \big(-550 + r\, [550 + r\, (-275 + 48\, r)]\big)\big]\Big) \,U_3^3 - 
	14\, U_1\, (1215\, U_1 + 193\, r\,U_2)\, U_3^4 \nonumber \\
	&+ &
	180\, (36\, U_1 + 7\, r\, U_2)\, U_3^6\bigg]\, z^{10} + 
	U_3\, \bigg[18\, (140 - 137\, r)\, U1^3 + 98\, r\, U_1^2 \,U_2\, U_3\nonumber \\
	& + &
	9\,\Big (-165 + 
	r\, \big[825 + r\, \big(-1650 + r\, [1650 + r \,(-825 + 164 \,r)]\big)\big]\Big)\, U_3^2 - 
	70\, U_1\, (243 \,U_1 + 32\, r \,U_2) \,U_3^3 \nonumber \\
	&+&
	1890\, (6\, U_1 + r\, U_2) \,U_3^5\bigg]\, z^{11} + 
	2\, U_3\, \bigg[-U_1^2\, (360\, U_1 + 7\, r\, U_2) + 30\, U_1\, (189 \,U_1 + 19\, r \,U_2)\, U_3^2\nonumber \\
	& -& 
	189\, (36\, U_1\, + 5\, r\, U_2)\, U_3^4 + 1485\, U_3^6\bigg] \,z^{12} + 
	2\, \bigg[45 \,U_1^3 - 2\, U_1\, (1215\, U_1 + 82\, r\, U_2)\, U_3^2 + 
	630\, (9\, U_1 + r\, U_2) \,U_3^4 \nonumber \\
	&- &
	2079\, U_3^6\bigg]\, z^{13} + 
	U3\, \bigg[U_1\, (1215\, U_1 + 41\, r\, U_2) - 540\, (12\, U_1 + r\, U_2)\, U3^2 + 
	4158\, U3^4\bigg] \,z^{14}\nonumber \\
	& - &
	135\, \bigg[U_1^2 - (18\, U_1 + r \,U_2)\, U_3^2 + 22\, U_3^4\bigg]\, z^{15} - 
	15\, U_3 \,\bigg[36\, U_1 + r \,U_2 - 99\, U_3^2\bigg]\, z^{16} \nonumber \\
	&+& 9\, \bigg[6\, U_1 - 55\, U_3^2\bigg]\, z^{17} + 
	99\, U_3\, z^{18} - 9\, z^{19},
\end{eqnarray}
\begin{eqnarray}
	V_7&=&-7\, r^4\, U_1^2\, U_2^2\, U_3^2 + 49\, r^5 \,U_1^2\, U_2^2\, U_3^2 \nonumber \\
	&- &
	147\, r^6\, U_1^2 \,U_2^2 \,U_3^2 + 245\, r^7\, U_1^2\, U_2^2\, U_3^2 - 
	245\, r^8\, U_1^2\, U_2^2\, U_3^2 + 147\, r^9\, U_1^2\, U_2^2\, U_3^2 - 
	49\, r^{10}\, U_1^2\, U_2^2\, U_3^2\nonumber \\
	& +&
	7\, r^{11}\, U_1^2\, U_2^2\, U_3^2 + 
	2\, r^4\, U_1^2\, U_2^3\, U_3^3 - 10 \,r^5\, U_1^2\, U_2^3\, U_3^3 + 
	20\, r^6\, U_1^2\, U_2^3\, U_3^3 - 20\, r^7\, U_1^2\, U_2^3\, U_3^3 \nonumber \\ 
	&+ &
	10\, r^8\, U_1^2\, U_2^3\, U_3^3 - 2\, r^9\, U_1^2\, U_2^3\, U_3^3  + 
	2\, r^3\, U_1^2\, U_2\, U_3^7\, \bigg[(1 - 6 r)\, U_2^2 + 23 \,r U_2\, U_3 + 3\, r\, U+3^2\bigg] \,z \nonumber \\
	&+ &
	r^3\, U_1\, U_3^6 \,\bigg[4\, (-3 + 8\, r)\, U_1\, U_2^3 - 
	125\, r \,U_1\, U_2^2\, U3 + \Big(r\, U_1 + \big[r^3 - 6\, (1 + 6\, r)\, U_1\big]\, U_2\Big)\, U_3^2\bigg]\, z^2 \nonumber \\
	&+&
	r^2\, U_1\, U_3^5\, \bigg[10\, (3 - 5\, r) r \,U_1\, U_2^3 + 
	180 \,r^2\, U_1\, U_2^2\, U_3 - \Big(r\, (2 + r + 2\, r^2)\, U_1 + 4\, r^4\, U_2\nonumber \\
	& - &
	6\, \big[1 + 6\, r\, (1 + 2\, r)\big]\, U_1 \,U_2\Big)\, U_3^2 + 2\, r^3\, U_3^4\bigg] \,z^3 + 
	r\, U_3^4\, \bigg[10\, r^2\, (-4 + 5\, r)\, U_1^2\, U_2^3 - 141\, r^3\, U_1^2\, U_2^2\, U_3\nonumber \\
	& +& 
	U_1\,\Big(r\, \big[3 + r\, \big(3 + r \,(-5 + 9 \,r)\big)\big]\, U_1 + 
	6\, \big[r^5 - \big(1 + r\, [6 + r\, (11 + 7\, r)]\big)\, U_1\big]\, U_2\Big)\, U_3^2 - 
	12\, r^4\, U_1\, U_3^4 + r^4 \,U_3^6\bigg]\, z^4\nonumber \\
	& + &
	r \,U_3^4\, \bigg[r^2\, U_1^2\, U_2^2\, (47\, r + 30 \,U_2) + 
	U_1\, \Big(2 \,\big[-3 + r\, \big(15 + r \,[-30 + r\, (30 + (-15 + r)\, r)]\big)\big]\, U_2\nonumber \\
	& + &
	U_1\, \big[-4 + 60\, U_2 + 
	r\, \big(-5 + 24\, U_2 + 
	r\, [9 + r\, (3 - 13\, r - 24\, U_2) + 36 \,U_2]\big)\big]\Big) \,U_3 + 
	30 \,r^4\, U_1\, U_3^3 - 7\, r^4\, U_3^5\bigg]\, z^5\nonumber \\
	& + &
	U_3^3\, \bigg[-10\, r^3\, U_1^2\, U_2^3 + 
	U_1\, (5\, U_1 + 
	r\, \Big(66 - 5\, r\, \big[66 + r\, \big(-132 + r \,[132 + r\, (-66 + 13\, r)]\big)\big]\nonumber \\
	& - &
	258\, U_1 + 6\, r\, \big[59 + 10\, (-3 + r)\, r\big] \,U_1\Big)\, U_2)\, U_3 - 
	r \,(-7 + r\, (8 + r\, (15 + 4 \,r))) \,U_1^2\, U_3^2\nonumber \\
	& -&
	40\, r^5 \,U_1\, U_3^3 + 
	21\, r^5\, U_3^5\bigg] \,z^6 + 
	U_3^3\, \bigg[140\, r^8 - 35\, r^9 + 2\, r^6\, (70 - 9\, U_1) + 3\, (2 - 15\, U_1)\, U_1 \nonumber \\
	&+& 
	6 \,r^7\, (-35 + 4\, U_1) + 3\, r^4\, U_1\, (70 - 11\, U_1 + 424\, U_2\, U_3) + 
	r^2\, U_1\, (126 - 107\, U_1 + 1272\, U_2 \,U_3\nonumber \\
	& -&
	738\, U_1\, U_2\, U_3) + 
	r\, U_1\, (-42 + 113\, U_1 + 6\, (-53 + 104 \,U+1) \,U_2\, U_3)\nonumber \\
	& + &
	3 \,r^3\, U_1\, \big[-70 - 636\, U_2 \,U_3 + 3 \,U_1\, (7 + 26\, U_2\, U_3)\big] + 
	r^5\, \big[-35 + 8\, U_1^2 - 6 \,U_1 \,(16 + 53\, U_2\, U_3)\big]\bigg]\, z^7\nonumber \\
	& + &
	U_3^3\, \bigg[-6 r\, \big[155 + r\, (-249 + 98\, r)\big]\, U_1^2\, U_2 + 
	U_1\,\Big(-60 + 12\, r\, (5 + (-5 + r)\, r) \big[5 + r\, (-5 + 4\, r)\big]\nonumber \\
	& + &
	180\, U_1 + 
	r \,\big[-316 + (197 - 49\, r) \,r\big] \,U_1\Big)\, U_3 + \Big(1 + 
	r\, \big[-5 + r\, \big(10 + r \,[-10 + r\, (5 + 34\, r)]\big)\big]\Big) \,U_3^3\nonumber \\
	& + &
	882\, r\, U_1\, U_2\, U_3^4\bigg]\, z^8 - 
	U_3^3 \,\bigg[U_1^2\, \big[420 + 
	r\, \big(-980 - 876\, U_2 + r\, (763 - 201\, r + 732 \,U_2)\big)\big]\nonumber \\
	& +&
	\Big(11 + 
	5\, r\, \big[-11 + r\, \big(22 + r\, [-22 + r\, (11 + 2\, r)]\big)\big]\Big)\, U_3^2\nonumber \\
	& + &
	2\, U_1\, \Big(-135 + 
	r \,\big[675 + r \,\big(-1350 + r\, [1350 + r\, (-675 + 134 \,r)]\big) + 
	777\, U_2\, U_3^3\big]\Big)\bigg] \,z^9\nonumber \\
	& + &
	U3^2 \,\bigg[6\, r\, (-85 + 81\, r)\, U_1^2 \,U_2 + 
	7\, \big[90 + r\, (-148 + 61\, r)\big]\, U_1^2\, U_3\nonumber \\
	& + &
	\Big(55 + 
	r\, \big[-275 + r\, \big(550 + r\, (-550 + (275 - 48\, r)\, r)\big)\big]\Big)\, U_3^2 + 
	1806\, r\, U_1\, U_2\, U_3^3 - 720\, U_1\, U_3^5\bigg]\, z^{10}\nonumber \\
	& + &
	U_3^2\, \bigg[U_1^2 \,\big[-630 - 521\, r^2 + 28\, r\, (41 + 6 \,U_2)\big]\nonumber \\
	& +& \Big(-165 + 
	r\, \big[825 + r\, \big(-1650 + r\, [1650 + r\, (-825 + 164 \,r)]\big)\big]\Big)\, U_3 + 
	126\, U_1\, U_3^2\, (-11\, r\, U_2 + 10 \,U_3^2)\bigg]\, z^{11}\nonumber \\
	& +& 
	2\, U_3\, \bigg[3\, U_3\, (70\, U_1^2 - 252\, U_1\, U_3^3 + 55\, U_3^5) + 
	r\, U_1\, \big[339 \,U_2\, U_3^2 - 2 \,U_1\, (6\, U_2 + 97\, U_3)\big]\bigg]\, z^{12}\nonumber \\
	& -& 
	2\, U_3\, \bigg[231 + 1155 \,r^4 - 231 \,r^5 + 210\, r^2\, (11 - 9 \,U_1) + 
	90 \,(-7 + U1)\, U_1 + 210 \,r^3\, (-11 + 3 \,U_1)\nonumber \\
	& + &
	r\, \big[-1155 + 2\, U_1\, (945 - 44 \,U_1 + 48\, U_2\, U_3)\big]\bigg]\, z^{13} + 
	3\, U_3\, \bigg[15 \,U_1^2 + 154 \,U_3^4 + 8\, U_1 \,(r\, U_2 - 30\, U_3^2)\bigg]\, z^{14} \nonumber \\
	&-& 
	5\, \bigg[U_1^2 - 54\, U_1\, U_3^2 + 66\, U_3^4\bigg]\, z^{15} + 
	15\, U_3\, \bigg[-4\, U_1 + 11 \,U_3^2\bigg]\, z^{16} + \bigg[6\, U_1 - 55\, U_3^2\bigg] z^{17} + 11 U_3 z^{18} - 
	z^{19},
\end{eqnarray}
	\begin{eqnarray}
		V_{8}&=& -7\, r^4\, U_1^2\, U_2^2 \,U_3^8 
		 +3 \,r^4 \,U_1^2\, U_2\, U_3^5\, \bigg[\big[13 + r\, (-26 + 11\, r)\big]\, U_2 + 
		2 \,\big[2 + r\, (-4 + 3\, r)\big]\, U_3\bigg]\, z\nonumber \\
		&+ &
		r^3\, U_1\, U_3^4\, \bigg[-2\, r \,(43 - 86\, r + 40\, r^2) \,U_1 \,U_2^2 
		- 12\, U_1\, U_2 \,U_3 + 
		6\, r\, (-6 + 11\, r)\, U_1\, U_2 \,U_3^2 + r\, (3\, U_1 + 2\, r^2\, U_2)\, U_3^3\bigg] \,z^2 \nonumber \\
		&+& 
		2\, r^2\, U_1 \,U_3^4 \,\bigg[3 \Big(2 + r\, \big[6 + r\, \big(-4 + 3\, r\, (-5 + 4\, r)\big)\big]\Big)\, U_1\, U_2 + 
		47\, r^2\, U_1\, U_2^2\, U_3 - 3\, r\, (U_1 + r^2\, U_1 + r^2\, U_2)\, U_3^2 \nonumber \\
		&+& 
		3\, r^3\, U_3^4\bigg]\, z^3 - 
		r \,U_3^4\, \bigg[-3\, r^4\, U_3^5 + 
		U_1^2\, \big[U_2\, \big(12 + 6\, r\, [8 + r\, (2 + 3\, (-4 + r)\, r)] + 47 \,r^3\, U_2\big) \nonumber \\
		&-& 
		3\, r\, (3 + r - 5\, r^2 + 7\, r^3) \,U_3\big] + 
		6\, r^4\, U_1\, U_3\, \big[(-4 + 3\, r)\, U_2 + 5 \,U_3^2\big]\bigg] \,z^4 - 
		2 \,r\, U_3^3\, \bigg[U_1 \Big(6 - 54 \,U_1 \nonumber \\
		&+& 
		r\, \big[-36 + 
		r\, \big(90 + r\, [r \,(r\, (-71 + 23\, r) - 27\, (-4 + U_1)) + 
		24 \,(-5 + U_1)] - 6\, U_1\big) + 60 \,U_1\big]\Big)\, U_2 \nonumber \\
		&+ &
		3\, (2 + r - 5 \,r^2 + r^3 + 3\, r^4)\, U_1^2\, U_3 - 30\, r^4\, U_1\, U_3^3 + 
		9 \,r^4\, U_3^5\bigg]\, z^5 \nonumber  \\
		&  + &
		3\, U_3^3 \,\bigg[-60 \,r^8 + 15\, r^9 + r^7 \,(90 - 20\, U_1) + 5\, U_1^2 + 
		20\, r^6\, (-3 + 2\, U_1) + r^4\, U_1\, (9 \,U_1 - 160\, U_2\, U_3) \nonumber \\
		& + &
		r\, U_1 \,\big[3 \,U_1 + 8\, (5 - 17\, U_1)\, U_2\, U_3\big] + 
		3\, r^3\, U_1\, \big[U_1 - 8\,(-10 + U_1)\, U_2\, U_3\big] + 
		r^2\, U_1\, \big[-17\, U_1 + 32\, (-5 + 3 \,U_1)\, U_2\, U_3\big] \nonumber \\
		& + &
		r^5\, \big[15 - 2 \,U_1\, (10 + U_1 - 24\, U_2\, U_3)\big]\bigg]\, z^6 - 
		6\, U_3^3\, \bigg[r\, U_1\, \big[86 - 140\, U_1 \nonumber \\
		& +& 
		r\, \big(-344 + r\, [516 + r\, (-344 + 87\, r) - 66\, U_1] + 198\, U_1\big)\big]\, U_2 - 
		U_1\, \big[3 - 20\, U_1 \nonumber \\
		&+ &
		r\, \big(-15 + 30\, r - 30 \,r^2 + 15\, r^3 + 2\, r^4 + 
		3\, [6 + (-2 + r)\, r] \,U_1\big)\big]\, U_3 + 10\, r^5\, U_3^3\bigg]\, z^7 \nonumber \\
		&+& 
		3\, U_3^3 \,\bigg[U_1 \,\big[-54 + 140\, U_1 + 
		r\, \big(270 - 280 \,U_1 + 
		r\, [-540 + r\, (540 - 270\, r + 52 \,r^2 - 43 \,U_1) + 185 \,U_1] \nonumber \\
		& + &
		20\, (-17 + 13\, r)\, U_1\, U_2\big)\big] + \Big(1 + 
		r\, \big[-5 + r\, \big(10 + r\, [-10 + r\, (5 + 14\, r)]\big)\big]\Big)\, U_3^2 + 
		416 \,r U_1\, U_2\, U_3^3\bigg]\, z^8 \nonumber \\
		&+ &
		6\, U_3^2\, \bigg[2\, (61 - 
		57\, r)\, r\, U_1^2\, U_2 + \big[-140 + (210 - 79\, r) \,r\big]\, U_1^2\, U_3 \nonumber \\
		& +& \big[-5 + 
		r\, \big(25 + r\, [-50 + (-10 + r)\, r (-5 + 2\, r)]\big)\big]\, U_3^2 - 
		310\, r\, U_1\, U_2\, U_3^3 + 108\, U_1\, U_3^5\bigg]\, z^9  \nonumber \\
		&-& 
		3\, U_3^2 \,\bigg[U_1^2 \,\big[-350 + r\, (616 - 269\, r + 96 \,U_2)\big] + \big[-45 + 
		r\, \big(15 + r \,(-15 + 4 \,r)\big) \big(15 + r \,(-15 + 11\, r)\big)\big]\, U_3 \nonumber \\
		& +& 
		8 \,U_1 \,U_3^2\, (-73\, r\, U_2 + 63\, U_3^2)\bigg]\, z^{10}  + 
		12\, U_3\, \bigg[-70\, U_1^2\, U_3 + 189\, U_1\, U_3^4 - 30\, U_3^6 \nonumber \\ 
		&+& 
		r\, U_1\, (4\, U_1\, U_2 + 63\, U_1\, U_3 - 85\, U_2\, U_3^2)\bigg]\, z^{11}  
		+ 
		6\, U_3 \,\bigg[105 - 525\, r + 525\, r^4 - 105\, r^5 \nonumber \\
		&+& 42\, r^2\, (25 - 27 \,U_1) + 
		14\, U_1 \,(-27 + 5\, U_1) + 42 \,r^3 (-25 + 9\, U_1) + 
		2 \,r\, U_1 \,(567 - 34\, U_1 + 28\, U_2\, U_3)\bigg]\, z^{12}\nonumber \\
		& - &
		12\, U_3\, \bigg[10\, U_1^2 + 4\, r\, U_1\, U_2 - 126\, U_1\, U_3^2 + 63\, U_3^4\bigg]\, z^{13}+ 
		3\, \bigg[5\, U_1^2 - 216\, U_1\, U_3^2 + 210\, U_3^4\bigg]\, z^{14} \nonumber \\
		& +& 
		18 \,U_3\, \bigg[9\, U_1 - 20\, U_3^2\bigg]\, z^{15}+ 9\, \bigg[-2\, U_1 + 15\, U_3^2bigg]\, z^{16} - 
		30\, U_3 \,z^{17} + 3 \,z^{18} ,
	\end{eqnarray}
\begin{eqnarray}
	V_9&=&7\, r^7 \,U_1^2\, U_2^2\, (U_2 - U_3) \,U_3^5 \nonumber \\
	&+ &
	r^4\, U_1^2\, U_2\, U_3^5\, \bigg[r^2\, U2 \,(-6 + 13\, r + 7\, U_2) + 
	6\, \big[1 + r\, (-2 + 3\, r)\big] \,U_3^2\bigg]\, z + 
	r^3\, U_1\, U_3^4 \,\bigg[-7\, r^3\, U_1 \,U_2^3\nonumber \\
	& +&
	12\, r^3\, U_1\, U_2^2\, U_3 - 
	6\, \big[1 + r\,\big (4 + r\, (-9 + 10\, r)\big)\big]\, U_1\, U_2\, U_3^2 + 
	r\, (3\, U_1\, + r^2\, U_2) \,U_3^4\bigg]\, z^2 \nonumber \\
	&+& 
	r^2\, U_1\, U_3^4\, \bigg[-6\, r^4\, U_1\, U_2^2 + 6\, \big[1 + r\, (4 + r)\big]\, U_1\, U_2\, U_3 - 
	60\, r^3\, U_1\, U_2\, U_3^2 \nonumber \\
	&+& 
	r\, \big[-3\, (2 + r\, + 2\, r^2)\, U_1 + 4\, r^2\, (-3 + 2\, r)\, U_2\big]\, U_3^3 + 
	6\, r^3\, U_3^5\bigg]\, z^3\nonumber \\
	& +& 
	3\, r\, U_3^4\, \bigg[r^4\, U_3^6 - 2\, r^4\, U_1\, U_3^2\, \big[(-10 + 9\, r)\, U_2 + 6\, U_3^2\big] + 
	U_1^2\, \Big(2\, \big[-1 - 4\, r + 3\, r^3\, (3 + (-3 + r)\, r)\big]\, U_2\nonumber \\
	& +& 
	r\, \big[3 + r\, \big(3 + r\, (-5 + 9\, r)\big)\big]\, U_3^2\Big)\bigg]\, z^4 + 
	r\, U_3^4 \,\bigg[2 \,\Big(-3 + 
	r\, \big[15 + r\, \big(-30 + r\, [30 + r\, (-75 + 61\, r)]\big)\big]\Big)\, U_1 \,U_2\, U_3\nonumber \\
	& + &
	90\, r^4\, U_1\, U_3^3 - 21\, r^4\, U_3^5 + 
	3\, U_1^2\, \Big(4\, \big[5 + r\, (-3 + r - 5\, r^2 + 6\, r^3)\big]\, U_2 + \big[-4 + 
	r\, \big(-5 + r\, [9 + (3 - 13\, r)\, r]\big)\big] \,U_3\Big)\bigg]\, z^5 \nonumber \\
	&+ &
	U_3^3\, \bigg[r\, U1\, \Big[66 - 258\, U_1 + 
	r\, \Big(-396 + 5\, r\, \big[198 + r\, \big(-264 + r\, [222 + r\, (-127 + 37\, r)]\big)\big]\nonumber \\
	& +& 
	612\, U_1 - 6\, r\, \big[89 + 4\, r\, (-10 + 3\, r)\big]\, U_1\Big)\Big]\, U2 + 15\, U_1^2\, U_3 - 
	3\, r\, \big[-7 + r\, \big(8 + r\, (15 + 4\, r)\big)\big] \,U_1^2\, U_3^2 \nonumber \\
	&-& 120\, r^5\, U_1\, U_3^3 + 
	63\, r^5\, U_3^5\bigg]\, z^6 - 
	3\, U_3^3\, \bigg[-140\, r^8 + 35\, r^9 + 6\, r^7\, (35 - 4 \,U_1) + 
	2 \,r^6\, (-70 + 9\, U_1) \nonumber \\
	&+& 3\, U_1\, (-2 + 15\, U_1) + 
	r^4\, U_1\, (-210 + 33\, U_1 - 424 \,U_2\, U_3)\nonumber \\
	& + &
	r^2\, U_1\, (-126 + 107\, U1 - 424\, U_2\, U_3 + 246 \,U_1\, U_2 \,U_3) + 
	3\, r^3 \,U_1\, \big[70 + 212\, U_2 \,U_3 - U_1 \,(21 + 26\, U_2\, U_3)\big]\nonumber \\
	& + &
	r^5\, \big[35 + 2 \,U_1\, (48 - 4\, U_1 + 63\, U_2\, U_3)\big] + 
	r \,U_1\, \big[42 + 106\, U_2\, U_3 - U_1\, (113 + 208\, U_2\, U_3)\big]\bigg]\, z^7\nonumber \\
	& + &
	3\, U_3^3\, \bigg[2\, r\, U_1\, \big[147 - 155\, U_1 + 
	r\, \big(-588 + r\, [r\, (-588 + 149\, r) - 98\, (-9 + U_1)] + 249\, U_1\big)\big]\, U_2\nonumber \\
	& + &
	U_1\, \big[-60 + 12\, r\, \big(5 + (-5 + r)\, r\big)\, \big(5 + r \,(-5 + 4 r)\big) + 180\, U_1 + 
	r\, \big(-316 + (197 - 49\, r)\, r\big)\, U_1\big]\, U_3\nonumber \\
	& +&
	\Big(1 + 
	r\, \big[-5 + r\, \big(10 + r\, [-10 + r\, (5 + 34\, r)]\big)\big]\Big)\, U_3^3\bigg]\, z^8 - 
	3 U_3^3\, \bigg[U_1\, \Big(30\, (-9 + 14 \,U_1)\nonumber \\
	& + &
	r \big[1350 - 980\, U_1 + 
	r\,\big(-2700 + r\, [2700 + 2\, r \,(-675 + 134 \,r) - 201 U_1] + 
	763\, U_1\big) \nonumber \\
	&+ &
	4\, (-73 + 61\, r)\, U_1\, U_2\big]\Big) +\Big (11 + 
	5\, r\, \big[-11 + r \,\big(22 + r \,[-22 + r \,(11 + 2\, r)]\big)\big]\Big) U_3^2 + 
	518 \,r \,U_1\, U_2\, U_3^3\bigg]\, z^9\nonumber \\
	& -& 
	3\, U_3^2\, \bigg[2 \,(85 - 81 \,r)\, r \,U_1^2\, U_2 - 
	7 \,\big[90 + r\, (-148 + 61\, r)\big] \,U_1^2\, U_3\nonumber \\
	& +&
	\Big(-55 + 
	r\, \big[275 + r\, \big(-550 + r\, [550 + r\, (-275 + 48\, r)]\big)\big]\Big) U_3^2 - 
	602 \,r \,U_1 \,U_2\, U_3^3 + 720\, U_1\, U_3^5\bigg]\, z^{10}\nonumber \\
	& + &
	3\, U_3^2\, \bigg[U_1^2 \,\big[-630 + r\, (1148 - 521\, r + 56\, U_2)\big]\nonumber \\
	& +& \Big(-165 + 
	r\, \big[825 + r\, \big(-1650 + r\, [1650 + r\, (-825 + 164\, r)]\big)\big]\Big)\, U_3 + 
	42\, U_1\, U_3^2\, (-11\, r\, U_2 + 30 \,U_3^2)\bigg]\, z^{11} \nonumber \\
	&+& 
	6\, U_3\, \bigg[3\, U_3\, (70 \,U_1^2 - 252 \,U_1\, U_3^3 + 55\, U_3^5) + 
	r\, U_1\, \big[113 \,U_2\, U_3^2 - 2\, U_1\, (2\, U_2 + 97\, U_3)\big]\bigg]\, z^{12}\nonumber \\
	& - &
	6\, U_3\, \bigg[231 + 1155\, r^4 - 231\, r^5 + 210\, r^2\, (11 - 9\, U_1) + 
	90\, (-7 + U_1)\, U_1 + 210\, r^3 \,(-11 + 3\, U_1) \nonumber \\
	&+ &
	r\, \big[-1155 + 2 \,U_1\, (945 - 44 \,U_1 + 16\, U_2\, U_3)\big]\bigg]\, z^{13} + 
	3\, U_3\, \bigg[45\, U_1^2 + 462\, U_3^4 + 8\, U_1\, (r\, U_2 - 90\, U_3^2)\bigg]\, z^{14} \nonumber \\
	&-& 
	15\, \bigg[U_1^2 - 54\, U_1\, U_3^2 + 66\, U_3^4\bigg]\, z^{15} + 
	45\, U_3\, \bigg[-4 \,U_1 + 11 \,U_3^2\bigg]\, z^{16} + 3\, \bigg[6\, U_1 - 55\, U_3^2\bigg]\, z^{17} + 
	33\, U_3\, z^{18} - 3\, z^{19},
\end{eqnarray}
	\begin{eqnarray}
		V_{10}&=&-7\, r^7 \,U_1^2\, U_2^2\, U_3^6 \nonumber \\
		&+& 
		r^6\, U_1^2\, U_2\, U_3^5 \,\bigg[7\, r\, U_2 + 6\, U_3^2\bigg]\, z + 
		r^4\, U_1^2\, U_3^6\, \bigg[-6\, r \,(1 + 2\, r)\, U_2 + U_3^2\bigg]\, z^2 \nonumber \\
		&- &
		r^3\, U_1\, U_3^5\, \bigg[6\, (-4 + r)\, r^2\, U_1 \,U_2+ (2 + r + 2 \,r^2)\, U_1\, U_3^2 + 
		6\, r^2\, U_2 \,U_3^3 - 2 \,r^2\, U_3^4\bigg]\, z^3\nonumber \\
		& +& 
		r^2\, U_3^4\, \bigg[6\, r^3\, (-6 + 5\, r) \,U_1^2\, U_2
		+ \big[3 + 
		r\, \big(3 + r\, (-5 + 9\, r)\big)\big]\, U_1^2\, U_3^2 + 30\, r^3\, U_1\, U_2\, U3^3\nonumber \\
		& -& 
		12\, r^3\, U_1 \,U_3^4 + r^3\, U_3^6\bigg] \,z^4 + 
		r\, U_3^4\, \bigg[24 \,r^4 \,U_1^2\, U_2 + \big[-4 + 
		r\, \big(-5 + r \,(9 + (3 - 13\, r)\, r)\big)\big]\, U_1^2\, U_3 \nonumber \\
		& -& 60\, r^4\, U_1\, U_2\, U_3^2 + 
		30\, r^4\, U_1\, U_3^3 - 7\, r^4\, U_3^5\bigg]\, z^5 + 
		U_3^3\, \bigg[-6\, r^5\, U_1^2\, U_2 + 5\, U_1^2\, U_3 \nonumber \\
		& + &
		r\, U_1\, \big[7\, U_1 - r \,\big(8 + r\, (15 + 4\, r)\big)\, U_1 + 60\, r^4\, U_2\big] \,U_3^2 - 
		40\, r^5\, U_1\, U_3^3 + 21 \,r^5\, U_3^5\bigg] \,z^6 \nonumber \\
		& +& 
		U_3^3\, \bigg[\Big(-45 + r\, \big[113 + r \,\big(-107 + r\, [63 + r\, (-33 + 8 \,r)]\big)\big]\Big)\, U_1^2 - 
		35\, r^5\, U_3^4 \nonumber \\
		&+ &
		6\, U_1\, U_3 \,\Big(-5\, r^5\, U_2 + U_3 + 
		r\, \big[-5 + r\, \big(10 + r\, [-10 + r\, (5 + 4 \,r)]\big)\big]\, U_3\Big)\bigg]\, z^7 \nonumber \\
		&+& 
		U_3^3\, \bigg[6\, r^5\, U_1\, U_2 + 
		12\, \big[-5 + r\, \big(5 + (-5 + r)\, r\big) \,\big(5 + r\, (-5 + 4\, r)\big)\big]\, U_1\, U_3 \nonumber \\
		&-& \big[-180 + 
		r\, \big(316 + r\, (-197 + 49\, r)\big)\big]\, U_1^2\, U_3 + \Big(1 + 
		r\, \big[-5 + r\, \big(10 + r \,[-10 + r\, (5 + 34\, r)]\big)\big]\Big)\, U_3^3\bigg]\, z^8 \nonumber \\
		&- &
		U_3^3\, \bigg[ U_1\, \big[30\, (-9 + 14\, U_1) + 
		r \,\big(1350 - 980 \,U_1 + 
		r\, [-2700 + r \,(2700 + 2 \,r \,(-675 + 134\, r) - 201\, U_1) \nonumber \\
		&+ &
		763\, U_1]\big)\big] 
		+ 11\, U_3^2 + 
		5\, r \,\big[-11 + r\, \big(22 + r\, [-22 + r\, (11 + 2\, r)]\big)\big] \,U_3^2\bigg] \,z^9 \nonumber \\
		& + &
		U_3^3 \,\bigg[55 - 330 \,r + 825\, r^2 - 1100\, r^3 + 825 \,r^4 - 323\, r^5 + 
		48\, r^6 + 7 \,\big[90 + r \,(-148 + 61\, r)\big]\, U_1^2 - 720\, U_1\, U_3^4\bigg]\, z^{10} \nonumber \\
		&-& 
		U_3^2\, \bigg[165 + 630\, U_1^2 + 
		r\, \Big(r \,\Big[2475 + r\, \big(-3300 + r\, [2475 + r \,(-989 + 164 \,r)]\big) + 
		521\, U_1^2\big] - 2\, (495 + 574\, U_1^2)\Big) \nonumber \\
		&-& 1260 \,U_1\, U_3^4\bigg]\, z^11 + 
		2\, U_3^2\,\bigg[ (210 - 194 \,r) U_1^2 - 756\, U_1\, U_3^3 + 165\, U_3^5\bigg] \,z^{12} + 
		2\, U_3\, \bigg[(-90 + 88\, r)\, U_1^2 \nonumber \\
		& +& 630\, U_1\, U_3^3 - 231\, U_3^5\bigg]\, z^{13} + 
		3\, U_3\, \bigg[15 \,U_1^2 - 240\, U_1\, U_3^2 + 154\, U_3^4\bigg]\, z^{14} - 
		5\,\bigg[U_1^2 - 54\, U_1\, U_3^2 + 66\, U_3^4\bigg]\, z^{15} \nonumber \\
		& +& 
		15\, U_3 \,\bigg[-4\, U_1 + 11\, U_3^2\bigg]\, z^{16} + \bigg[6\, U_1 - 55\, U_3^2\bigg]\, z^{17} + 11 \,U_3\, z^{18} -
		z^{19},
	\end{eqnarray}
\begin{eqnarray}
	S_1&=&3 r^3\, U_1\, U_2\, U_3^2 \,\bigg[1188\, z^7 + 
	U_2\, U_3\, (U_3 - z)\, z\, (48\, U_3^3 - 240\, U_3^2\, z + 295\, U_3\, z^2 - 295\, z^3)\nonumber \\
	& + &
	U_1 \,U_3 \,\big[16\, (U_2 - 3\, U_3)\, U_3^3 - 
	80\, (U_2 - 3 \,U_3)\, U_3^2\, z + (52\, U_2 - 297\, U_3) \,U_3\, z^2 - 
	9\, (4\,U_2 + 15\, U_3) \,z^3 + 834 \,z^4\big]\bigg] \nonumber \\
	&-& 
	96\, r\, U_3\, z^2 \,\bigg[2\, U_2\, (U_3 - z)^5 \,z^3\, (7\, U_3^2 + 15\, z^2) + 
	U_1\, U_2 \,(U_3 - z)^4\, z \,(2 \,U_3^3 + 28\, U_3^2\, z + 15 \,U_3 \,z^2 + 42\, z^3)\nonumber \\
	& + &
	U_1^2\, \big[2\, U_2 \,U_3^5 + 8\, (U_2 - U_3)\, U_3^4\, z + 
	U_3^3\, (-33\, U_2 + 16\, U_3)\, z^2 + U_3^2\, (55\, U_2 + 32\, U_3)\, z^3 - 
	U_3\, (77\, U2 + 216\, U_3) \,z^4\nonumber \\
	& +&
	 3 \,(192 + 23\, U_2)\, U_3 \,z^5 - 
	12\, (2\, U_2 + 33\, U_3)\, z^6 + 388 \,z^7\big]\bigg]\nonumber \\
	& -& 
	384\, z^4\, \bigg[(U_3 - z)^7 \,z^2\, (U_3^2 + 3\, z^2) + 
	2\, U1\, (U_3 - z)^6\, z\, (2\, U_3^2 - U_3\, z + 6 \,z^2)\nonumber \\
	& + &
	U_1^2\, (3\, U_3^5 - 17 \,U_3^4 \,z + 49\, U_3^3\, z^2 - 9 \,z^7 + 
	5\, U_3^2\, z^3\, (-19 + 25\, z) + U_3\, z^5\, (-103 + 47\, z))\bigg] \nonumber \\
	&+ &
	r^2\, U_3^2\, z\, \bigg[-384\, U2^2\, (U_3 - z)^3\, z^3\, (U_3^2 + 3\, z^2) + 
	U_1^2\, \Big(-32 \,U_2^2 \,U_3\, (U_3 - z)^2\, (2 \,U_3 + 15\, z)\nonumber \\
	& +& 
	3\, U_2\, \big[16\, U_3^4\, (3 + 4\, U_3) - 16 \,U_3^3\, (21 + 16\, U_3)\, z + 
	U_3^2 \,(825 + 1568\, U_3)\, z^2 - 3\, U_3 \,(377 + 1184\, U3)\, z^3 \nonumber \\
	&+&
	4667\, U3\, z^4 - 3337 \,z^5\big] - 
	384\, z\, \big[U_3^5 - 3\, U_3^4 \,z + 5\, U_3^3 \,z^2 - 7\, U_3^2\, z^3 + 
	16\, U_3 \,z^4 - 52\, z^5\big]\Big)\nonumber \\
	& + &
	3 \,U_1\, U_2\, z\, \Big(-32\, U_2\, (U_3 - z)^2\, (2\, U_3^3 + 23\, U_3\, z^2 - 18\, z^3)\nonumber \\
	& + &
	3 \,\big[16\, U_3^6 - 128\, U_3^5\, z + 387\, U_3^4\, z^2 - 652\, U_3^3 \,z^3 + 
	674\, U_3^2\, z^4 + 99\, (-4 + z)\, z^5\big]\Big)\bigg],
\end{eqnarray}
\begin{eqnarray}
  S_2&=&-57\, U_4^2\, z\, (U_1 + z - z^2) + 57\, r^3\, (U_1 + z^2)\nonumber \\
  & + &
  r^2 \,\bigg[-114\, U_1 + z \,(55\, U_2 \,U_3 - 171\, U_4\,z)\bigg] + 
  r \bigg[-20\, U_1\,U_2\, U_3 + U_4\, z\, (-55\, U_2\, U_3 + 171 \,U_4\, z)\nonumber \\
  & + &
   57\, U_1\, (1 + z - 2\, z^2)\bigg],	
\end{eqnarray}
\begin{eqnarray}
	S_3&=&	-3\, r^4\, U_1 \,U_2\, U_3^4\, \bigg[16\, U_3^4 - 96\, U_3^3 \,z + 211\, U_3^2\, z^2 - 262\, U_3\, z^3 + 
	163\, z^4\bigg]\nonumber \\
	& -& 
	64\, r^2\, U_1\, U_3^2\, z^3\, \bigg[U_3^3\, (U_2^2 + 10\, U_2\, U_3 - 4\, U_3^2) - 
	4 \,U_3^2\, (U_2^2 + 7\, U_2 \,U_3 - 2\, U_3^2)\, z + 
	U_3\, (6 \,U_2^2 + 43 \,U_2\, U_3 - 8 \,U_3^2) \,z^2\nonumber \\ &+& 
	U_3\, \big[-U_2\, (55 + 4\, U_2) + 32\, U_3\big]\, z^3 + (U_2^2 - 92\, U_3 + 
	47\, U_2\, U_3) \,z^4 + 4\, (4\, U_2 + 15\, U_3)\, z^5 - 128\, z^6\bigg]\nonumber \\
	& -& 
	32\, r^3\, U_1 \,U_3^3\, z\, \bigg[-2\, U_2\, U_3^5 + U_3^3\, (5\, U_2^2 + 6\, U_2\, U_3 + 8\, U_3^2) \,z - 
	U_3^2\, (U_2\, (-3 + 17\, U_2) + 45\, U_2\, U_3 + 32\, U_3^2)\, z^2\nonumber \\
	& +& 
	U_3\, \big[U_2\, (-6 + 23\, U_2) + 124\, U_2 \,U_3 + 72\, U_3^2\big] \,z^3 - (17\, U_2^2 + 
	174\, U_2 \,U_3 + 88\, U_3^2) \,z^4 + 2\, (89\, U_2 + 8\, U_3)\, z^5 + 72\, z^6\bigg] \nonumber \\
	&+& 
	64 \,r\, U_3\, z^4\, \bigg[U_2\, (U_3 - z)^5 \,z\, (7\, U_3^2 + 15\, z^2) + 
	U_1\, U_2\, \big[7\, U_3^4 - 22\, U_3^3 \,z + U_3^2\, (27 - 28\, z)\, z^2 + 
	U_3 \,(37 - 30\, z)\, z^4 + 9 \,z^6\big] \nonumber \\
	&-& 
	4\, U_1\, \big[U_3^5 - 2\, U_3^4\, z - 5\, U_3^3\, z^2 + U_3^2\, (29 - 41\, z)\, z^3 + 
	U_3\, (89 - 51\, z)\, z^5 + 43\, z^7\big]\bigg]\nonumber \\
	& +& 
	256\, z^5\, \bigg[(U_3 - z)^7 \,z\, (U_3^2 + 3 \,z^2) + 
	U_1 \,\big[U_3^5 - 6 \,U_3^4 \,z + 2\, U_3^3\, (9 - 19\, z)\, z^2 + 
	6\, U_3^2\, (10 - 11 \,z)\, z^4\nonumber \\
	& +& 2\, U_3\, (23 - 9\, z) \,z^6 + 3\, z^8\big]\bigg],
\end{eqnarray}
\begin{eqnarray}
	S_4&=&-3\, r^6\, U_1 \,U_2\, U_3^5\, z \,\bigg[35\, U_3^2 - 169\, U_3 \,z + 364 \,z^2\bigg]\nonumber \\
	& + &
	r^5\, U_1\, U_2\, U_3^3\, \bigg[32\, U_3^5\, (-3\, U_2 + 4\, U_3) + (153 + 416\, U_2 - 
	768\, U_3)\, U_3^4\, z \nonumber \\
	&+ &
	3\, U_3^3\, (-281 - 192\, U_2 + 576\, U_3)\, z^2 + 
	3\, (671 + 96\, U_2 - 640\, U_3)\, U_3^2\, z^3\nonumber \\
	& + &
	4\, U_3\, (40\, U_2 + 199\, U_3) \,z^4 - 
	4645 \,U_3\, z^5 - 585\, z^6\bigg] + 
	256 \,z^6\, \bigg[(U_3 - z)^8 \,z\, (U_3^2 + 3\, z^2)\nonumber \\ &+& 
	U_1\, \big[U_3^6 - 7\, U_3^5 \,z + 24\, U_3^4\, z^2 - 3 \,z^9 + 
	14\, U_3^3\, z^3\,(-4 + 7 z) + 14\, U_3^2\, z^5\, (-9 + 8\, z) + 
	U_3\, z^7\, (-64 + 21\, z)\big]\bigg]\nonumber \\
	& - &
	r^4\, U_1\, U_2 \,U_3^3\, z\, \bigg[16 \,U_3^3\, (-21 + 4\, U_2 - 48\, z) \,z - (873 + 
	256\, U_2)\, z^5\nonumber \\
	& +&
	 16 \,U_3^4\, (3 + 8 \,z) + 
	U_3^2 \,z^2 \,(921 - 192 \,U_2 + 3413\, z) + 
	U_3\, z^3 \,\big[32\, U_2\, (7 + 5\, z) + z\, (-13733 + 10656\, z)\big]\bigg]\nonumber \\
	& + &
	32\, r^3 \,U_1\, U_3^3\, z^2\, \bigg[-8 \,z\, (U_3^6 - 5 \,U_3^5\, z + 13\, U_3^4\, z^2 - 
	20\, U_3^3\, z^3 + 13\, U_3^2\, z^4 + 7 \,U_3\, z^5 - 24\, z^6) \nonumber \\
	&+& 
	U_2\, \big[2 \,U_3^4 - 12\, U_3^3\, z + 108\, U_3^2\, z^2 + 
	U_3\, z^3\, (-338 + 483\, z - 372\, z^2) + 3\, z^4\, (-3 + 92\, z^2)\big]\bigg]\nonumber \\
	& - &
	64 \,r^2\, U_1 \,U_3^2\, z^4\, \bigg[-4\, \big[U_3^6 - 3\, U_3^5\, z + 4\, U_3^4\, z^2 - 
	10 \,U_3^3\, z^3 + 31\, U_3^2\, z^4 - 83\, z^7 + U_3\, z^5 \,(-79 + 47\, z)\big]\nonumber \\
	& + &
	U_2\, \Big(10 \,U_3^3 + U_3\, (74 - 31\, z)\, z^4 - 46\, z^6 - 
	2\, U_3^2\, z\, \big[12 + 7\, z\, (-3 + 7 \,z)\big]\Big)\bigg]\nonumber \\
	& +& 
	64\, r\, U_3\, z^5\, \bigg[U_2\, (U_3 - z)^6\, z\, (7 \,U_3^2 + 15\, z^2) - 
	4\, U_1\, \big[U_3^6 - 3\, U_3^5\, z - 3\, U_3^4\, z^2 + 34\, U_3^3\, z^3 - 61\, z^8\nonumber \\
	& +& 
	2\, U_3\, z^6\, (-103 + 47\, z) + 2\, U_3^2 \,z^4 \,(-54 + 65\, z)\big] + 
	U_1\, U_2\, \Big(-9\, z^7 + 5 \,U_3^2\, z^3\, (-11 + 13\, z) + U_3\, z^5 (-67 + 39\, z)\nonumber \\
	& + &
	U_3^3\, \big[7 + z\, (-29 + 49\, z)\big]\Big)\bigg],
\end{eqnarray}
\begin{eqnarray}
S_5&=&r\, U_3^2\, \bigg[-U_2 + U_3\bigg] - 
U_3 \,\bigg[U_2 - 2\, r\, U_2 + 2\, r\, U_3\bigg]\, z + \bigg[-2 + (7 - 5\, r)\, r + 
2\, U_2\, U_3\bigg]\, z^2 - \bigg[U_2 - 2\, U_3\bigg]\, z^3,
\end{eqnarray}
\begin{eqnarray}
	S_6&=&r^3 \,U2\, U3^6 + \bigg[1 - 6\, r\bigg]\, r^2\, U2 \,U3^5\, z + 
	2\, r^2\, \bigg[-3 + 7\, r\bigg]\, U2\, U3^4 \,z^2 + 
	r\, U3^3\, \bigg[U2 + (11 - 16\, r)\, r\, U2 - 2\, U3^3\bigg]\, z^3 \nonumber \\
	&+& 
	U3 \,\bigg[(1 + r \,(-7 + r + 3 \,r^2))\, U2\, U3 + 10\, r\, U3^4\bigg]\, z^4 + 
	U3^3\, \bigg[(-5 + 13\, r)\, U2 + 4\, (1 - 7\, r) \,U3\bigg]\, z^5\nonumber \\
	& - &
	2\, \bigg[(-5 + 2\, (7 - 5\, r)\, r)\, U2\, U3 + 8\, (1 - 3\, r)\, U3^3\bigg]\, z^6 + 
	U3\, \bigg[(-10 + 13\, r)\, U2 + (24 - 46\, r) \,U3\bigg] \,z^7\nonumber \\
	 &+&
	  \bigg[-16 + 38\, r - 22\, r^2 + 
	5\, U2\, U3\bigg]\, z^8 - \bigg[U2 - 4\, U3\bigg]\, z^9,
\end{eqnarray}
\begin{eqnarray}
S_7&=&r^3\, U_2\, U_3^5 + \bigg[1 - 6\, r\bigg]\, r^2\, U_2\, U_3^4\, z + 
2\, r^2\, \bigg[-3 + 7\, r\bigg]\, U_2\, U_3^3\, z^2 - 
r\, U_3^2 \,\bigg[r \,(-13 + 17\, r) \,U_2 + U_3^3\bigg]\, z^3\nonumber \\
& +& 
r\, U_3\, \bigg[2\, r\, (-5 + 4\, r)\, U_2 + 5\, U_3^3\bigg]\, z^4 + 
2\, U_3^2\, \bigg[r\, (U_2 - 7 \,U_3) + U_3\bigg]\, z^5 + 
2 \,\bigg[r\, (-2 + 3\, r)\, U_2 + 4 \,(-1 + 3\, r)\, U_3^2\bigg]\, z^6 \nonumber \\
&+& \bigg[2\, r\, U_2 + 12\, U_3 - 
23\, r\, U_3\bigg]\, z^7 + \bigg[-8 + 11\, r\bigg]\, z^8 + 2\, z^9.
\end{eqnarray}
Here, we  have utilized the following short-hand symbolizations: 
\begin{eqnarray}
	K &=&\frac{U_1}{U_3\,r\,z} \Big((m_{Q}^2 \, U_3 + m_{Q'}^2 \, r \Big),\nonumber\\
	D &=&\frac{U_3}{U_1^2} \Big((m_{Q}^2 \, U_3 + m_{Q'}^2 \, r)\,U_1 + s \,r\,z\,U_3 \Big),\nonumber\\
	U_1&=& r^2 + r\, (-1 + z) + (-1 + z) \, z,\nonumber\\
	U_2&=&1 -r - z,\nonumber\\
	U_3&=&1 -r,\nonumber\\
	U_4&=&1 -z.
\end{eqnarray}
%


\end{document}